\documentclass[a4paper,11pt]{article}
\pdfoutput=1 

\usepackage{jheppub} 
\usepackage{physics}
\usepackage[T1]{fontenc} 
\usepackage{appendix}
\usepackage{amsmath, amssymb, graphicx}
\usepackage{amsfonts}
\usepackage{stmaryrd}
\usepackage{mathtools}
\usepackage{float}
\usepackage{amsthm, color, xcolor, hyperref, slashed, subfigure, ytableau}
\usepackage{bbm}
\ytableausetup{boxsize=0.8em}

\def \bal#1\eal  {\begin{align} #1 \end{align}}
\def \bga#1\ega  {\begin{gather} #1 \end{gather}}
\def\({\left(}
\def\){\right)}
\def\[{\left[}
\def\]{\right]}
\def\<{\left\langle}
\def\>{\right\rangle}

\newcommand{\bim} {\begin{itemize}[noitemsep]}

\newcommand{\eim}{\end{itemize}}
\newcommand{\beq} {\begin{equation}}
\newcommand{\eeq} {\end{equation}}
\newcommand{\bc}{\begin{center}}
\newcommand{\ec}{\end{center}}

\title{Rank polarization and dynamical autonomy in a projective rotor model}

\author[a, b]{Shannon Wang,}
\author[b]{David Wang}

\affiliation[a]{Department of Physics, University of California, Santa Barbara, CA 93106, USA}
\affiliation[b]{Kuaige, Hacienda Heights, CA 91745, USA}

\emailAdd{shannonwang@physics.ucsb.edu}
\emailAdd{david.wang@kuaige.org}

\abstract{
We study how a higher-rank classical collective order parameter can select one dominant projective eigenline and, in a $p=2$ realization, support a dynamically autonomous low-energy director sector.  In a fully connected projective-rotor model, the collective second moment is a positive semidefinite tensor $T$ of fixed trace.  Conventional projective ordering drives its effective rank toward one, selecting a single dominant eigenline.  Rank polarization alone, however, does not determine dynamical autonomy: extrinsic-curvature criteria depend on the chosen state-space metric, and the equilibrium tensor does not determine the transverse free-energy Hessian.  A two-harmonic model makes this logical independence explicit by producing families with identical $T$ and $r_{\rm eff}$ but different transverse quadratic forms.  We then construct rotationally invariant locking interactions that leave a prescribed ordered tensor exactly fixed while increasing every non-Goldstone restoring coefficient as $\Theta(\lambda)$ within a globally controlled stable regime.  For a phenomenological inertial realization with a finite positive kinetic matrix, the lowest transverse normal-mode frequency grows as $\Theta(\lambda^{1/2})$; more generally, the low-frequency transverse susceptibility is suppressed as $O(\lambda^{-1})$, including in overdamped dynamics.  The construction therefore proves that spectral selection and infrared transverse suppression are independent data.  For $p>2$ the selected eigenline has a $(p-1)$-dimensional director manifold; the one-dimensional Goldstone sector used in the dynamical construction is specific to $p=2$.  Promoting the selected internal eigenline to a physical clock requires additional soldering and integrability structure.
}

\begin{document} 
\maketitle
\flushbottom

\section{Introduction}
\label{sec:introduction}

Time occupies an asymmetric position in fundamental physics.  In ordinary
quantum mechanics it enters as an external parameter, whereas in general
relativity temporal structure belongs to the dynamical spacetime geometry.
This tension underlies the canonical problem of time and has motivated
relational and emergent approaches in which evolution is reconstructed from
correlations, internal clocks, modular flow, or other dynamical degrees of
freedom
\cite{Isham1993,UnruhWald1989,PageWootters1983,Rovelli1990,
ConnesRovelli1994,HohnSmithLock2021}.
A logically prior question is less often isolated: even if time is not
fundamental, why should its effective replacement contain only one
distinguished temporal direction?

Many frameworks introduce a preferred temporal structure directly.
Einstein--aether theory contains a dynamical timelike vector
\cite{JacobsonMattingly2001}, while Newton--Cartan and
Ho\v{r}ava--Lifshitz constructions employ a clock one-form or preferred
foliation
\cite{ChristensenHartongObersRollier2014,HartongObers2015}.
Preferred directions may also arise through spontaneous Lorentz breaking or
condensate backgrounds
\cite{KosteleckySamuel1989,ArkaniHamedEtAl2004}.  More directly, spontaneous
time--space asymmetry has been explored as an order-parameter phenomenon
\cite{Wetterich2005,Wetterich2022}.  Matrix-model approaches to emergent
spacetime provide a related spectral diagnostic: the eigenvalues of a
positive moment-of-inertia tensor are used to determine how many directions
become dynamically extended
\cite{KimNishimuraTsuchiya2012,Nishimura2012}.
Here we ask a different question: can collective ordering select a single
dominant internal eigenline, and can the remaining non-Goldstone collective
deformations become dynamically inaccessible at low energy?

We study a minimal realization based on the positive semidefinite tensor
\begin{equation}
    T^{AB}
    =
    \frac1N\sum_{i=1}^{N}u_i^A u_i^B,
    \qquad
    u_i\sim-u_i,
    \label{eq:intro-T}
\end{equation}
with projective directors $u_i\in\mathbb{RP}^{p-1}$ and
$\Tr T=1$.  The indices $A,B$ are internal collective indices, not spacetime
indices.  We therefore do not assume at this stage a clock one-form,
Lorentzian signature, causal cone, or arrow of time.

We characterize the spectrum of $T$ by the inverse participation ratio
\begin{equation}
    r_{\rm eff}
    =
    \frac{(\Tr T)^2}{\Tr T^2}.
    \label{eq:intro-reff}
\end{equation}
Strong polarization corresponds to
\begin{equation}
    T^{AB}
    \simeq
    \Lambda\,\tau^A\tau^B,
    \qquad
    r_{\rm eff}-1\ll1.
    \label{eq:intro-rank-one}
\end{equation}
Because $T$ is invariant under $\tau\to-\tau$, this selects a projective line
rather than an oriented vector.

The ordering mechanism itself is conventional.  Fully connected projective
rotors are the mean-field analogue of familiar nematic ordering models
\cite{MaierSaupe1958,MaierSaupe1960,LebwohlLasher1972,BallMajumdar2010},
and their large-$N$ saddle is a Bingham distribution
\cite{Bingham1974}.  In the exactly solvable $p=2$ case the ordered branch
appears continuously at $\kappa_c=4$ and approaches
$r_{\rm eff}=1$ as $\kappa\to\infty$.  Thus spectral polarization provides a
controlled mechanism for selecting one dominant tensor direction.

The central point of this paper is that spectral polarization is not the
same as dynamical autonomy.  A natural first attempt is to diagnose autonomy
through the extrinsic geometry of the ordered orbit in a state-space metric.
Fisher geometry supplies a canonical statistical metric on parametric
families \cite{Efron1975,AmariNagaoka2000}, while the Bures metric supplies a
natural geometry on positive trace-one matrices
\cite{Hubner1992,BraunsteinCaves1994}.  For the same strongly polarized
$p=2$ family, however, Hilbert--Schmidt, Fisher, and Bures metrics give
qualitatively different geodesic-curvature limits.  Moreover, the absolute
Fisher curvature can be driven to zero by extensivity alone.  Extrinsic
flattening therefore cannot define autonomy unless the relevant metric is
itself derived from the physical dynamics.

This shifts the question from geometry to response.  Effective
one-dimensionality requires two independent conditions,
\begin{equation}
    \boxed{
    r_{\rm eff}-1\ll1,
    \qquad
    \frac{\Omega_{\rm IR}}{\Delta_\perp}\ll1,
    }
    \label{eq:intro-two-small}
\end{equation}
where $\Delta_\perp$ is the lowest physical excitation frequency transverse
to the ordered orbit.  The first condition selects a direction; the second
determines whether the remaining collective directions participate in
infrared dynamics.

To demonstrate their independence, we introduce a two-harmonic projective
rotor.  Its first moment $M_1$ fixes $T$, while a second moment $M_2$ changes
the physical fluctuation Hessian without changing the tensor spectrum.
We then add rotationally invariant locking interactions whose value and
first variation vanish on a prescribed ordered orbit.  Along a positive
strong-locking trajectory,
\begin{equation}
    T_\star=\mathrm{constant},
    \qquad
    r_{\rm eff}(T_\star)=\mathrm{constant},
    \qquad
    K_\perp=\lambda L_\perp+O(1),
    \quad
    L_\perp>0.
    \label{eq:intro-static-summary}
\end{equation}
For a finite positive nonsingular normal kinetic matrix this implies
\begin{equation}
    \boxed{
    \Delta_\perp=\Theta(\lambda^{1/2}),
    \qquad
    \sup_{|\nu|\leq\Omega_{\rm IR}}
    \left\|
        \widehat{\mathcal G}_\perp(\nu)
    \right\|
    =
    O(\lambda^{-1}),
    }
    \label{eq:intro-dynamic-summary}
\end{equation}
and tangent-induced normal leakage is likewise $O(\lambda^{-1})$ for fixed
low-frequency driving.  Polarization therefore selects the direction,
whereas an independently tunable transverse gap protects it dynamically.

The temporal interpretation is deliberately limited.  The construction
establishes spectral selection of one internal eigenline and, in the $p=2$
realization studied dynamically below, a one-dimensional director orbit with
parametrically suppressed non-Goldstone response.  The microscopic evolution
parameter remains external to this construction: motion of the director as a
function of that parameter is not motion ``along'' the selected eigenline.
Promoting the eigenline to a physical clock therefore requires additional
soldering and integrability conditions; orienting it into an arrow of time
requires further nonequilibrium structure.

The paper is organized as follows.  Section~\ref{sec:polarization} develops
the projective ordering model and its rank polarization.
Section~\ref{sec:geometry} shows why state-space extrinsic geometry alone
does not furnish a metric-independent autonomy criterion.
Section~\ref{sec:twoharmonic} derives the physical even and odd fluctuation
Hessians of the two-harmonic model.
Section~\ref{sec:locking} constructs a stable fixed-polarization family with
parametrically increasing transverse stiffness.
Section~\ref{sec:dynamics} derives the resulting gap and low-frequency
response.  Section~\ref{sec:discussion} discusses the interpretation,
limitations, and additional structures required for a physical clock.


\section{Projective ordering and rank polarization}
\label{sec:polarization}

The first ingredient required for an effectively one-direction collective
sector is spectral polarization.  In this section we isolate this equilibrium
problem from the dynamical question studied later.  We consider $N$
projective directors
\begin{equation}
    u_i\in\mathbb{RP}^{p-1},
    \qquad
    u_i\sim-u_i,
\end{equation}
represented by unit vectors $u_i\in S^{p-1}$ modulo sign, and define the
empirical second-moment tensor
\begin{equation}
    T_N^{AB}
    =
    \frac{1}{N}
    \sum_{i=1}^N
    u_i^A u_i^B.
    \label{eq:empirical-T}
\end{equation}
By construction,
\begin{equation}
    T_N\succeq0,
    \qquad
    \Tr T_N=1.
    \label{eq:T-positive-trace}
\end{equation}
The eigenvalue spectrum of $T_N$ therefore provides a natural measure of how
many collective directions participate appreciably in the state.

\subsection{Collective tensor and effective rank}
\label{subsec:effective-rank}

Let the eigenvalues of $T$ be
\begin{equation}
    \lambda_1,\ldots,\lambda_p,
    \qquad
    \lambda_a\geq0,
    \qquad
    \sum_{a=1}^p\lambda_a=1.
\end{equation}
We characterize their spectral participation by
\begin{equation}
    \boxed{
    r_{\rm eff}(T)
    =
    \frac{(\Tr T)^2}{\Tr T^2}
    =
    \frac{1}{\sum_{a=1}^p\lambda_a^2}.
    }
    \label{eq:reff-def}
\end{equation}
This is the inverse participation ratio of the eigenvalue spectrum.  It
satisfies
\begin{equation}
    1\leq r_{\rm eff}\leq p,
    \label{eq:reff-bounds}
\end{equation}
with
\begin{equation}
    r_{\rm eff}=1
    \quad\Longleftrightarrow\quad
    \rank T=1,
\end{equation}
and
\begin{equation}
    r_{\rm eff}=p
    \quad\Longleftrightarrow\quad
    T=\frac1p\mathbbm{1}.
\end{equation}
Thus $r_{\rm eff}$ interpolates continuously between a rank-one tensor and
the isotropic configuration.  We use it only as a spectral participation
measure; it should not itself be identified with the dimension of a
spacetime or configuration manifold.

For a uniaxial state with director $n$,
\begin{equation}
    T
    =
    \frac1p\mathbbm{1}
    +
    S\left(
        nn^{\mathsf T}
        -
        \frac1p\mathbbm{1}
    \right),
    \label{eq:uniaxial-T}
\end{equation}
the eigenvalues are
\begin{equation}
    \lambda_\parallel
    =
    \frac{1+(p-1)S}{p},
    \qquad
    \lambda_\perp
    =
    \frac{1-S}{p},
    \label{eq:uniaxial-eigenvalues}
\end{equation}
where $\lambda_\perp$ has multiplicity $p-1$.  Hence
\begin{equation}
    \boxed{
    r_{\rm eff}(S)
    =
    \frac{p}{1+(p-1)S^2}.
    }
    \label{eq:reff-uniaxial}
\end{equation}
The strong-ordering limit $S\to1$ is therefore precisely the
effective-rank-one limit
\begin{equation}
    r_{\rm eff}\to1.
\end{equation}

At finite $N$, an exactly $O(p)$-invariant Gibbs state with no
symmetry-breaking field satisfies
\begin{equation}
    \langle T_N\rangle
    =
    \frac1p\mathbbm{1},
    \label{eq:finiteN-isotropic}
\end{equation}
even when typical configurations are strongly ordered.  Throughout the
paper, polarization therefore refers to the empirical tensor, its
thermodynamic eigenvalue distribution, or the usual symmetry-broken
quasi-average state.  The corresponding order of limits is reviewed in
Appendix~\ref{app:meanfield}.

\subsection{Mean-field projective rotor and the Bingham saddle}
\label{subsec:meanfield-projective}

As a minimal mechanism for generating spectral polarization, consider the
fully connected projective rotor Hamiltonian
\begin{equation}
    H_0
    =
    -\frac{J}{2N}
    \sum_{i,j=1}^N
    (u_i\cdot u_j)^2.
    \label{eq:projective-H}
\end{equation}
Since
\begin{equation}
    \sum_{i,j}
    (u_i\cdot u_j)^2
    =
    N^2\Tr T_N^2,
\end{equation}
this may be written
\begin{equation}
    \boxed{
    H_0
    =
    -\frac{JN}{2}\Tr T_N^2.
    }
    \label{eq:H-T}
\end{equation}
The interaction therefore favors concentration of the spectrum of $T_N$.

This is the fully connected version of familiar nematic ordering models,
used here as a controlled benchmark rather than as a novel ordering
mechanism
\cite{MaierSaupe1958,MaierSaupe1960,LebwohlLasher1972,
BallMajumdar2010}.  With
\begin{equation}
    \kappa=\beta J,
\end{equation}
the large-$N$ variational free energy for a normalized one-site density
$\rho(u)$ is
\begin{equation}
    \mathcal F[\rho]
    =
    \int d\mu(u)\,
        \rho(u)\log\rho(u)
    -
    \frac{\kappa}{2}
    \Tr T[\rho]^2,
    \label{eq:variational-free-energy}
\end{equation}
where
\begin{equation}
    T[\rho]
    =
    \int d\mu(u)\,
        \rho(u)\,uu^{\mathsf T}.
    \label{eq:T-rho}
\end{equation}
Its stationary points are Bingham distributions,
\begin{equation}
    \boxed{
    \rho_\star(u)
    =
    \frac{
        \exp\!\left[
            \kappa\,u^{\mathsf T}T_\star u
        \right]
    }{
        Z(T_\star)
    },
    \qquad
    T_\star=T[\rho_\star].
    }
    \label{eq:Bingham-saddle}
\end{equation}
The variational derivation, uniaxial reduction, and equivalent
rate-function formulation are given in
Appendix~\ref{app:meanfield}.

The $p=2$ case is analytically transparent and will serve as our principal
equilibrium benchmark.

\subsection{The exactly solvable \texorpdfstring{$p=2$}{p=2} model}
\label{subsec:p2}

For $p=2$, write
\begin{equation}
    u=(\cos\theta,\sin\theta)
\end{equation}
and introduce the doubled projective angle
\begin{equation}
    x=2\theta,
    \qquad
    \psi=2\phi,
\end{equation}
where $\phi$ denotes the director angle.  The collective tensor takes the
form
\begin{equation}
    T
    =
    \frac12
    \begin{pmatrix}
        1+m\cos\psi & m\sin\psi\\
        m\sin\psi   & 1-m\cos\psi
    \end{pmatrix}.
    \label{eq:p2-T}
\end{equation}
Its eigenvalues are
\begin{equation}
    \lambda_\pm
    =
    \frac{1\pm m}{2},
    \label{eq:p2-eigenvalues}
\end{equation}
so that
\begin{equation}
    \boxed{
    r_{\rm eff}
    =
    \frac{2}{1+m^2}.
    }
    \label{eq:p2-reff}
\end{equation}

The one-site distribution is the von Mises distribution
\begin{equation}
    \rho(x|r,\psi)
    =
    \frac{
        \exp[r\cos(x-\psi)]
    }{
        2\pi I_0(r)
    },
    \label{eq:vonMises}
\end{equation}
with
\begin{equation}
    r=\frac{\kappa m}{2}.
    \label{eq:r-kappa-m}
\end{equation}
Defining
\begin{equation}
    A(r)
    =
    \frac{I_1(r)}{I_0(r)},
    \label{eq:A-def}
\end{equation}
the self-consistency condition becomes
\begin{equation}
    \boxed{
    m
    =
    A\!\left(
        \frac{\kappa m}{2}
    \right).
    }
    \label{eq:p2-self-consistency}
\end{equation}

A convenient mean-field auxiliary potential whose stationary points
reproduce Eq.~\eqref{eq:p2-self-consistency} is
\begin{equation}
    \boxed{
    f_\kappa(m)
    =
    -\frac{\kappa}{4}
    +
    \frac{\kappa m^2}{4}
    -
    \log I_0\!\left(
        \frac{\kappa m}{2}
    \right).
    }
    \label{eq:p2-free-energy}
\end{equation}
Its relation to the constrained moment free energy is discussed in
Appendix~\ref{app:meanfield}.  Expanding about the isotropic state,
\begin{equation}
\begin{split}
    f_\kappa(m)
    =
    -\frac{\kappa}{4}
    &+
    \frac{\kappa}{4}
    \left(
        1-\frac{\kappa}{4}
    \right)m^2
    +
    \frac{\kappa^4}{1024}m^4
\\
    &-
    \frac{\kappa^6}{36864}m^6
    +
    O(m^8).
\end{split}
    \label{eq:p2-Landau}
\end{equation}
The quadratic coefficient changes sign at
\begin{equation}
    \boxed{
    \kappa_c=4,
    }
    \label{eq:p2-kappac}
\end{equation}
and the positive quartic term gives a continuous mean-field transition.
Close to the transition,
\begin{equation}
    m^2
    =
    \frac{\kappa-4}{2}
    +
    O\!\left(
        (\kappa-4)^2
    \right).
    \label{eq:p2-near-critical}
\end{equation}

The strong-coupling regime is more important for the present application.
Solving Eq.~\eqref{eq:p2-self-consistency} asymptotically gives
\begin{equation}
    m
    =
    1
    -
    \frac{1}{\kappa}
    -
    \frac{3}{2\kappa^2}
    -
    \frac{9}{2\kappa^3}
    +
    O(\kappa^{-4}),
    \label{eq:p2-m-strong}
\end{equation}
and therefore
\begin{equation}
    \boxed{
    r_{\rm eff}-1
    =
    \frac{1}{\kappa}
    +
    \frac{2}{\kappa^2}
    +
    \frac{6}{\kappa^3}
    +
    O(\kappa^{-4}).
    }
    \label{eq:p2-reff-strong}
\end{equation}
Thus
\begin{equation}
    \kappa\to\infty
    \qquad\Longrightarrow\qquad
    r_{\rm eff}\to1.
    \label{eq:p2-rankone-limit}
\end{equation}

\begin{figure}[t]
    \centering
    \includegraphics[width=0.92\linewidth]{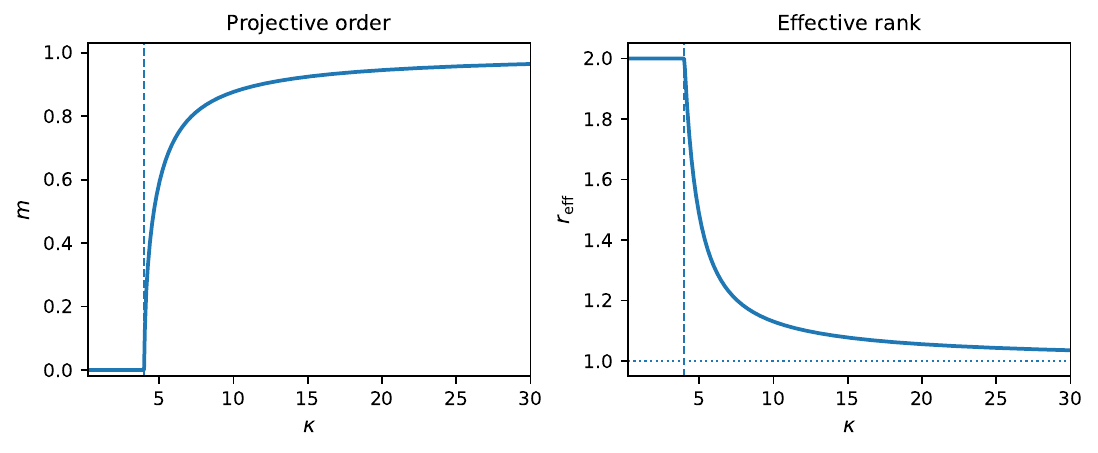}
    \caption{
    Rank polarization in the exactly solvable $p=2$ projective-rotor model.
    The order parameter $m$ turns on continuously at
    $\kappa_c=4$, while the effective rank decreases from
    $r_{\rm eff}=2$ in the isotropic phase toward
    $r_{\rm eff}=1$ at strong coupling.
    }
    \label{fig:rank-polarization-p2}
\end{figure}

The full crossover is shown in Fig.~\ref{fig:rank-polarization-p2}.

Exact rank one is not required at finite coupling.  The relevant statement
is that the subdominant eigenvalue can be made parametrically small.  Whether
the corresponding transverse degrees of freedom are also dynamically
inaccessible is a separate question.

\subsection{General dimension and the strong-polarization limit}
\label{subsec:general-p}

The $p=2$ transition is special, but the approach to effective rank one is
not.  For a uniaxial state define
\begin{equation}
    P_p(x)
    =
    \frac{
        px^2-1
    }{
        p-1
    },
    \qquad
    x=n\cdot u.
    \label{eq:Pp-def}
\end{equation}
Its first nontrivial isotropic moments are
\begin{equation}
    \left\langle
        P_p^2
    \right\rangle_0
    =
    \frac{
        2
    }{
        (p-1)(p+2)
    },
    \label{eq:P2-moment}
\end{equation}
and
\begin{equation}
    \boxed{
    \left\langle
        P_p^3
    \right\rangle_0
    =
    \frac{
        8(p-2)
    }{
        (p-1)^2(p+2)(p+4)
    }.
    }
    \label{eq:P3-moment}
\end{equation}
As shown in Appendix~\ref{app:meanfield}, these imply that the isotropic
saddle loses local stability at
\begin{equation}
    \boxed{
    \kappa_\ast
    =
    \frac{p(p+2)}{2}.
    }
    \label{eq:general-spinodal}
\end{equation}
The cubic invariant vanishes only for $p=2$.  For $p>2$ the mean-field
Landau functional generically contains a cubic term, so the isotropic
spinodal need not coincide with the coexistence transition.

For example, for $p=3$ the uniaxial coexistence point is
\begin{equation}
    \kappa_c
    \simeq
    6.81218849,
    \qquad
    S_c
    \simeq
    0.42902903,
    \label{eq:p3-coexistence}
\end{equation}
for which
\begin{equation}
    r_{\rm eff}(S_c)
    \simeq
    2.19277118.
    \label{eq:p3-reff-jump}
\end{equation}
The one-variable coexistence calculation is given explicitly in
Appendix~\ref{app:meanfield}.

At strong coupling, the uniaxial branch obeys
\begin{equation}
    S
    =
    1-\frac{p}{2\kappa}
    +
    O(\kappa^{-2}),
    \label{eq:general-S-strong}
\end{equation}
and hence
\begin{equation}
    \boxed{
    r_{\rm eff}-1
    =
    \frac{p-1}{\kappa}
    +
    O(\kappa^{-2}).
    }
    \label{eq:general-reff-strong}
\end{equation}
The derivation is given in Appendix~\ref{app:meanfield}.  Thus the
effective-rank-one limit is not peculiar to $\mathbb{RP}^1$: strong
projective ordering produces a single dominant tensor direction throughout
the uniaxial family.

We will not require a classification of the complete phase diagram for
general $p$, nor assume that every higher-dimensional saddle is globally
uniaxial.  The $p=2$ model supplies an analytically controlled benchmark,
while $p=3$ demonstrates that the polarization mechanism persists when the
microscopic projective space has more than one angular degree of freedom.

For general $p$, however, a rank-one tensor selects an \emph{eigenline}, not
a one-dimensional Goldstone manifold.  The orientation of that line lies in
$\mathbb{RP}^{p-1}$ and therefore carries $p-1$ Goldstone coordinates in the
symmetry-broken theory.  The dynamical construction in
Secs.~\ref{sec:twoharmonic}--\ref{sec:dynamics} specializes to $p=2$, where
$\mathbb{RP}^{1}$ is one dimensional.

\subsection{Polarization is not dynamical autonomy}
\label{subsec:polarization-not-autonomy}

The equilibrium analysis above establishes a limited but important result.
A conventional projective ordering interaction can generate a positive
semidefinite collective tensor for which
\begin{equation}
    r_{\rm eff}-1\ll1,
\end{equation}
or equivalently
\begin{equation}
    T
    \simeq
    \Lambda\,nn^{\mathsf T}
\end{equation}
up to parametrically small subleading eigenvalues.  The ordered phase
therefore selects one dominant projective direction.

Nothing in this statement determines the dynamical cost of departing from
the polarized sector.  The spectrum of $T$ specifies an equilibrium order
parameter, whereas transverse response is determined by the free-energy
Hessian together with the kinetic or dissipative dynamics.  These are
distinct data.

The distinction should not be confused with a claim that the undeformed
one-harmonic branch is dynamically soft.  When that branch is embedded in
the two-harmonic collective description of Sec.~\ref{sec:twoharmonic} and
endowed with the illustrative inertial normalization
$\mu_1=\mu_2=1$ used in Sec.~\ref{sec:dynamics}, the lowest generalized
normal frequency grows rapidly with polarization:
\begin{equation}
    \Delta_\perp
    \simeq
    2.18,\ 3.86,\ 17.55,\ 34.70
    \qquad
    \text{at}
    \qquad
    m=0.90,\ 0.95,\ 0.99,\ 0.995,
    \label{eq:base-gap-slaving}
\end{equation}
respectively.  In the strongly polarized regime this growth is approximately
proportional to $(r_{\rm eff}-1)^{-1}$.  Thus the simplest base model
correlates polarization with stiffness.  The result established below is
instead one of \emph{independence}: this correlation is not fixed by $T$,
and the transverse quadratic form can be changed while the equilibrium
tensor is held fixed.

It is useful to state this distinction schematically as
\begin{equation}
    \boxed{
    \text{spectral polarization}
    \neq
    \text{dynamical autonomy}.
    }
    \label{eq:polarization-not-autonomy}
\end{equation}
Rank polarization answers whether one collective eigenline is selected.  It
does not determine the cost of exciting non-Goldstone deformations of the
ordered state.

A natural first attempt is to formulate the second question geometrically:
does the ordered orbit become asymptotically geodesic in a natural
state-space metric as $r_{\rm eff}\to1$?  We show in the next section that
this criterion is not metric independent.  That failure motivates the
dynamical construction developed in the remainder of the paper.


\section{Why geometry alone does not define autonomy}
\label{sec:geometry}

Section~\ref{sec:polarization} established that projective ordering can
select one dominant eigen-direction of the collective tensor,
\begin{equation}
    T
    \simeq
    \Lambda\,nn^{\mathsf T},
    \qquad
    r_{\rm eff}-1\ll1.
\end{equation}
This is an equilibrium statement.  A separate question is whether the
ordered sector becomes dynamically self-contained: if the collective state
moves along its orientation direction, are transverse degrees of freedom
necessarily excited?

A tempting geometric criterion is to identify the ordered configurations
with a submanifold of a state space and ask whether that submanifold becomes
asymptotically geodesic as polarization increases.  For a one-dimensional
orientation orbit this amounts to asking whether its geodesic curvature, or
equivalently the norm of its second fundamental form, vanishes in the
rank-one limit.

We show below that this does not define a metric-independent notion of
autonomy.  For the same strongly polarized family, different natural
state-space metrics give qualitatively different extrinsic-curvature limits.
Moreover, even within Fisher geometry, the absolute curvature can be made
small merely by increasing the number of statistically independent copies.
The physical criterion used in the remainder of the paper must therefore be
dynamical rather than purely geometric.

\subsection{Fisher geometry of the ordered family}
\label{subsec:fisher-geometry}

Consider the exactly solvable $p=2$ family of
Sec.~\ref{subsec:p2}.  In terms of the doubled projective angle $x$ and
orientation coordinate $\psi$, the one-site equilibrium distribution is
\begin{equation}
    \rho(x|r,\psi)
    =
    \frac{
        e^{r\cos(x-\psi)}
    }{
        2\pi I_0(r)
    },
    \label{eq:fisher-vonmises}
\end{equation}
with
\begin{equation}
    m=A(r),
    \qquad
    A(r)=\frac{I_1(r)}{I_0(r)}.
    \label{eq:fisher-A}
\end{equation}
On the mean-field saddle, $r=\kappa m/2$, but in this section
$(r,\psi)$ are regarded simply as coordinates on the statistical family.

The Fisher metric of the von Mises family is
\cite{Efron1975,AmariNagaoka2000}
\begin{equation}
    \boxed{
    ds_F^2
    =
    A'(r)\,dr^2
    +
    rA(r)\,d\psi^2 ,
    }
    \label{eq:fisher-metric}
\end{equation}
where
\begin{equation}
    A'(r)
    =
    1-\frac{A(r)}{r}-A(r)^2.
    \label{eq:A-prime}
\end{equation}
A derivation from the score functions is given in
Appendix~\ref{app:geometry}.

For a diagonal metric
\begin{equation}
    ds^2=E(r)\,dr^2+G(r)\,d\psi^2,
\end{equation}
the geodesic curvature of a fixed-$r$ orbit is
\begin{equation}
    k_g
    =
    \frac{|G'(r)|}
         {2G(r)\sqrt{E(r)}} .
    \label{eq:general-kg}
\end{equation}
Hence the orientation orbit in the Fisher metric has
\begin{equation}
    \boxed{
    k_F(r)
    =
    \frac{
        A(r)+rA'(r)
    }{
        2rA(r)\sqrt{A'(r)}
    }.
    }
    \label{eq:fisher-kg}
\end{equation}

Thus Fisher geometry gives a definite answer once that metric has been
chosen.  The question is whether that answer can be regarded as a physical
property of polarization itself.

\subsection{The strong-polarization limit is horocyclic}
\label{subsec:fisher-horocycle}

Strong polarization corresponds to
\begin{equation}
    r\rightarrow\infty,
    \qquad
    m=A(r)\rightarrow1.
\end{equation}
Using the large-$r$ expansion
\begin{equation}
    A(r)
    =
    1
    -
    \frac{1}{2r}
    -
    \frac{1}{8r^2}
    -
    \frac{1}{8r^3}
    +
    O(r^{-4}),
    \label{eq:A-large-r}
\end{equation}
one finds
\begin{equation}
    \boxed{
    k_F(r)
    \longrightarrow
    \frac{1}{\sqrt2}.
    }
    \label{eq:fisher-kg-limit}
\end{equation}
The ordered orbit therefore does \emph{not} become geodesic in Fisher
geometry as the collective tensor approaches rank one.

The limiting value has a useful intrinsic interpretation.  The Gaussian
curvature of the Fisher manifold obeys
\begin{equation}
    \boxed{
    K_F
    \longrightarrow
    -\frac12,
    }
    \label{eq:fisher-K-limit}
\end{equation}
and hence
\begin{equation}
    \boxed{
    \frac{k_F}{\sqrt{-K_F}}
    \longrightarrow1.
    }
    \label{eq:horocycle-ratio}
\end{equation}
The strongly polarized region is therefore asymptotically hyperbolic, while
the orientation orbits approach horocycles rather than geodesics.  The
appearance of hyperbolic geometry is consistent with the classical
Fisher--Rao geometry of Gaussian location--scale families
\cite{AtkinsonMitchell1981,CostaSantosStrapasson2015}.  The full curvature
calculation is collected in Appendix~\ref{app:geometry}.

The radial Fisher distance also diverges,
\begin{equation}
    L_F(r)
    =
    \int^r dt\,\sqrt{A'(t)}
    =
    \frac{1}{\sqrt2}\log r+O(1),
    \label{eq:fisher-radial-distance}
\end{equation}
so perfect polarization lies at infinite Fisher distance.  The rank-one
boundary is therefore not reached by a simple geometric straightening of
the orientation orbit.

\subsection{Metric dependence of the ordered orbit}
\label{subsec:metric-dependence}

The nonvanishing Fisher result would still be meaningful as a criterion of
autonomy if the physical problem uniquely selected the Fisher metric.  It
does not.

Write the $p=2$ collective tensor as
\begin{equation}
    T(m,\psi)
    =
    \frac12
    \begin{pmatrix}
        1+m\cos\psi & m\sin\psi\\
        m\sin\psi   & 1-m\cos\psi
    \end{pmatrix}.
    \label{eq:T-m-psi-geometry}
\end{equation}
The sequence of physical states under consideration is
\begin{equation}
    m\longrightarrow1,
    \qquad
    r_{\rm eff}
    =
    \frac{2}{1+m^2}
    \longrightarrow1.
    \label{eq:same-polarized-sequence}
\end{equation}

With the flat Hilbert--Schmidt metric inherited from the space of symmetric
matrices,
\begin{equation}
    ds_{\rm HS}^2
    =
    \Tr(dT^2)
    =
    \frac12
    \left(
        dm^2+m^2d\psi^2
    \right),
    \label{eq:HS-metric}
\end{equation}
the fixed-$m$ orientation orbit has
\begin{equation}
    \boxed{
    k_{\rm HS}
    =
    \frac{\sqrt2}{m}
    \longrightarrow
    \sqrt2.
    }
    \label{eq:HS-kg}
\end{equation}

Alternatively, treating the positive trace-one tensor as a
density-matrix-like object and equipping it with the Bures metric
\cite{Uhlmann1976,Hubner1992,BraunsteinCaves1994} gives
\begin{equation}
    ds_B^2
    =
    \frac14
    \left[
        \frac{dm^2}{1-m^2}
        +
        m^2d\psi^2
    \right],
    \label{eq:Bures-metric}
\end{equation}
and therefore
\begin{equation}
    \boxed{
    k_B
    =
    \frac{2\sqrt{1-m^2}}{m}
    \longrightarrow0.
    }
    \label{eq:Bures-kg}
\end{equation}
Moreover,
\begin{equation}
    k_B^2
    =
    4\,\frac{1+m^2}{m^2}
    \left(r_{\rm eff}-1\right)
    \sim
    8\left(r_{\rm eff}-1\right),
    \label{eq:Bures-slaved}
\end{equation}
so in this geometry the apparent flattening is directly slaved to spectral
polarization.

\begin{figure}[t]
    \centering
    \includegraphics[width=\linewidth]{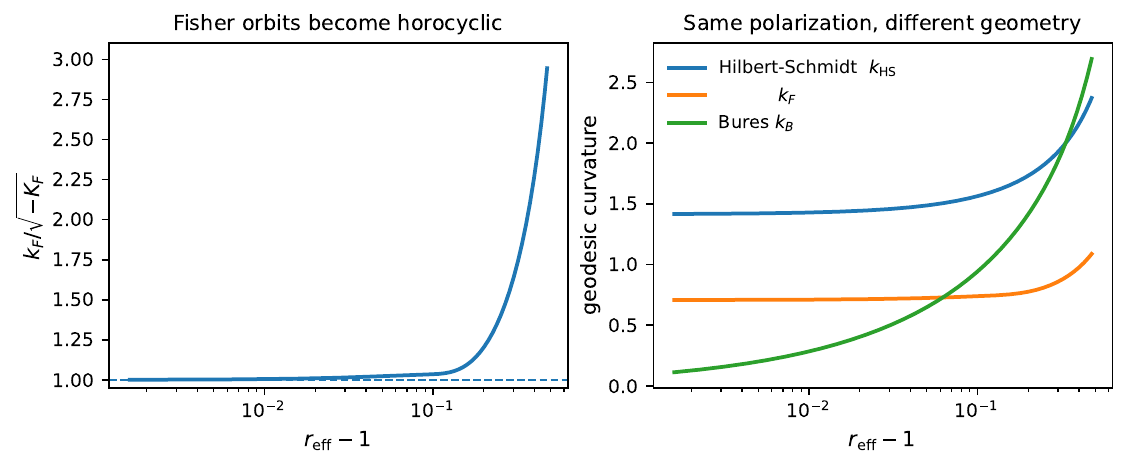}
    \caption{
    Geometry of the strongly polarized $p=2$ family.
    Left: in Fisher geometry,
    $k_F/\sqrt{-K_F}\to1$ as $r_{\rm eff}\to1$, so the orientation
    orbits approach horocycles rather than geodesics.
    Right: the same sequence of increasingly rank-one tensors has
    qualitatively different geodesic-curvature limits in
    Hilbert--Schmidt, Fisher, and Bures geometry.
    Extrinsic flattening is therefore not determined by spectral
    polarization alone.
    }
    \label{fig:geometry-comparison}
\end{figure}

The same sequence of increasingly rank-one states therefore gives
\begin{equation}
    \boxed{
    k_{\rm HS}\longrightarrow\sqrt2,
    \qquad
    k_F\longrightarrow\frac{1}{\sqrt2},
    \qquad
    k_B\longrightarrow0.
    }
    \label{eq:metric-comparison}
\end{equation}
Overall metric normalizations can change the nonzero constants, but cannot
turn a vanishing limit into a nonvanishing one. The geometric comparison
is summarized in Fig.~\ref{fig:geometry-comparison}.

Extrinsic flattening is therefore not a property of the polarized state
alone.  It is a property of the pair
\begin{equation}
    (\mathcal M,g)
\end{equation}
consisting of an ambient state space and a metric on it.  Unless the
microscopic theory physically selects this geometry, the condition
\begin{equation}
    \|B\|\longrightarrow0
\end{equation}
cannot by itself define dynamical autonomy.

\subsection{Extensivity is not autonomy}
\label{subsec:fisher-extensivity}

There is a second problem with using the absolute magnitude of extrinsic
curvature.  For $N$ statistically independent copies of a model, Fisher
information is extensive:
\begin{equation}
    g_F^{(N)}
    =
    N g_F^{(1)}.
    \label{eq:fisher-extensive}
\end{equation}
A constant rescaling of the metric by $N$ rescales inverse lengths by
$N^{-1/2}$, so
\begin{equation}
    k_F^{(N)}
    =
    \frac{1}{\sqrt N}
    k_F^{(1)},
    \label{eq:kg-N-scaling}
\end{equation}
while
\begin{equation}
    K_F^{(N)}
    =
    \frac1N
    K_F^{(1)}.
\end{equation}
Consequently,
\begin{equation}
    \boxed{
    \frac{
        k_F^{(N)}
    }{
        \sqrt{|K_F^{(N)}|}
    }
    =
    \frac{
        k_F^{(1)}
    }{
        \sqrt{|K_F^{(1)}|}
    }.
    }
    \label{eq:N-invariant-ratio}
\end{equation}

Thus $k_F^{(N)}\to0$ as $N\to\infty$ even at a fixed ordered state.  This
cannot represent increasing dynamical autonomy; it is simply a consequence
of the extensive normalization of the information metric.

Neither a condition such as
\begin{equation}
    \|B\|\ll1
\end{equation}
nor a large-$N$ limit in which $\|B\|\to0$ therefore has an invariant
physical interpretation without an independently specified length or
energy scale.

\subsection{From geometric bending to dynamical response}
\label{subsec:geometry-to-dynamics}

The preceding calculations do not undermine the distinction between
spectral polarization and dynamical autonomy.  They show instead that the
latter cannot be identified with the extrinsic curvature of an arbitrarily
chosen state-space embedding.

Geodesic curvature controls normal acceleration only for dynamics tied to
the metric being used.  In a physical system, transverse response is
determined by the actual equations of motion.  If $X^a$ denote coordinates
along an ordered sector and $y^\alpha$ transverse coordinates, the
linearized normal dynamics takes the schematic form
\begin{equation}
    M_{\alpha\beta}\ddot y^\beta
    +
    \Gamma_{\alpha\beta}\dot y^\beta
    +
    K_{\alpha\beta}y^\beta
    =
    F_\alpha[X,\dot X,\ddot X]
    +
    O(y^2),
    \label{eq:normal-dynamics-schematic}
\end{equation}
where $M$, $\Gamma$, and $K$ are physical inertia, damping, and stiffness
matrices.

The non-Goldstone frequencies are determined by
\begin{equation}
    K e_\alpha
    =
    \omega_{\perp,\alpha}^2
    M e_\alpha,
    \label{eq:normal-generalized-eigenvalue}
\end{equation}
and in the adiabatic regime the leading normal response is
\begin{equation}
    y^\alpha
    \simeq
    (K^{-1})^{\alpha\beta}
    F_\beta.
    \label{eq:adiabatic-normal-response}
\end{equation}
The suppression of transverse motion is therefore controlled jointly by the
physical kinetic geometry, the restoring quadratic form, and the resulting
response operator, rather than by the smallness of extrinsic curvature in a
freely chosen state-space metric.

This motivates the operational notion used below: an ordered sector is
dynamically autonomous over a specified low-frequency window when
tangential driving produces parametrically small physical normal response.

The next step is therefore to construct a model in which the spectral
polarization of $T$ and the transverse restoring spectrum are independently
controllable.  We do so by adding a second projective harmonic.  The
resulting model contains shape information invisible to $r_{\rm eff}$ and
supplies a physical normal Hessian whose spectrum can be varied without
changing the ordered tensor itself.

\section{Two-harmonic order and the transverse free-energy Hessian}
\label{sec:twoharmonic}

The preceding section showed that the extrinsic curvature of the ordered
orbit does not furnish a metric-independent notion of autonomy.  We now
replace that geometric criterion by a physical one.  The relevant
transverse data will be obtained from the Hessian of the thermodynamic free
energy and, once a kinetic structure is specified, from the corresponding
normal-mode spectrum.

The single-harmonic $p=2$ rotor is too restrictive for this purpose: apart
from the common orientation, its equilibrium distribution contains only one
concentration parameter.  We therefore introduce a second projective
harmonic.  This adds shape information that is invisible to the collective
tensor and allows states with identical spectral polarization to possess
different physical transverse stiffnesses.

\subsection{Two-harmonic mean-field model}
\label{subsec:twoharmonic-model}

Let
\begin{equation}
    x_i=2\theta_i
\end{equation}
denote the doubled projective angle, and define
\begin{equation}
    M_1
    =
    \frac1N
    \sum_{i=1}^N e^{ix_i},
    \qquad
    M_2
    =
    \frac1N
    \sum_{i=1}^N e^{2ix_i}.
    \label{eq:M1-M2-def}
\end{equation}
We consider
\begin{equation}
    \boxed{
    H
    =
    -\frac{J_1N}{2}|M_1|^2
    -
    \frac{J_2N}{2}|M_2|^2 .
    }
    \label{eq:twoharmonic-H}
\end{equation}
Competing first and second angular harmonics also occur in generalized
Hamiltonian mean-field models
\cite{TelesBenettiPakterLevin2012}; here they are used for the different
purpose of separating tensor polarization from transverse thermodynamic
stiffness.

With
\begin{equation}
    \kappa_1=\beta J_1,
    \qquad
    \kappa_2=\beta J_2,
\end{equation}
the large-$N$ free-energy functional is
\begin{equation}
\begin{split}
    \mathcal F[\rho]
    =
    &\int_0^{2\pi}
    \frac{dx}{2\pi}\,
    \rho(x)\log\rho(x)
\\
    &-
    \frac{\kappa_1}{2}|M_1[\rho]|^2
    -
    \frac{\kappa_2}{2}|M_2[\rho]|^2 .
\end{split}
    \label{eq:twoharmonic-functional}
\end{equation}

On the aligned reflection-symmetric branch,
\begin{equation}
    M_1
    =
    m e^{i\psi},
    \qquad
    M_2
    =
    q e^{2i\psi}.
    \label{eq:aligned-moments}
\end{equation}
Using rotational symmetry to set $\psi=0$, the one-site distribution is
\begin{equation}
    \boxed{
    \rho_{a,b}(x)
    =
    \frac{
        e^{a\cos x+b\cos2x}
    }{
        Z(a,b)
    },
    }
    \label{eq:twoharmonic-density}
\end{equation}
where
\begin{equation}
    \Phi(a,b)
    =
    \log Z(a,b),
\end{equation}
and
\begin{equation}
    \boxed{
    m=\partial_a\Phi,
    \qquad
    q=\partial_b\Phi.
    }
    \label{eq:m-q-Phi}
\end{equation}
At a mean-field saddle,
\begin{equation}
    \boxed{
    a=\kappa_1m,
    \qquad
    b=\kappa_2q.
    }
    \label{eq:twoharmonic-self-consistency}
\end{equation}

The significance of the second harmonic is immediate.  For $p=2$,
\begin{equation}
    T
    =
    \frac12
    \begin{pmatrix}
        1+\Re M_1 & \Im M_1\\
        \Im M_1   & 1-\Re M_1
    \end{pmatrix},
    \label{eq:T-from-M1}
\end{equation}
so
\begin{equation}
    \boxed{
    r_{\rm eff}
    =
    \frac{2}{1+|M_1|^2}
    =
    \frac{2}{1+m^2}.
    }
    \label{eq:reff-twoharmonic}
\end{equation}
The higher harmonic $q$ does not enter $T$.  It therefore parametrizes
equilibrium shape information invisible to the tensor spectrum.

\subsection{Thermodynamic fluctuation Hessians}
\label{subsec:physical-hessians}

For reflection-even fluctuations, define the covariance matrix
\begin{equation}
    C_e
    =
    \begin{pmatrix}
        \operatorname{Var}(\cos x)
        &
        \operatorname{Cov}(\cos x,\cos2x)
        \\[1mm]
        \operatorname{Cov}(\cos x,\cos2x)
        &
        \operatorname{Var}(\cos2x)
    \end{pmatrix}_{a,b}.
    \label{eq:Ce-def}
\end{equation}
Legendre duality of the two-harmonic exponential family
\cite{WainwrightJordan2008} gives the physical even Hessian
\begin{equation}
    \boxed{
    K_e
    =
    C_e^{-1}
    -
    \begin{pmatrix}
        \kappa_1&0\\
        0&\kappa_2
    \end{pmatrix}.
    }
    \label{eq:Ke-def}
\end{equation}
The derivation, including the rate function
\begin{equation}
    I(m,q)
    =
    \sup_{a,b}
    \left[
        am+bq-\Phi(a,b)
    \right],
\end{equation}
is given in Appendix~\ref{app:hessians}.

Equation~\eqref{eq:Ke-def} is the thermodynamic transverse quadratic form in
the even sector.  Unlike the extrinsic curvature of
Sec.~\ref{sec:geometry}, it is fixed by the free-energy functional of the
specified model.  Its signature at a stationary point is coordinate
invariant, but ordinary matrix eigenvalues depend on the normalization of
the moment coordinates.  Physical normal frequencies require a kinetic
metric and are introduced only in Sec.~\ref{sec:dynamics}.  Local stability
in the even sector requires
\begin{equation}
    \boxed{
    K_e>0.
    }
    \label{eq:even-stability}
\end{equation}
Its two eigenmodes are generically mixtures of polarization amplitude and
higher-harmonic shape.

The odd sector describes phase fluctuations.  Define
\begin{equation}
    C_o
    =
    \begin{pmatrix}
        \langle\sin^2x\rangle
        &
        \langle\sin x\,\sin2x\rangle
        \\[1mm]
        \langle\sin x\,\sin2x\rangle
        &
        \langle\sin^22x\rangle
    \end{pmatrix}_{a,b}.
    \label{eq:Co-def}
\end{equation}
Reflection symmetry separates even and odd fluctuations, and the odd
Hessian is
\begin{equation}
    \boxed{
    K_o
    =
    C_o^{-1}
    -
    \begin{pmatrix}
        \kappa_1&0\\
        0&\kappa_2
    \end{pmatrix}.
    }
    \label{eq:Ko-def}
\end{equation}

To identify its physical content, write
\begin{equation}
    M_1
    =
    m e^{i\psi_1},
    \qquad
    M_2
    =
    q e^{i\psi_2},
\end{equation}
and define
\begin{equation}
    \boxed{
    \psi=\psi_1,
    \qquad
    \chi=\psi_2-2\psi_1.
    }
    \label{eq:psi-chi}
\end{equation}
The common angle $\psi$ shifts under a global rotation, whereas $\chi$ is
invariant.  Thus $\psi$ is the tangent Goldstone coordinate and $\chi$ the
relative-phase normal coordinate \cite{GoldstoneSalamWeinberg1962}.

An infinitesimal common rotation produces the odd moment vector
\begin{equation}
    v_G
    =
    \begin{pmatrix}
        m\\
        2q
    \end{pmatrix}.
    \label{eq:Goldstone-vector}
\end{equation}
Rotational invariance gives the exact Ward identity
\begin{equation}
    \boxed{
    K_o v_G=0.
    }
    \label{eq:Goldstone-Ward}
\end{equation}
The remaining odd eigenvalue,
\begin{equation}
    \lambda_o^+>0,
    \label{eq:odd-positive-eigenvalue}
\end{equation}
is the restoring stiffness of the relative-phase mode.  A detailed
derivation of the block decomposition, the Ward identity, and its relation
to $\chi$ is given in Appendix~\ref{app:hessians}.

The aligned phase therefore contains one tangent Goldstone mode and three
non-Goldstone collective modes:
\begin{equation}
    \boxed{
    \psi
    \quad\oplus\quad
    \{
        \text{two even amplitude--shape modes},
        \ \chi
    \}.
    }
    \label{eq:mode-decomposition}
\end{equation}

\subsection{Fixed-polarization families}
\label{subsec:fixed-polarization}

We can now ask directly whether the tensor spectrum determines this
transverse free-energy quadratic form.

Fix
\begin{equation}
    m=m_\star,
    \qquad
    0<m_\star<1.
    \label{eq:mstar-fixed}
\end{equation}
Then
\begin{equation}
    \boxed{
    r_{\rm eff}
    =
    \frac{2}{1+m_\star^2}
    =
    \mathrm{constant}.
    }
    \label{eq:fixed-reff}
\end{equation}

For each chosen $b$, determine $a=a(b)$ from
\begin{equation}
    \partial_a\Phi(a,b)
    =
    m_\star,
    \label{eq:a-fixed-m}
\end{equation}
and then set
\begin{equation}
    q(b)
    =
    \partial_b\Phi(a(b),b),
\end{equation}
\begin{equation}
    \boxed{
    \kappa_1(b)
    =
    \frac{a(b)}{m_\star},
    \qquad
    \kappa_2(b)
    =
    \frac{b}{q(b)}.
    }
    \label{eq:kappas-fixed-m}
\end{equation}
Appendix~\ref{app:hessians} shows that this branch is locally unique and
that
\begin{equation}
    \frac{dq}{db}
    =
    \frac{
        \det C_e
    }{
        (C_e)_{11}
    }
    >0.
    \label{eq:q-monotone-fixed-m}
\end{equation}
Thus the higher-harmonic shape changes continuously while $M_1$, and hence
the complete tensor $T$, remains fixed up to its arbitrary common
orientation.

Along this family,
\begin{equation}
    q,\qquad
    C_e,\qquad
    C_o,\qquad
    K_e,\qquad
    K_o
\end{equation}
vary even though $r_{\rm eff}$ does not.

For example, take
\begin{equation}
    m_\star=0.9,
\end{equation}
so that
\begin{equation}
    r_{\rm eff}
    =
    \frac{2}{1+0.9^2}
    \simeq
    1.10497238.
    \label{eq:m09-reff}
\end{equation}
Three representative states, numerically identified as global minima of the full two-harmonic mean-field free energy, are shown in
Table~\ref{tab:fixed-m-spectrum}; the global-minimum check is described in
Appendix~\ref{app:global-minimum-check}.

\begin{table}[t]
\centering
\begin{tabular}{c c c c c c c}
\hline
$b$
&
$q$
&
$\kappa_1$
&
$\kappa_2$
&
$k_{e,\mathrm{coord}}^-$
&
$k_{e,\mathrm{coord}}^+$
&
$\lambda_o^+$
\\
\hline
$1.0$
&
$0.6925$
&
$3.3306$
&
$1.4440$
&
$3.8555$
&
$70.9149$
&
$62.8272$
\\
$2.7$
&
$0.8246$
&
$2.0907$
&
$3.2742$
&
$6.6456$
&
$15.9761$
&
$79.0556$
\\
$6.0$
&
$0.9181$
&
$1.8008$
&
$6.5354$
&
$4.8978$
&
$68.2930$
&
$166.2429$
\\
\hline
\end{tabular}
\caption{
Representative states numerically identified as global minima of the two-harmonic model at fixed
$m_\star=0.9$, and therefore fixed
$r_{\rm eff}\simeq1.10497$.
The quantities $k_{e,\mathrm{coord}}^\pm$ and $\lambda_o^+$ are ordinary
matrix eigenvalues in the canonical real moment-coordinate normalization
$(\Re M_1,\Re M_2,\Im M_1,\Im M_2)$.  Their numerical values are therefore
coordinate dependent, while the positivity statements are not.  The table
demonstrates that the transverse free-energy quadratic form changes while
the collective tensor is held fixed; physical frequencies are defined only
after a kinetic metric is specified in Sec.~\ref{sec:dynamics}.
}
\label{tab:fixed-m-spectrum}
\end{table}

All three examples satisfy
\begin{equation}
    k_{e,\mathrm{coord}}^\pm>0,
    \qquad
    \lambda_o^+>0,
\end{equation}
so all three are strictly locally stable; the independent global
minimization reported in Appendix~\ref{app:global-minimum-check} finds no
lower competing state.

We therefore obtain
\begin{equation}
    \boxed{
    T
    \ \text{does not determine the transverse free-energy Hessian}.
    }
    \label{eq:T-not-normal-spectrum}
\end{equation}
In particular,
\begin{equation}
    r_{\rm eff}
    =
    \mathrm{constant}
\end{equation}
does not imply fixed amplitude, shape, or relative-phase stiffness.

This is the physical analogue of the metric-dependence result of
Sec.~\ref{sec:geometry}.  There the extrinsic curvature changed when the
ambient metric was changed.  Here the distinction is dynamical: even within
a specified microscopic thermodynamic theory, the equilibrium tensor does
not encode all of the fluctuation data needed to determine departures from
the ordered sector.

\subsection{Toward parametric separation of scales}
\label{subsec:toward-locking}

The two-harmonic model establishes finite independence of spectral
polarization and transverse stiffness.  It does not yet establish the
stronger limit needed for an autonomous low-energy sector.

First, not every point of the fixed-$m$ family is stable.  Depending on the
microscopic couplings, an even shape mode or the relative-phase mode may
soften.

Second, the original two couplings do not provide a controlled limit in
which
\begin{equation}
    r_{\rm eff}
    =
    \mathrm{constant}
\end{equation}
while every non-Goldstone restoring coefficient tends to infinity.

We therefore seek interactions that leave a prescribed ordered state
\begin{equation}
    (m_\star,q_\star)
\end{equation}
fixed while modifying only the second and higher variations of the free
energy.  In the next section we construct nonnegative locking interactions
that vanish, together with their first derivatives, on the target
orientation orbit.

They leave
\begin{equation}
    m_\star,\qquad
    q_\star,\qquad
    T_\star,\qquad
    r_{\rm eff}(T_\star)
\end{equation}
unchanged while allowing the complete non-Goldstone restoring spectrum to
be made parametrically large.

\section{Fixed-polarization locking and parametric stability}
\label{sec:locking}

Section~\ref{sec:twoharmonic} showed that the collective tensor does not
determine its transverse fluctuation spectrum.  At fixed $T$ and
$r_{\rm eff}$, higher-harmonic structure can change both the even and odd
restoring forces.  That establishes independence at finite coupling.  We now
strengthen it to a controlled parametric separation of scales.

Starting from a stable aligned ordered state, we construct rotationally
invariant interactions whose value and first variation vanish on its entire
orientation orbit, but whose second variation is positive in the transverse
directions.  The equilibrium tensor therefore remains fixed while the
non-Goldstone restoring spectrum can be made arbitrarily stiff.

Throughout this section $f$ denotes the dimensionless free energy.  It is
useful to distinguish the dimensionless locking couplings from microscopic
energy scales.  Writing $G_i$ for dimensionful interaction strengths, we
define $\lambda_i=\beta G_i$; the microscopic $G_i$ need not depend on
temperature.

\subsection{Locking interactions around a prescribed ordered state}
\label{subsec:locking-interactions}

Let
\begin{equation}
    M_{1\star}
    =
    m_\star e^{i\psi},
    \qquad
    M_{2\star}
    =
    q_\star e^{2i\psi}
    \label{eq:locking-target}
\end{equation}
be a stable aligned solution of the two-harmonic model, and define
\begin{equation}
    c_\star
    =
    \frac{q_\star}{m_\star^2}.
    \label{eq:cstar-def}
\end{equation}
The full orientation orbit satisfies
\begin{equation}
    |M_1|^2=m_\star^2,
    \qquad
    M_2=c_\star M_1^2.
    \label{eq:target-constraints}
\end{equation}

We add
\begin{equation}
    \boxed{
    f_{\rm lock}
    =
    \frac{\lambda_1}{8}
    \left(
        |M_1|^2-m_\star^2
    \right)^2
    +
    \frac{\lambda_2}{2}
    \left|
        M_2-c_\star M_1^2
    \right|^2 .
    }
    \label{eq:locking-potential}
\end{equation}

Equivalently, the corresponding contribution to the microscopic mean-field
Hamiltonian may be written
\begin{equation}
    H_{\rm lock}
    =
    N\left[
        \frac{G_1}{8}
        \left(
            |M_1|^2-m_\star^2
        \right)^2
        +
        \frac{G_2}{2}
        \left|
            M_2-c_\star M_1^2
        \right|^2
    \right],
    \qquad
    \lambda_i=\beta G_i.
    \label{eq:locking-Hamiltonian}
\end{equation}
Because $M_1$ and $M_2$ are collective moments, these are higher-body
mean-field interactions.  They are deliberately engineered as a constructive
independence test; in a more microscopic theory analogous terms could arise,
for example, after integrating out additional gapped degrees of freedom.
Both terms are invariant under
\begin{equation}
    M_1\longrightarrow e^{i\alpha}M_1,
    \qquad
    M_2\longrightarrow e^{2i\alpha}M_2,
    \label{eq:locking-rotation}
\end{equation}
so the Goldstone rotation remains an exact symmetry.

On the target orbit,
\begin{equation}
    \boxed{
    f_{\rm lock}|_\star=0,
    \qquad
    df_{\rm lock}|_\star=0.
    }
    \label{eq:locking-zero-first}
\end{equation}
Consequently, varying $\lambda_1$ and $\lambda_2$ leaves
\begin{equation}
    \boxed{
    m_\star,\qquad
    q_\star,\qquad
    T_\star,\qquad
    r_{\rm eff}(T_\star)
    }
    \label{eq:locking-fixed-data}
\end{equation}
exactly unchanged.

The locking potential is a proof-of-principle interaction rather than a
claim of microscopic uniqueness.  Its purpose is to determine whether
spectral polarization and transverse restoring scales are independent
physical data.

\subsection{Preservation of a global minimum}
\label{subsec:global-minimum}

The construction has a simple global property.  Let $f_0[\rho]$ denote the
unmodified free energy and suppose that $\rho_\star$ is a global minimum:
\begin{equation}
    f_0[\rho]
    \geq
    f_0[\rho_\star].
    \label{eq:base-global-min}
\end{equation}
For
\begin{equation}
    \lambda_1,\lambda_2\geq0,
    \label{eq:positive-locking}
\end{equation}
the locking potential obeys
\begin{equation}
    f_{\rm lock}[\rho]\geq0,
    \qquad
    f_{\rm lock}[\rho_\star]=0.
\end{equation}
Therefore
\begin{equation}
    f[\rho]
    =
    f_0[\rho]+f_{\rm lock}[\rho]
    \geq
    f_0[\rho_\star]
    =
    f[\rho_\star],
\end{equation}
and hence
\begin{equation}
    \boxed{
    \rho_\star
    \text{ global minimum of }f_0
    \quad\Longrightarrow\quad
    \rho_\star
    \text{ global minimum of }f
    }
    \label{eq:global-preservation}
\end{equation}
throughout the nonnegative quadrant.

The minimum need not be unique.  What matters here is that increasing
positive locking couplings cannot destabilize the prescribed ordered orbit
in favor of a lower-free-energy state.

\subsection{Exact quadratic stability domain}
\label{subsec:locking-stability-domain}

Let
\begin{equation}
    A
    \equiv
    K_e^{(0)}
    =
    \begin{pmatrix}
        A_{11}&A_{12}\\
        A_{12}&A_{22}
    \end{pmatrix}
    \label{eq:base-even-A}
\end{equation}
denote the even Hessian of the unmodified model at
$(m_\star,q_\star)$, and let
\begin{equation}
    O\equiv K_o^{(0)}
    \label{eq:base-odd-O}
\end{equation}
denote the odd Hessian.  Define
\begin{equation}
    \eta
    =
    \frac{2q_\star}{m_\star},
    \qquad
    w
    =
    \begin{pmatrix}
        -\eta\\
        1
    \end{pmatrix}.
    \label{eq:eta-w-def}
\end{equation}

The quadratic expansion of
Eq.~\eqref{eq:locking-potential}, derived in
Appendix~\ref{app:stability}, gives
\begin{equation}
    \boxed{
    K_e
    =
    A
    +
    \lambda_1m_\star^2
    e_1e_1^{\mathsf T}
    +
    \lambda_2ww^{\mathsf T},
    }
    \label{eq:locked-even-Hessian}
\end{equation}
with $e_1=(1,0)^{\mathsf T}$.

Rotational invariance gives
\begin{equation}
    v_G
    =
    \begin{pmatrix}
        m_\star\\
        2q_\star
    \end{pmatrix},
    \qquad
    Ov_G=0,
    \qquad
    w^{\mathsf T}v_G=0.
    \label{eq:base-Goldstone}
\end{equation}
Since the non-Goldstone odd sector is one dimensional, we may write
\begin{equation}
    O
    =
    \gamma ww^{\mathsf T},
\end{equation}
and hence
\begin{equation}
    \boxed{
    K_o
    =
    (\gamma+\lambda_2)
    ww^{\mathsf T}.
    }
    \label{eq:locked-odd-Hessian}
\end{equation}
The Goldstone mode therefore remains exactly massless.

The complete strict quadratic stability domain, after removing that
symmetry-required zero mode, is
\begin{equation}
    \boxed{
    \lambda_2
    >
    \max\{-A_{22},-\gamma\},
    \qquad
    \lambda_1
    >
    \lambda_{1,\mathrm{crit}}(\lambda_2),
    }
    \label{eq:complete-stability-domain}
\end{equation}
where
\begin{equation}
    \boxed{
    \lambda_{1,\mathrm{crit}}(\lambda_2)
    =
    \frac1{m_\star^2}
    \left[
        \frac{
            (A_{12}-\eta\lambda_2)^2
        }{
            A_{22}+\lambda_2
        }
        -
        A_{11}
        -
        \eta^2\lambda_2
    \right].
    }
    \label{eq:lambda1-critical}
\end{equation}
The Schur-complement derivation and the full quadratic expansion are given in
Appendix~\ref{app:stability}.

The boundary
\begin{equation}
    \lambda_2=-\gamma
\end{equation}
softens the relative-phase mode, while
\begin{equation}
    \lambda_1
    =
    \lambda_{1,\mathrm{crit}}(\lambda_2)
\end{equation}
softens one of the even amplitude--shape modes.

\subsection{A guaranteed globally stable quadrant}
\label{subsec:stable-quadrant}

For the parametric construction below, the full stability boundary is less
important than a simple sufficient region.

Suppose the base state is already strictly locally stable:
\begin{equation}
    A>0,
    \qquad
    \gamma>0.
    \label{eq:base-strictly-stable}
\end{equation}
For
\begin{equation}
    \lambda_1,\lambda_2\geq0,
\end{equation}
the additional quadratic forms in
Eqs.~\eqref{eq:locked-even-Hessian} and
\eqref{eq:locked-odd-Hessian} are positive semidefinite.  Hence
\begin{equation}
    \boxed{
    \lambda_1,\lambda_2\geq0
    \quad\Longrightarrow\quad
    \text{local stability is preserved}.
    }
    \label{eq:positive-quadrant-local}
\end{equation}

If the base state is also a global minimum, Sec.~\ref{subsec:global-minimum}
gives the stronger result
\begin{equation}
    \boxed{
    \lambda_1,\lambda_2\geq0
    \quad\Longrightarrow\quad
    \rho_\star
    \text{ remains a global minimum}.
    }
    \label{eq:positive-quadrant-global}
\end{equation}

\begin{figure}[t]
    \centering
    \includegraphics[width=0.92\linewidth]{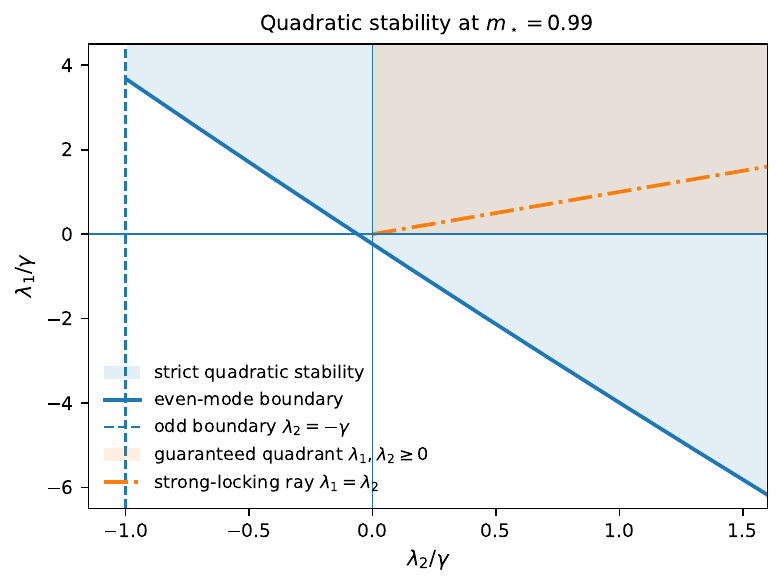}
    \caption{
    Stability domain of the locking construction for the strongly polarized
    $m_\star=0.99$ benchmark.
    The axes show the dimensionless couplings
    $\lambda_1/\gamma$ and $\lambda_2/\gamma$.
    The solid curve marks the even-mode stability boundary, while the dashed
    vertical line marks softening of the relative-phase mode.
    If the base state is a global minimum, the nonnegative quadrant is
    guaranteed to preserve it.  For this benchmark $\gamma\simeq2.16\times10^4$, so the displayed axes
    correspond to locking couplings of order $10^4$; the figure is a
    stability map rather than an asymptotic-scaling plot.  The strong-locking
    ray $\lambda_1=\lambda_2$ moves deeper into the stable region as the
    locking scale is increased.
    }
    \label{fig:locking-stability-domain}
\end{figure}

The strong-locking limit therefore lies inside a guaranteed stable region,
rather than approaching a critical surface.  The exact quadratic stability
domain, the globally controlled nonnegative quadrant, and a representative
strong-locking trajectory are shown in
Fig.~\ref{fig:locking-stability-domain}.

\subsection{Parametric separation at fixed polarization}
\label{subsec:parametric-separation}

Consider the ray
\begin{equation}
    \boxed{
    \lambda_1=\alpha_1\lambda,
    \qquad
    \lambda_2=\alpha_2\lambda,
    \qquad
    \alpha_1,\alpha_2>0,
    }
    \label{eq:strong-locking-ray}
\end{equation}
with
\begin{equation}
    \lambda\longrightarrow\infty.
\end{equation}
Along this trajectory,
\begin{equation}
    \boxed{
    m=m_\star,
    \qquad
    q=q_\star,
    \qquad
    T=T_\star,
    \qquad
    r_{\rm eff}=r_{\rm eff}(T_\star).
    }
    \label{eq:fixed-along-ray}
\end{equation}

The even Hessian takes the asymptotic form
\begin{equation}
    K_e
    =
    \lambda L_e+O(1),
    \label{eq:Ke-asymptotic}
\end{equation}
with
\begin{equation}
    L_e
    =
    \alpha_1 m_\star^2
    e_1e_1^{\mathsf T}
    +
    \alpha_2 ww^{\mathsf T}.
    \label{eq:Le-def}
\end{equation}
Its determinant is
\begin{equation}
    \boxed{
    \det L_e
    =
    \alpha_1\alpha_2 m_\star^2>0,
    }
    \label{eq:Le-det}
\end{equation}
and its trace is positive.  Hence
\begin{equation}
    L_e>0,
\end{equation}
so both even restoring coefficients grow linearly:
\begin{equation}
    \boxed{
    k_{e,\pm}
    =
    \Theta(\lambda).
    }
    \label{eq:even-stiffness-scaling}
\end{equation}

The relative-phase stiffness follows from
Eq.~\eqref{eq:locked-odd-Hessian}:
\begin{equation}
    k_o^+
    =
    (\gamma+\alpha_2\lambda)
    (1+\eta^2)
    =
    \boxed{
    \alpha_2(1+\eta^2)\lambda+O(1).
    }
    \label{eq:odd-stiffness-scaling}
\end{equation}
Thus every non-Goldstone restoring coefficient obeys
\begin{equation}
    \boxed{
    k_{\perp,\alpha}
    =
    \Theta(\lambda),
    }
    \label{eq:all-stiffness-linear}
\end{equation}
while the common orientation remains exactly gapless.

We therefore obtain the central equilibrium result:
\begin{equation}
    \boxed{
    \begin{gathered}
        T_\star=\mathrm{constant},
        \qquad
        r_{\rm eff}(T_\star)=\mathrm{constant},
        \\[1mm]
        k_{\perp,\alpha}
        \longrightarrow\infty
        \quad
        \text{for every non-Goldstone normal mode}.
    \end{gathered}
    }
    \label{eq:central-static-result}
\end{equation}
Spectral polarization and transverse stiffness are parametrically
independent.

The explicit eigenvalues of $L_e$ are given in
Appendix~\ref{app:stability}.  Once a finite positive kinetic matrix is
specified, the result implies
\begin{equation}
    \omega_{\perp,\alpha}
    =
    \Theta(\lambda^{1/2}),
    \label{eq:frequency-scaling-preview}
\end{equation}
as shown dynamically in Sec.~\ref{sec:dynamics}.

\subsection{A strongly polarized benchmark}
\label{subsec:locking-benchmark}

As a concrete example, choose
\begin{equation}
    m_\star=0.99,
    \qquad
    \kappa_2=0.
    \label{eq:benchmark-m}
\end{equation}
The corresponding one-harmonic saddle has
\begin{equation}
    q_\star\simeq0.960600,
    \qquad
    \eta
    =
    \frac{2q_\star}{m_\star}
    \simeq1.940606,
\end{equation}
with the detailed base parameters listed in
Appendix~\ref{app:stability}.  Its tensor is already strongly polarized:
\begin{equation}
    \boxed{
    r_{\rm eff}
    =
    \frac{2}{1+0.99^2}
    \simeq
    1.010050.
    }
    \label{eq:benchmark-reff}
\end{equation}

Along the simple ray
\begin{equation}
    \lambda_1=\lambda_2=\lambda,
    \label{eq:benchmark-ray}
\end{equation}
the leading restoring coefficients are
\begin{align}
    k_{e,-}
    &=
    0.175958\,\lambda+O(1),
    \\
    k_{e,+}
    &=
    5.570095\,\lambda+O(1),
    \\
    k_o^+
    &=
    4.765952\,\lambda+O(1).
    \label{eq:benchmark-stiffnesses}
\end{align}
In particular,
\begin{equation}
    \boxed{
    k_{\perp,\min}
    =
    0.175958\,\lambda+O(1),
    }
    \label{eq:benchmark-min-stiffness}
\end{equation}
while $r_{\rm eff}$ remains exactly fixed.

The $O(1)$ base Hessian is numerically important for this strongly polarized
benchmark.  At $\lambda=0$ its softest non-Goldstone coordinate-Hessian
eigenvalue is approximately $308.17$, corresponding to
$\sqrt{308.17}\simeq17.55$ in the illustrative unit-inertia normalization.
The leading locking contribution $0.175958\,\lambda$ reaches this scale only
near
\begin{equation}
    \lambda_{\rm cross}
    \simeq
    \frac{308.17}{0.175958}
    \simeq
    1.75\times10^3.
    \label{eq:benchmark-crossover}
\end{equation}
Thus the asymptotic coefficients in
Eq.~\eqref{eq:benchmark-stiffnesses} should not be interpreted as accurate
finite-$\lambda$ frequencies below the crossover.  The full Hessian is used
in the dynamical figures of Sec.~\ref{sec:dynamics}.

The locking scale therefore does not improve the spectral polarization of
the tensor.  It changes an independent property: the physical cost of
leaving the selected one-direction sector.

The remaining question is dynamical.  A large restoring spectrum yields an
autonomous infrared sector only if it suppresses actual transverse response
under motion along the Goldstone direction.  We address that question in
Sec.~\ref{sec:dynamics}.

\section{Dynamical leakage and low-energy autonomy}
\label{sec:dynamics}

The preceding sections established two distinct equilibrium properties.
Projective ordering can produce a collective tensor with one dominant
eigenline, while the locking construction of Sec.~\ref{sec:locking} allows
every non-Goldstone transverse stiffness to be made parametrically large
without changing that tensor.  We now specialize to $p=2$, for which the
director manifold $\mathbb{RP}^{1}$ is one dimensional.  The dynamical
question is whether motion on this orientation orbit induces parametrically
small excursions into its non-Goldstone transverse modes.

To avoid assuming the interpretation we ultimately seek to derive, we
distinguish the microscopic evolution parameter from an emergent physical
clock.  Throughout this section,
\begin{equation}
    \dot X\equiv\frac{dX}{ds},
\end{equation}
where $s$ parametrizes the underlying microscopic dynamics.  It is not
identified with the projective direction selected by $T$ or with a spacetime
clock coordinate.

The claim established below is correspondingly narrower: for a specified
collective dynamics, the $p=2$ director orbit can become dynamically
autonomous in the infrared.  The microscopic parameter $s$ remains the
evolution parameter; the angle that orients the selected eigenline is a
collective coordinate evolving as a function of $s$.

\subsection{Collective dynamics and the microscopic evolution parameter}
\label{subsec:collective-dynamics}

We do not derive the collective dynamics from the microscopic rotor
Hamiltonian.  Fully connected rotor models can instead lead to kinetic/Vlasov
dynamics, while stochastic order-parameter dynamics belongs to different
dynamic universality classes
\cite{AntoniRuffo1995,CampaDauxoisRuffo2009,HohenbergHalperin1977}.
For an explicit inertial realization we adopt the phenomenological ansatz
\begin{equation}
    \boxed{
    L_{\rm kin}
    =
    \frac{N}{2}
    \left[
        \mu_1|\dot M_1|^2
        +
        \mu_2|\dot M_2|^2
    \right],
    }
    \label{eq:collective-kinetic-ansatz}
\end{equation}
with $\mu_1,\mu_2>0$.  Writing
\begin{equation}
    M_1
    =
    m e^{i\psi},
    \qquad
    M_2
    =
    q e^{i(2\psi+\chi)},
    \label{eq:dynamical-M1-M2}
\end{equation}
identifies $\psi$ as the Goldstone orientation coordinate and
\begin{equation}
    y^\alpha
    =
    (\delta m,\delta q,\chi)
    \label{eq:normal-coordinates}
\end{equation}
as the three non-Goldstone collective coordinates.

On the aligned slice $\chi=0$, the tangent kinetic coefficient is
\begin{equation}
    \boxed{
    {\cal I}(m,q)
    =
    \mu_1m^2
    +
    4\mu_2q^2.
    }
    \label{eq:orientation-inertia}
\end{equation}
More generally, in coordinates adapted to the ordered orbit, the quadratic
collective dynamics may be written locally as
\begin{equation}
    \boxed{
    L_{\rm coll}
    =
    \frac{N}{2}
    \left[
        {\cal I}(y)\dot\psi^2
        +
        \dot y^{\mathsf T}
        M_\perp
        \dot y
    \right]
    -
    Nf(y)
    +
    \cdots ,
    }
    \label{eq:general-collective-L}
\end{equation}
with
\begin{equation}
    M_\perp>0.
\end{equation}
Dissipation may be represented by a positive semidefinite matrix
$\Gamma_\perp$.

The detailed phase-sector kinetic reduction is given in
Appendix~\ref{app:dynamics}.  Because the original $(\psi,\chi)$ coordinates
have a kinetic cross term, we first orthogonalize the Goldstone and
relative-phase directions there.  From this point onward, for notational
economy, $\psi$ denotes that kinetically orthogonalized Goldstone coordinate
$\widetilde\psi$.  This removes the kinetic cross term at the reference
state, but it does \emph{not} eliminate a physical normal component of a
generic external drive.  Such components are retained explicitly through
$F_\perp^{\rm ext}$ below.  The subsequent scaling results require only that
$M_\perp$ remain finite, positive, and nonsingular in the strong-locking
limit.

\subsection{Normal-mode spectrum and the transverse gap}
\label{subsec:normal-gap}

Expand the locked free energy about the ordered orbit:
\begin{equation}
    f(y)
    =
    f_\star
    +
    \frac12
    y^{\mathsf T}K_\perp y
    +
    O(y^3).
    \label{eq:normal-free-energy-expansion}
\end{equation}
After removing the Goldstone direction, the physical normal frequencies are
defined by
\begin{equation}
    \boxed{
    K_\perp e_\alpha
    =
    \omega_{\perp,\alpha}^2
    M_\perp e_\alpha.
    }
    \label{eq:normal-frequency-eigenproblem}
\end{equation}
We denote the lowest one by
\begin{equation}
    \boxed{
    \Delta_\perp
    =
    \min_\alpha
    \omega_{\perp,\alpha}.
    }
    \label{eq:transverse-gap}
\end{equation}

Along the strong-locking ray of
Sec.~\ref{subsec:parametric-separation},
\begin{equation}
    K_\perp
    =
    \lambda L_\perp+O(1),
    \qquad
    L_\perp>0.
    \label{eq:Kperp-strong-locking}
\end{equation}
If
\begin{equation}
    M_\perp(\lambda)
    \longrightarrow
    M_\perp^{(\infty)}>0,
    \label{eq:Mperp-limit}
\end{equation}
this condition is automatic for any kinetic metric that is a smooth function
of the equilibrium state alone, because the locking trajectory leaves that
state $(m_\star,q_\star)$ exactly fixed.  Then
\begin{equation}
    \boxed{
    \omega_{\perp,\alpha}^2
    =
    \lambda\nu_\alpha+O(1),
    \qquad
    \nu_\alpha>0,
    }
    \label{eq:omega-strong-locking}
\end{equation}
and therefore
\begin{equation}
    \boxed{
    \Delta_\perp
    =
    \Theta(\lambda^{1/2}).
    }
    \label{eq:gap-sqrt-lambda}
\end{equation}

Together with Sec.~\ref{sec:locking},
\begin{equation}
    \boxed{
    r_{\rm eff}
    =
    \mathrm{constant},
    \qquad
    \Delta_\perp
    \longrightarrow\infty.
    }
    \label{eq:fixed-rank-growing-gap}
\end{equation}
Spectral polarization and the dynamical excitation scale of the transverse
sector are therefore parametrically independent.

\subsection{Tangential motion and physical normal forcing}
\label{subsec:tangent-normal-force}

The orientation inertia generally varies across the transverse directions.
Write
\begin{equation}
    {\cal I}(y)
    =
    {\cal I}_\star
    +
    g^{\mathsf T}y
    +
    O(y^2),
    \qquad
    g_\alpha
    =
    \left.
    \partial_\alpha{\cal I}
    \right|_\star.
    \label{eq:I-expansion}
\end{equation}
The linearized normal equation is then
\begin{equation}
    \boxed{
    M_\perp\ddot y
    +
    \Gamma_\perp\dot y
    +
    K_\perp y
    =
    \frac12g\,\dot\psi^2
    +
    F_\perp^{\rm ext}
    +
    \cdots .
    }
    \label{eq:physical-normal-equation}
\end{equation}

This equation also reconciles the geometric and dynamical viewpoints.  For
the kinetic metric of Eq.~\eqref{eq:general-collective-L},
$\Gamma^\alpha_{\psi\psi}=-\tfrac12(M_\perp^{-1}g)^\alpha$ at the reference
orbit.  The term $\tfrac12g\dot\psi^2$ is therefore the normal acceleration
associated with the orbit's extrinsic bending in the physical kinetic
metric.  Leakage is controlled not by that curvature alone, but by its
competition with the transverse restoring operator $K_\perp$.

For the explicit kinetic realization
Eq.~\eqref{eq:collective-kinetic-ansatz},
\begin{equation}
    \boxed{
    g_e
    =
    \begin{pmatrix}
        2\mu_1m_\star\\
        8\mu_2q_\star
    \end{pmatrix}
    }
    \label{eq:explicit-g-even}
\end{equation}
in the even $(m,q)$ sector.  The centrifugal term
$\tfrac12 g\dot\psi^2$ has no linear component along the relative-phase
coordinate in these adapted coordinates.  This does not mean that a generic
external phase drive leaves $\chi$ unforced.  In particular, define
\begin{equation}
    c_{\rm kin}
    =
    \frac{2\mu_2q_\star^2}{{\cal I}_\star}.
    \label{eq:ckin}
\end{equation}
Since locally
$\widetilde\psi=\psi_1+c_{\rm kin}\chi$, a generalized torque $\tau$ coupled
to the original phase $\psi_1$ decomposes in the adapted coordinates as
\begin{equation}
    \delta W
    =
    \tau\,\delta\widetilde\psi
    -
    c_{\rm kin}\tau\,\delta\chi,
    \qquad
    F_\chi^{\rm ext}
    =
    -c_{\rm kin}\tau.
    \label{eq:phase-drive-normal-force}
\end{equation}
Thus only a drive matched to rigid motion along the kinetic Goldstone
direction has $F_\perp^{\rm ext}=0$; a generic phase drive produces a linear
normal forcing as well.

\subsection{Adiabatic leakage}
\label{subsec:adiabatic-leakage}

For steady motion
\begin{equation}
    \dot\psi=\Omega
\end{equation}
in the regime
\begin{equation}
    \Omega\ll\Delta_\perp,
    \label{eq:adiabatic-condition}
\end{equation}
the leading normal displacement is
\begin{equation}
    \boxed{
    y_{\rm leak}
    =
    \frac{\Omega^2}{2}
    K_\perp^{-1}g
    +
    O(\Omega^4).
    }
    \label{eq:adiabatic-leakage}
\end{equation}
Since
\begin{equation}
    K_\perp^{-1}
    =
    O(\lambda^{-1}),
\end{equation}
the strong-locking family gives
\begin{equation}
    \boxed{
    y_{\rm leak}
    =
    O\!\left(
        \frac{\Omega^2}{\lambda}
    \right)
    =
    O\!\left(
        \frac{\Omega^2}{\Delta_\perp^2}
    \right).
    }
    \label{eq:leakage-lambda-scaling}
\end{equation}

This is the first operational autonomy result.  At fixed spectral
polarization, the actual normal displacement induced by tangent motion can
be made parametrically small.

Equivalently, adiabatic elimination of the transverse modes produces only
$O(\lambda^{-1})$ corrections to the tangent effective theory; the explicit
effective-action calculation is given in Appendix~\ref{app:dynamics}.

\subsection{Frequency-dependent response and resonant breakdown}
\label{subsec:frequency-response}

Define the normal response operator
\begin{equation}
    \boxed{
    {\cal G}_\perp(\nu)
    =
    \left[
        K_\perp
        -
        \nu^2M_\perp
        +
        i\nu\Gamma_\perp
    \right]^{-1}.
    }
    \label{eq:normal-response-operator}
\end{equation}
For a periodic \emph{rigid tangent drive}, chosen so that
$F_\perp^{\rm ext}=0$ in the adapted coordinates,
\begin{equation}
    \psi(s)
    =
    \epsilon\cos\omega s,
\end{equation}
the normal sector is driven at zero frequency and at $2\omega$, with
\begin{equation}
    \boxed{
    y_{2\omega}
    =
    -
    \frac{\epsilon^2\omega^2}{4}
    {\cal G}_\perp(2\omega)\,g.
    }
    \label{eq:oscillatory-leakage}
\end{equation}

A generic phase drive can also contain a normal component already at the
fundamental frequency.  For example, a torque coupled to the original
$M_1$ phase gives Eq.~\eqref{eq:phase-drive-normal-force} and hence
\begin{equation}
    \chi_\omega
    =
    {\cal G}_\chi(\omega)
    F_{\chi,\omega}^{\rm ext}.
    \label{eq:linear-relative-phase-response}
\end{equation}
For an inertial phase modulation whose required torque scales as
$\tau_\omega=O(\epsilon\omega^2)$, the infrared bound below gives
\begin{equation}
    \boxed{
    \chi_\omega
    =
    O\!\left(
        \frac{\epsilon\omega^2}{\lambda}
    \right).
    }
    \label{eq:linear-relative-phase-scaling}
\end{equation}
This contribution is linear in the drive amplitude and can dominate the
quadratic centrifugal response for sufficiently small $\epsilon$.  Both
effects, however, are governed by the same $O(\lambda^{-1})$ transverse
susceptibility.

At frequencies well below the transverse spectrum this reduces to the same
$K_\perp^{-1}$-suppressed response as the adiabatic result.  In a weakly
damped system, however, the tangent-induced response is enhanced when
\begin{equation}
    2\omega
    \simeq
    \omega_{\perp,\alpha}.
    \label{eq:tangent-resonance}
\end{equation}
The transverse degrees of freedom have therefore not disappeared.  They
have been displaced to a higher excitation scale.

Dynamical autonomy in this construction is consequently an infrared
statement, not exact decoupling at arbitrary energy.

\subsection{Operational autonomy and the strong-locking limit}
\label{subsec:operational-autonomy}

To state the infrared result invariantly, define the mass-normalized
susceptibility
\begin{equation}
    \widehat{\cal G}_\perp(\nu)
    =
    \left[
        \widehat K_\perp
        -
        \nu^2\mathbbm{1}
        +
        i\nu\widehat\Gamma_\perp
    \right]^{-1},
    \label{eq:mass-normalized-response}
\end{equation}
where the hats denote the corresponding mass-normalized operators.

For every fixed infrared window
\begin{equation}
    |\nu|\leq\Omega_{\rm IR},
\end{equation}
the strong-locking limit obeys
\begin{equation}
    \boxed{
    \sup_{|\nu|\leq\Omega_{\rm IR}}
    \left\|
        \widehat{\cal G}_\perp(\nu)
    \right\|
    =
    O(\lambda^{-1}).
    }
    \label{eq:uniform-susceptibility-scaling}
\end{equation}
The operator-norm proof is given in
Appendix~\ref{app:dynamics}.  More generally, the same scaling holds for an
infrared window that grows with $\lambda$, provided
\begin{equation}
    \frac{
        \Omega_{\rm IR}(\lambda)
    }{
        \Delta_\perp(\lambda)
    }
    \longrightarrow0.
    \label{eq:growing-IR-window}
\end{equation}

The susceptibility conclusion is not restricted to the inertial ansatz.
For purely relaxational dynamics,
\begin{equation}
    \Gamma_\perp\dot y+K_\perp y=F_\perp,
\end{equation}
the response operator is
\begin{equation}
    {\cal G}^{\rm od}_\perp(\nu)
    =
    \left[
        K_\perp+i\nu\Gamma_\perp
    \right]^{-1},
\end{equation}
and the same positivity argument gives
\begin{equation}
    \sup_{|\nu|\leq\Omega_{\rm IR}}
    \left\|
        {\cal G}^{\rm od}_\perp(\nu)
    \right\|
    =
    O(\lambda^{-1}).
    \label{eq:overdamped-susceptibility}
\end{equation}
What is specific to the inertial realization is the centrifugal forcing
$\tfrac12g\dot\psi^2$ and the interpretation of $\Delta_\perp$ as a normal
oscillation frequency.

\begin{figure}[t]
    \centering
    \includegraphics[width=0.82\linewidth]{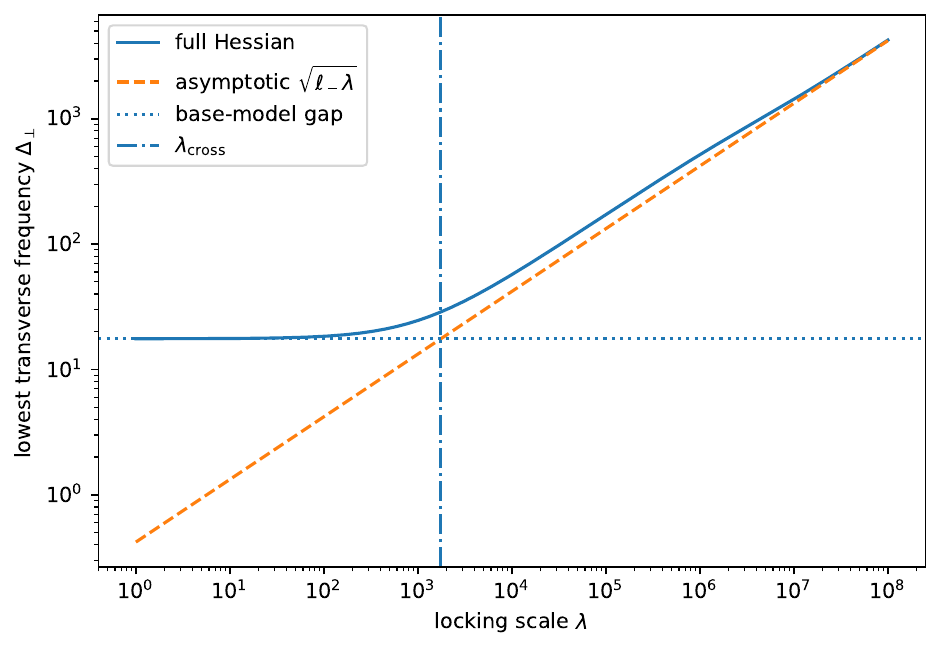}
    \caption{
    Lowest transverse normal frequency for the $m_\star=0.99$ benchmark
    with the full Hessian
    $K_\perp=K_\perp^{(0)}+\lambda L_\perp$ and illustrative
    mass normalization $M_\perp=\mathbbm{1}$.
    The gap begins at the substantial base-model value
    $\Delta_\perp\simeq17.55$, crosses over when the locking contribution
    becomes comparable to the soft base stiffness, and approaches the
    asymptotic $\lambda^{1/2}$ law only at large locking scale.
    }
    \label{fig:dynamical-gap-full}
\end{figure}

\begin{figure}[t]
    \centering
    \includegraphics[width=0.82\linewidth]{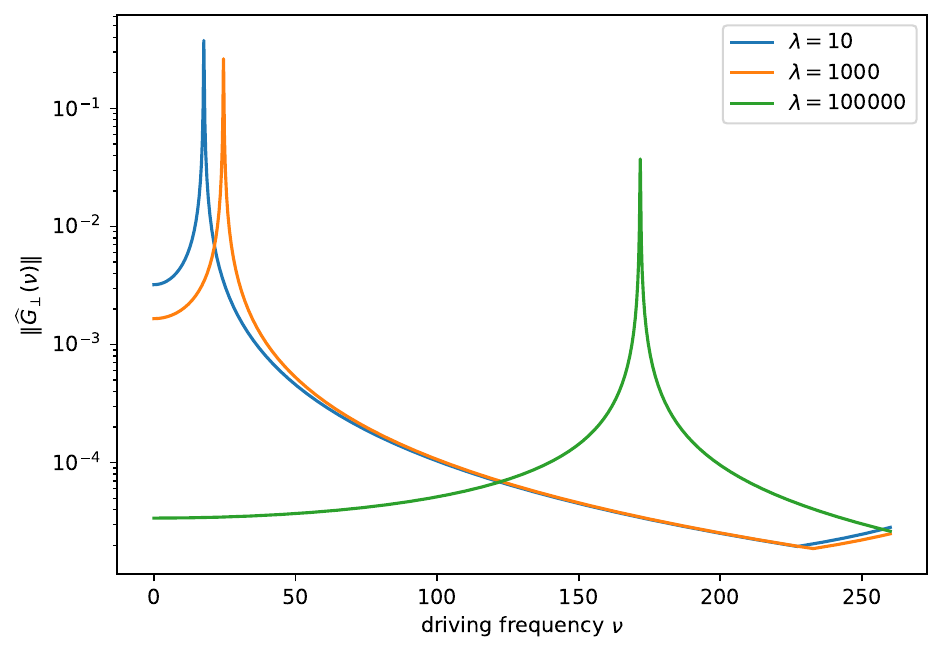}
    \caption{
    Representative mass-normalized response norm for the same full-Hessian
    benchmark, using $M_\perp=\mathbbm{1}$ and
    $\Gamma_\perp=0.15\,\mathbbm{1}$.
    Increasing $\lambda$ suppresses the infrared susceptibility and moves
    the lowest transverse resonance to higher frequency.  These kinetic
    values are illustrative; the $O(\lambda^{-1})$ infrared susceptibility
    bound does not require this particular inertial normalization.
    }
    \label{fig:dynamical-response-full}
\end{figure}

Thus the complete transverse susceptibility, not merely one particular
response coefficient, vanishes uniformly below the moving transverse scale.
Figures~\ref{fig:dynamical-gap-full} and
\ref{fig:dynamical-response-full} illustrate the inertial benchmark using
the \emph{full} Hessian, including the substantial base stiffness at
$m_\star=0.99$.

Combining the equilibrium and dynamical results gives
\begin{equation}
    \boxed{
    \begin{gathered}
        T_\star
        =
        \mathrm{constant},
        \qquad
        r_{\rm eff}(T_\star)
        =
        \mathrm{constant},
        \\[1mm]
        \Delta_\perp
        =
        \Theta(\lambda^{1/2}),
        \\[1mm]
        \sup_{|\nu|\leq\Omega_{\rm IR}}
        \left\|
            \widehat{\cal G}_\perp(\nu)
        \right\|
        =
        O(\lambda^{-1}),
        \\[1mm]
        y_{\rm leak}
        =
        O(\lambda^{-1})
        \quad
        \text{for fixed low-frequency driving}.
    \end{gathered}
    }
    \label{eq:dynamical-autonomy-summary}
\end{equation}

The central distinction may therefore be summarized as
\begin{equation}
    \boxed{
    \text{polarization selects the direction, while the gap protects it
    dynamically.}
    }
    \label{eq:selection-protection}
\end{equation}

In the regime
\begin{equation}
    r_{\rm eff}-1\ll1,
    \qquad
    \frac{\Omega_{\rm IR}}{\Delta_\perp}\ll1,
    \label{eq:effective-one-direction-regime}
\end{equation}
the $p=2$ director orbit has parametrically suppressed response into all
non-Goldstone collective modes throughout the accessible infrared window.
This is the precise sense in which the ordered sector is dynamically
autonomous in the present paper.

The distinction between the selected eigenline and its orientation variable
is essential.  The coordinate $\psi(s)$ parametrizes rotations of that
eigenline as the microscopic system evolves in $s$; it is not a coordinate
along the eigenline itself.  Neither object has yet been identified with a
spacetime clock, and no clock one-form, foliation, Lorentzian signature,
causal cone, or arrow of time has been derived.  Promoting the selected
eigenline to physical time requires additional structure, discussed in
Sec.~\ref{sec:discussion}.

\section{Discussion}
\label{sec:discussion}

The construction developed in this paper separates two properties that need
not coincide when a collective tensor approaches rank one.  Spectral
polarization selects one dominant projective direction, while dynamical
autonomy requires physical transverse response to remain parametrically
small throughout the low-frequency regime of interest.

The two-harmonic model first shows finite independence: states with the same
$M_1$, and hence the same tensor $T$, can possess different higher-harmonic
structure and different fluctuation spectra.  The locking construction then
makes the distinction parametric.  Along the globally controlled positive
locking ray,
\begin{equation}
    T_\star=\mathrm{constant},
    \qquad
    r_{\rm eff}(T_\star)=\mathrm{constant},
    \qquad
    K_\perp=\lambda L_\perp+O(1),
    \quad
    L_\perp>0,
    \label{eq:discussion-static-summary}
\end{equation}
so every non-Goldstone restoring coefficient is $\Theta(\lambda)$.  For a
finite positive nonsingular normal kinetic matrix,
\begin{equation}
    \Delta_\perp=\Theta(\lambda^{1/2}),
    \qquad
    \sup_{|\nu|\leq\Omega_{\rm IR}}
    \left\|
        \widehat{\mathcal G}_\perp(\nu)
    \right\|
    =
    O(\lambda^{-1}),
    \label{eq:discussion-dynamic-summary}
\end{equation}
and slow tangent motion produces normal leakage of order
$O(\Omega^2/\lambda)$.  Thus the same equilibrium tensor may describe either
a comparatively soft ordered state or a parametrically protected infrared
sector.

\subsection{From geometric autonomy to dynamical autonomy}

The distinction above is the principal reason for abandoning a purely
extrinsic-geometric definition of autonomy.

For the ordered $p=2$ family, the same sequence of increasingly polarized
states has qualitatively different extrinsic curvature in different natural
ambient metrics.  In the Hilbert--Schmidt metric the orientation curvature
remains finite and tends to $\sqrt{2}$; in the Fisher metric it tends to
$1/\sqrt{2}$ and the orbit becomes asymptotically horocyclic; in the Bures
metric it vanishes.  Moreover, the raw Fisher curvature of an $N$-fold
product family scales as $N^{-1/2}$ purely because the information metric is
extensive.

These results do not imply that state-space geometry is irrelevant.
They imply instead that a geometric criterion acquires physical content only
after the relevant geometry has been tied to a specified dynamics.  In the
present construction the quantities that control autonomy are the actual
kinetic and restoring operators of the collective theory,
\begin{equation}
    M_\perp,
    \qquad
    \Gamma_\perp,
    \qquad
    K_\perp,
\end{equation}
together with the response they generate.  The relevant small quantity is
not the bending of an arbitrarily chosen embedding but the physical
susceptibility of the transverse sector.

This change of viewpoint is closely analogous to the usual logic of
effective field theory and slow--fast reduction
\cite{AppelquistCarazzone1975,Fenichel1979}.  A degree of freedom need not disappear
microscopically in order to be absent from a low-energy description.  It is
sufficient that its excitation scale be large compared with the frequencies
being probed.  In the present model,
\begin{equation}
    \Omega_{\rm IR}\ll\Delta_\perp
\end{equation}
defines precisely such a regime.

\subsection{Effective temporal dimensionality}

The motivating interpretation of the collective tensor is temporal.  If the
microscopic structure underlying physical evolution is tensorial, then a
rank-polarized phase of the form
\begin{equation}
    T^{AB}
    \simeq
    \Lambda\,n^A n^B
\end{equation}
contains one dominant projective direction.  The results of this paper show
that the remaining collective directions need not vanish.  They may instead
remain present as massive or stiff transverse modes whose excitation is
suppressed below a dynamical gap.

For the $p=2$ model, this suggests a useful pair of conditions for a
rank-polarized state with an infrared-autonomous director sector:
\begin{equation}
    \boxed{
    r_{\rm eff}-1\ll1,
    \qquad
    \frac{\Omega_{\rm IR}}{\Delta_\perp}\ll1.
    }
    \label{eq:discussion-two-conditions}
\end{equation}
The first condition states that one tensor eigenline dominates the equilibrium
spectrum.  The second states that non-Goldstone departures from the director
orbit are dynamically inaccessible in the regime of interest.

The two conditions answer different questions.  Spectral polarization asks
which direction is selected.  The normal gap asks whether the additional
directions participate in low-energy dynamics.  The undeformed one-harmonic branch happens to correlate stronger polarization
with larger stiffness, as noted in Sec.~\ref{subsec:polarization-not-autonomy}.
The fixed-$T$ families of Secs.~\ref{sec:twoharmonic} and
\ref{sec:locking} show that this correlation is not enforced by the tensor
spectrum.  Dynamically soft polarized states can be reached, for example,
near the stability boundaries of Fig.~\ref{fig:locking-stability-domain};
whether broad such regimes arise in more microscopic models is an open
question.

This interpretation differs from postulating a fundamental preferred vector
or clock form.  The projective direction is selected by collective ordering,
and its low-energy isolation is controlled by an independent fluctuation
scale.  At the same time, the present model does not yet derive a spacetime
clock.  The index $A$ remains an internal collective index, and the
microscopic evolution parameter $s$ introduced in
Sec.~\ref{sec:dynamics} is not identified with the selected projective
direction.

Accordingly, the strongest statement established here is
\begin{equation}
    \boxed{
    \begin{gathered}
    \text{a tensorial collective system can select one dominant internal eigenline,}\\
    \text{while its $p=2$ director orbit becomes autonomous against
    non-Goldstone response in the infrared.}
    \end{gathered}
    }
    \label{eq:discussion-main-claim}
\end{equation}
This does not yet explain why physical evolution itself should be
one-dimensional: the microscopic parameter $s$ has been presupposed, and
$\psi(s)$ describes rotation of the selected eigenline rather than motion
along it.  Promoting the eigenline to physical time is a separate problem.

\subsection{What remains to obtain a clock}

Several additional structures are required before the selected direction can
be identified with an ordinary clock.

First, the internal projective direction must be soldered to a field on a
base manifold.  One would need a local tensor
\begin{equation}
    T_{\mu\nu}(x)
    \simeq
    \Lambda(x)\,
    \tau_\mu(x)\tau_\nu(x)
    \label{eq:discussion-soldering}
\end{equation}
whose dominant eigenline defines a candidate clock one-form $\tau$.

Second, the existence of a rank-one clock tensor does not imply
integrability.  A local temporal foliation requires a Frobenius condition
such as
\begin{equation}
    \tau\wedge d\tau=0,
    \label{eq:discussion-frobenius}
\end{equation}
while a locally exact clock requires the stronger condition
\begin{equation}
    d\tau=0,
    \qquad
    \tau=dt
\end{equation}
locally.  These are conditions on the variation of the selected direction
over a base manifold and are logically distinct from the state-space
polarization and dynamical gap studied here.

Third, the tensor
\begin{equation}
    T\sim\tau\otimes\tau
\end{equation}
is invariant under
\begin{equation}
    \tau\longrightarrow-\tau.
\end{equation}
The present mechanism therefore selects an unoriented temporal axis, not an
arrow of time.  Any orientation of histories, entropy production, or
thermodynamic arrow must arise from additional dynamical structure.

Finally, neither Lorentzian signature nor a causal cone follows merely from
rank polarization.  If the tensor is eventually to participate in the
effective spacetime geometry, the signature and coupling to matter must be
derived rather than assumed.

These distinctions are not deficiencies of the construction but part of its
logical structure.  The properties conventionally grouped under the word
``time'' need not have a common microscopic origin.

\subsection{Low-energy validity and breakdown}

The emergence described here is explicitly scale dependent.  The transverse
modes remain part of the microscopic theory.  They are absent from the
infrared description only because their response is suppressed when
\begin{equation}
    \omega\ll\Delta_\perp.
\end{equation}
Near a normal resonance,
\begin{equation}
    \omega
    \sim
    \omega_{\perp,\alpha},
\end{equation}
or for sufficiently strong nonlinear driving, the rank-polarized
low-energy description can fail.

This leads naturally to two distinct ways in which effective
one-dimensionality can break down:
\begin{align}
    &\text{depolarization:}
    &
    r_{\rm eff}-1
    &\not\ll1,
    \\
    &\text{gap failure:}
    &
    \omega/\Delta_\perp
    &\not\ll1.
\end{align}
The first removes the spectral preference for one direction.  The second can
occur even when the equilibrium tensor remains strongly polarized.

The distinction suggests that a more complete phase diagram need not consist
simply of an isotropic phase and a rank-one phase.  The engineered family contains dynamically soft polarized states near its
stability boundaries, even though the undeformed one-harmonic branch becomes
stiffer as it polarizes.  Determining whether extended soft-but-polarized
regimes occur in more microscopic or spatially local realizations is an
important open question.

\subsection{Limitations and scope}

The present model is deliberately minimal.  It is fully connected, so it
does not address spatial fluctuations, defects, finite-dimensional critical
behavior, or propagating collective modes on a physical base space.  The
locking terms of Eq.~\eqref{eq:locking-potential} are engineered
proof-of-principle interactions designed to isolate the distinction between
polarization and stiffness.  A more microscopic realization in which the
same locking structure arises dynamically, for example by integrating out
additional gapped fields, would be desirable.

The collective kinetic term
Eq.~\eqref{eq:collective-kinetic-ansatz} is a phenomenological inertial
realization rather than a derivation from the microscopic rotor Hamiltonian.
Hamiltonian mean-field dynamics and overdamped order-parameter dynamics lead
to different equations of motion
\cite{AntoniRuffo1995,CampaDauxoisRuffo2009,HohenbergHalperin1977}.
The strongest result here---the $O(\lambda^{-1})$ suppression of static and
low-frequency transverse susceptibility---persists in both inertial and
purely relaxational realizations.  The strong-locking limit is also closely
related to the classical problem of motion under increasingly strong
constraining forces \cite{RubinUngar1957,BornemannSchuette1997,FroeseHerbst2001}.

The analysis is classical and mean-field.  A quantum realization would
replace the classical normal spectrum by the excitation spectrum of the
corresponding many-body Hamiltonian and would permit additional geometric
structures, including an adiabatic Berry connection for gapped fast modes.
Such a connection is not assumed here.  In the minimal real projective
sector, nontrivial Berry curvature does not follow from rank polarization
alone.

We have also not coupled the collective tensor to gravity or ordinary
matter.  Consequently no claim is made here about observational constraints,
cosmological evolution, dark sectors, or modifications of causal structure.
Such questions require a local field theory and a specified coupling between
the tensorial sector and spacetime.

\subsection{Outlook}

The results of this paper suggest a hierarchy of questions rather than a
single emergence mechanism:
\begin{equation}
    \boxed{
    \begin{array}{c}
    \text{spectral polarization}
    \\[1mm]
    \Downarrow
    \\[-1mm]
    \text{one dominant projective eigenline}
    \\[2mm]
    +\ \text{suppression of non-Goldstone modes}
    \\[1mm]
    \Downarrow
    \\[-1mm]
    \text{infrared-autonomous director sector}
    \\[2mm]
    +\ \text{soldering and integrability}
    \\[1mm]
    \Downarrow
    \\[-1mm]
    \text{clock structure}
    \\[2mm]
    +\ \text{orientation and nonequilibrium dynamics}
    \\[1mm]
    \Downarrow
    \\[-1mm]
    \text{arrow of time}.
    \end{array}
    }
    \label{eq:discussion-hierarchy}
\end{equation}

The present work addresses the first two stages.  Its main result is that
rank polarization does not determine autonomy, but that the latter can be
generated independently through a physical separation of scales:
\begin{equation}
    \boxed{
    \text{rank polarization}
    +
    \text{transverse suppression}
    \longrightarrow
    \text{an infrared-autonomous ordered sector}.
    }
    \label{eq:discussion-final}
\end{equation}

The next question is whether the selected projective direction can be
promoted to an integrable clock one-form on a base manifold.  Once a gapped
slow--fast decomposition exists, one may also ask whether adiabatic
elimination of the transverse quantum sector induces a nontrivial Berry
connection over the resulting clock variables
\cite{Berry1984,KolodrubetzEtAl2017}.  These structures are
logically independent: clock nonintegrability and quantum geometric
holonomy need not coincide.

In this way the tensorial viewpoint changes the problem of temporal
dimensionality.  The familiar scalar clock variable need not be fundamental.
It may instead arise only after a higher-rank collective structure has
selected one eigenline, dynamically suppressed non-Goldstone departures
from its director orbit, and satisfied further geometric conditions required
of a clock.


\section*{Acknowledgments}

The authors used generative AI systems, including OpenAI ChatGPT and
Anthropic Claude, to assist with algebraic and numerical cross-checks,
literature discovery, adversarial review, and manuscript development and
editing. The authors independently reviewed the calculations, references,
arguments, and conclusions and take responsibility for the content of the
manuscript.  

\appendix

\section{Projective mean-field theory in general dimension}
\label{app:meanfield}

This appendix collects details of the equilibrium projective mean-field
theory used in Sec.~\ref{sec:polarization}.  We derive the large-$N$
variational problem, its Bingham saddle, the uniaxial reduction in arbitrary
dimension, the expansion about the isotropic state, and the
strong-polarization asymptotics.

Throughout the appendix, $d\mu_p(u)$ denotes the normalized
$O(p)$-invariant measure on $S^{p-1}$.  Since all observables considered
below are invariant under $u\rightarrow-u$, integration over
$S^{p-1}$ is equivalent to integration over
$\mathbb{RP}^{p-1}$ with its normalized invariant measure.

\subsection{Large-\texorpdfstring{$N$}{N} variational principle and the
Bingham saddle}
\label{app:meanfield-variational}

Consider the fully connected Hamiltonian
\begin{equation}
    H_0
    =
    -\frac{J}{2N}
    \sum_{i,j=1}^N
    (u_i\cdot u_j)^2,
    \qquad
    u_i\in S^{p-1},
    \qquad
    u_i\sim-u_i.
    \label{eq:app-H0}
\end{equation}
Introducing the empirical second moment
\begin{equation}
    T_N
    =
    \frac{1}{N}
    \sum_{i=1}^N
    u_i u_i^{\mathsf T},
    \label{eq:app-TN}
\end{equation}
one has
\begin{equation}
\begin{split}
    \Tr T_N^2
    &=
    \frac{1}{N^2}
    \sum_{i,j}
    \Tr
    \left(
        u_i u_i^{\mathsf T}
        u_j u_j^{\mathsf T}
    \right)
\\
    &=
    \frac{1}{N^2}
    \sum_{i,j}
    (u_i\cdot u_j)^2.
\end{split}
\end{equation}
Hence
\begin{equation}
    H_0
    =
    -\frac{JN}{2}\Tr T_N^2.
    \label{eq:app-H0-T}
\end{equation}

At large $N$, a configuration may be characterized by a normalized
one-site probability density $\rho(u)$ relative to $d\mu_p$:
\begin{equation}
    \int d\mu_p(u)\,\rho(u)=1.
\end{equation}
Its second moment is
\begin{equation}
    T[\rho]
    =
    \int d\mu_p(u)\,
    \rho(u)\,
    uu^{\mathsf T}.
    \label{eq:app-T-rho}
\end{equation}
The dimensionless free-energy functional, up to a
$\rho$-independent constant, is
\begin{equation}
    \boxed{
    \mathcal F_\kappa[\rho]
    =
    \int d\mu_p(u)\,
        \rho(u)\log\rho(u)
    -
    \frac{\kappa}{2}
        \Tr T[\rho]^2,
    }
    \qquad
    \kappa=\beta J.
    \label{eq:app-functional}
\end{equation}
This is the usual mean-field entropy--interaction variational problem for
projective or nematic order
\cite{MaierSaupe1958,MaierSaupe1960,BallMajumdar2010}.

Varying $\rho$ subject to normalization gives
\begin{equation}
    0
    =
    \delta\mathcal F_\kappa
    =
    \int d\mu_p(u)\,
    \delta\rho(u)
    \left[
        1+\log\rho(u)
        -
        \kappa\,u^{\mathsf T}Tu
        -
        \zeta
    \right],
\end{equation}
where $\zeta$ is a Lagrange multiplier.  Thus every stationary density has
the form
\begin{equation}
    \boxed{
    \rho_T(u)
    =
    \frac{
        \exp\!\left[
            \kappa\,u^{\mathsf T}Tu
        \right]
    }{
        Z(T)
    },
    }
    \label{eq:app-Bingham}
\end{equation}
where
\begin{equation}
    Z(T)
    =
    \int d\mu_p(u)\,
    \exp\!\left[
        \kappa\,u^{\mathsf T}Tu
    \right],
\end{equation}
together with the self-consistency condition
\begin{equation}
    T
    =
    \int d\mu_p(u)\,
        \rho_T(u)\,
        uu^{\mathsf T}.
    \label{eq:app-T-selfconsistent}
\end{equation}
Equation~\eqref{eq:app-Bingham} is a Bingham distribution
\cite{Bingham1974}.

Because adding a multiple of the identity to the matrix in the exponent
changes only the normalization, the Bingham natural parameter is defined
only modulo
\begin{equation}
    A\longrightarrow A+c\,\mathbbm{1}.
\end{equation}
The fixed-trace condition
\begin{equation}
    \Tr T=1
\end{equation}
provides a convenient representative of this equivalence.

\subsection{Uniaxial reduction}
\label{app:meanfield-uniaxial}

For a uniaxial configuration with director $n$,
\begin{equation}
    T
    =
    \frac{1}{p}\mathbbm{1}
    +
    S
    \left(
        nn^{\mathsf T}
        -
        \frac{1}{p}\mathbbm{1}
    \right),
    \label{eq:app-uniaxial-T}
\end{equation}
where $S$ is the scalar order parameter.  Define
\begin{equation}
    x=n\cdot u
\end{equation}
and the normalized quadratic invariant
\begin{equation}
    \boxed{
    P_p(x)
    =
    \frac{p x^2-1}{p-1}.
    }
    \label{eq:app-Pp}
\end{equation}
It obeys
\begin{equation}
    P_p(1)=1,
    \qquad
    \langle P_p\rangle_0=0,
\end{equation}
where $\langle\cdots\rangle_0$ denotes the uniform measure.

Using
\begin{equation}
    x^2
    =
    \frac{1+(p-1)P_p(x)}{p},
\end{equation}
one obtains
\begin{equation}
\begin{split}
    u^{\mathsf T}Tu
    &=
    \frac1p
    +
    S
    \left(
        x^2-\frac1p
    \right)
\\
    &=
    \frac1p
    +
    \frac{p-1}{p}
    S\,P_p(x).
\end{split}
    \label{eq:app-uTu}
\end{equation}
The isotropic constant may be absorbed into the partition function.  The
uniaxial saddle therefore takes the form
\begin{equation}
    \rho_h(u)
    =
    \frac{
        e^{hP_p(n\cdot u)}
    }{
        Z_p(h)
    },
    \label{eq:app-rho-h}
\end{equation}
with
\begin{equation}
    \boxed{
    h
    =
    \kappa
    \frac{p-1}{p}
    S.
    }
    \label{eq:app-h-kappa-S}
\end{equation}
Here
\begin{equation}
    Z_p(h)
    =
    \left\langle
        e^{hP_p}
    \right\rangle_0.
    \label{eq:app-Zp}
\end{equation}

Self-consistency is equivalent to
\begin{equation}
    \boxed{
    S
    =
    \frac{d}{dh}
    \log Z_p(h)
    =
    \langle P_p\rangle_h.
    }
    \label{eq:app-selfconsistency}
\end{equation}

For functions depending only on $x=n\cdot u$, the invariant measure reduces
to
\begin{equation}
    d\mu_p(x)
    =
    C_p
    (1-x^2)^{(p-3)/2}\,dx,
    \qquad
    -1\leq x\leq1,
    \label{eq:app-xmeasure}
\end{equation}
where
\begin{equation}
    C_p
    =
    \frac{
        \Gamma(p/2)
    }{
        \sqrt{\pi}\,
        \Gamma((p-1)/2)
    }.
\end{equation}
Thus
\begin{equation}
    Z_p(h)
    =
    C_p
    \int_{-1}^{1}
    dx\,
    (1-x^2)^{(p-3)/2}
    e^{hP_p(x)}.
    \label{eq:app-Zp-integral}
\end{equation}

The eigenvalues of Eq.~\eqref{eq:app-uniaxial-T} are
\begin{equation}
    \lambda_\parallel
    =
    \frac{1+(p-1)S}{p},
    \qquad
    \lambda_\perp
    =
    \frac{1-S}{p},
    \label{eq:app-uniaxial-eigenvalues}
\end{equation}
with $\lambda_\perp$ of multiplicity $p-1$.  Hence
\begin{equation}
    \Tr T^2
    =
    \frac{
        1+(p-1)S^2
    }{p},
\end{equation}
and
\begin{equation}
    \boxed{
    r_{\rm eff}(S)
    =
    \frac{
        p
    }{
        1+(p-1)S^2
    }.
    }
    \label{eq:app-reff-S}
\end{equation}

\subsection{Constrained free energy and Legendre form}
\label{app:meanfield-rate}

It is useful to formulate the uniaxial problem directly in terms of the
order parameter $S$.  The relation between molecular-field and Landau free
energies has a substantial liquid-crystal literature
\cite{KatrielKventselLuckhurstSluckin1986}.  Define the rate function
\begin{equation}
    I_p(S)
    =
    \sup_h
    \left[
        hS-\log Z_p(h)
    \right].
    \label{eq:app-rate}
\end{equation}
At an interior point, the maximizing $h$ satisfies
Eq.~\eqref{eq:app-selfconsistency}, and therefore
\begin{equation}
    \frac{dI_p}{dS}=h.
\end{equation}

For a uniaxial tensor,
\begin{equation}
    \Tr T^2
    =
    \frac{
        1+(p-1)S^2
    }{p}.
\end{equation}
Up to the $S$-independent isotropic energy
$-\kappa/(2p)$, the constrained mean-field free energy is
\begin{equation}
    \boxed{
    \Delta f_\kappa(S)
    =
    I_p(S)
    -
    \frac{
        \kappa(p-1)
    }{
        2p
    }
    S^2.
    }
    \label{eq:app-constrained-free-energy}
\end{equation}
Stationarity gives
\begin{equation}
    h
    =
    \kappa
    \frac{p-1}{p}
    S,
\end{equation}
recovering Eq.~\eqref{eq:app-h-kappa-S}.

The constrained free energy
Eq.~\eqref{eq:app-constrained-free-energy} should be distinguished from
auxiliary-field potentials obtained by substituting the mean-field source
directly into $\log Z_p$.  The two functions need not coincide away from
stationary points, but they have the same physical stationary states and
the same free energy there.  The constrained form is particularly convenient
for fluctuation and stability questions because $S$ is the physical moment
being held fixed.

\subsection{Isotropic moments and the Landau expansion}
\label{app:meanfield-landau}

The uniform moments of a Cartesian component of a unit vector in
$p$ dimensions are
\begin{equation}
    \left\langle x^{2n}\right\rangle_0
    =
    \frac{
        (2n-1)!!
    }{
        p(p+2)\cdots(p+2n-2)
    }.
    \label{eq:app-x-even-moments}
\end{equation}
In particular,
\begin{equation}
    \langle x^2\rangle_0
    =
    \frac1p,
\end{equation}
\begin{equation}
    \langle x^4\rangle_0
    =
    \frac{3}{p(p+2)},
\end{equation}
and
\begin{equation}
    \langle x^6\rangle_0
    =
    \frac{15}{p(p+2)(p+4)}.
\end{equation}

Using Eq.~\eqref{eq:app-Pp}, one finds
\begin{equation}
\begin{split}
    \left\langle P_p^2\right\rangle_0
    &=
    \frac{
        p^2\langle x^4\rangle_0
        -
        2p\langle x^2\rangle_0
        +
        1
    }{
        (p-1)^2
    }
\\
    &=
    \boxed{
    \frac{
        2
    }{
        (p-1)(p+2)
    }}.
\end{split}
    \label{eq:app-P2}
\end{equation}
Similarly,
\begin{equation}
\begin{split}
    \left\langle P_p^3\right\rangle_0
    &=
    \frac{
        p^3\langle x^6\rangle_0
        -
        3p^2\langle x^4\rangle_0
        +
        3p\langle x^2\rangle_0
        -
        1
    }{
        (p-1)^3
    }
\\
    &=
    \boxed{
    \frac{
        8(p-2)
    }{
        (p-1)^2(p+2)(p+4)
    }}.
\end{split}
    \label{eq:app-P3}
\end{equation}

Let
\begin{equation}
    \mu_2
    =
    \langle P_p^2\rangle_0,
    \qquad
    \mu_3
    =
    \langle P_p^3\rangle_0.
\end{equation}
Since $\langle P_p\rangle_0=0$, the cumulant expansion gives
\begin{equation}
    \log Z_p(h)
    =
    \frac{\mu_2}{2}h^2
    +
    \frac{\mu_3}{6}h^3
    +
    O(h^4).
    \label{eq:app-logZ-small-h}
\end{equation}
Therefore
\begin{equation}
    S
    =
    \mu_2 h
    +
    \frac{\mu_3}{2}h^2
    +
    O(h^3).
    \label{eq:app-S-small-h}
\end{equation}
Inverting,
\begin{equation}
    h
    =
    \frac{S}{\mu_2}
    -
    \frac{
        \mu_3
    }{
        2\mu_2^3
    }
    S^2
    +
    O(S^3).
    \label{eq:app-h-small-S}
\end{equation}
Substitution into the Legendre transform yields
\begin{equation}
    I_p(S)
    =
    \frac{
        S^2
    }{
        2\mu_2
    }
    -
    \frac{
        \mu_3
    }{
        6\mu_2^3
    }
    S^3
    +
    O(S^4).
    \label{eq:app-I-small-S}
\end{equation}
Hence
\begin{equation}
\begin{split}
    \Delta f_\kappa(S)
    =
    &
    \frac12
    \left[
        \frac1{\mu_2}
        -
        \kappa\frac{p-1}{p}
    \right]
    S^2
\\
    &
    -
    \frac{
        \mu_3
    }{
        6\mu_2^3
    }
    S^3
    +
    O(S^4).
\end{split}
    \label{eq:app-Landau-general}
\end{equation}

The isotropic state loses local stability when the quadratic coefficient
vanishes:
\begin{equation}
    \kappa_\ast
    =
    \frac{
        p
    }{
        (p-1)\mu_2
    }.
\end{equation}
Using Eq.~\eqref{eq:app-P2},
\begin{equation}
    \boxed{
    \kappa_\ast
    =
    \frac{
        p(p+2)
    }{2}.
    }
    \label{eq:app-kappa-spinodal}
\end{equation}

The cubic coefficient may also be written explicitly:
\begin{equation}
    -
    \frac{
        \mu_3
    }{
        6\mu_2^3
    }
    =
    -
    \frac{
        (p-2)(p-1)(p+2)^2
    }{
        6(p+4)
    }.
    \label{eq:app-cubic-coeff}
\end{equation}
It vanishes only for
\begin{equation}
    p=2.
\end{equation}
Thus the continuous $p=2$ transition is exceptional within the uniaxial
mean-field family.  For $p>2$, the symmetry permits a cubic contribution and
the isotropic spinodal need not coincide with the thermodynamic coexistence
point.

\subsection{The \texorpdfstring{$p=2$}{p=2} specialization}
\label{app:meanfield-p2}

For $p=2$,
\begin{equation}
    P_2(\cos\theta)
    =
    2\cos^2\theta-1
    =
    \cos 2\theta.
\end{equation}
Writing the doubled projective angle as
\begin{equation}
    x=2\theta,
\end{equation}
the partition function becomes
\begin{equation}
    Z_2(h)
    =
    \int_0^{2\pi}
    \frac{dx}{2\pi}
    e^{h\cos x}
    =
    I_0(h),
    \label{eq:app-Z2}
\end{equation}
and
\begin{equation}
    S
    =
    \frac{
        I_1(h)
    }{
        I_0(h)
    }.
    \label{eq:app-p2-S}
\end{equation}
Identifying $S=m$, Eq.~\eqref{eq:app-h-kappa-S} gives
\begin{equation}
    h=\frac{\kappa m}{2},
\end{equation}
and therefore
\begin{equation}
    m
    =
    \frac{
        I_1(\kappa m/2)
    }{
        I_0(\kappa m/2)
    },
\end{equation}
as used in Sec.~\ref{subsec:p2}.

Because all odd moments of $\cos x$ vanish under the isotropic measure, the
Landau expansion contains only even powers of $m$.  The quadratic
instability occurs at
\begin{equation}
    \kappa_\ast=4,
\end{equation}
and coincides with the continuous transition.

For comparison with the auxiliary-field potential used in
Sec.~\ref{subsec:p2}, one may write
\begin{equation}
    f_{\rm aux}(m)
    =
    -\frac{\kappa}{4}
    +
    \frac{\kappa m^2}{4}
    -
    \log I_0\!\left(
        \frac{\kappa m}{2}
    \right).
    \label{eq:app-p2-aux}
\end{equation}
Its stationary condition is exactly the self-consistency equation above.
The constrained free energy
Eq.~\eqref{eq:app-constrained-free-energy} and
$f_{\rm aux}$ are different off-shell parameterizations of the
mean-field variational problem, but agree on the physical stationary
branches.

\subsection{The \texorpdfstring{$p=3$}{p=3} coexistence point}
\label{app:meanfield-p3}

For $p=3$,
\begin{equation}
    P_3(x)
    =
    \frac{
        3x^2-1
    }{2},
\end{equation}
and the reduced invariant measure is simply
\begin{equation}
    d\mu_3(x)
    =
    \frac12\,dx.
\end{equation}
The partition function is therefore
\begin{equation}
    Z_3(h)
    =
    \frac12
    \int_{-1}^{1}
    dx\,
    \exp\!\left[
        \frac{h}{2}
        (3x^2-1)
    \right].
    \label{eq:app-Z3-integral}
\end{equation}
For $h>0$ this may be expressed as
\begin{equation}
    Z_3(h)
    =
    e^{-h/2}
    \frac{
        \sqrt{\pi}
    }{
        2\sqrt{3h/2}
    }
    \operatorname{erfi}
    \!\left(
        \sqrt{\frac{3h}{2}}
    \right).
    \label{eq:app-Z3-erfi}
\end{equation}

The ordered branch satisfies
\begin{equation}
    S(h)
    =
    \frac{d}{dh}
    \log Z_3(h)
\end{equation}
and
\begin{equation}
    \kappa
    =
    \frac{
        3h
    }{
        2S(h)
    }.
    \label{eq:app-kappa-p3-h}
\end{equation}
At coexistence, the ordered and isotropic constrained free energies are
equal:
\begin{equation}
    I_3(S)
    -
    \frac{\kappa}{3}S^2
    =
    0.
    \label{eq:app-p3-coexistence-condition}
\end{equation}
Using
\begin{equation}
    I_3(S)
    =
    hS-\log Z_3(h)
\end{equation}
and Eq.~\eqref{eq:app-kappa-p3-h}, this reduces to the convenient
one-variable condition
\begin{equation}
    \boxed{
    \log Z_3(h)
    =
    \frac12 h S(h).
    }
    \label{eq:app-p3-coexistence-h}
\end{equation}

Numerical solution gives
\begin{equation}
    \boxed{
    \kappa_c
    \simeq
    6.81218849,
    \qquad
    S_c
    \simeq
    0.42902903.
    }
    \label{eq:app-p3-values}
\end{equation}
Using Eq.~\eqref{eq:app-reff-S},
\begin{equation}
    \boxed{
    r_{\rm eff}(S_c)
    \simeq
    2.19277118.
    }
    \label{eq:app-p3-reff}
\end{equation}
The isotropic spinodal,
\begin{equation}
    \kappa_\ast
    =
    \frac{3(3+2)}{2}
    =
    \frac{15}{2},
\end{equation}
lies above the coexistence value, as expected for a first-order
mean-field transition.

No result below depends on the complete phase structure for arbitrary
$p$.  The $p=3$ calculation is included only to demonstrate explicitly
that the projective ordering mechanism is not restricted to
$\mathbb{RP}^{1}$ and that the continuous $p=2$ onset is nongeneric.

\subsection{Strong-polarization asymptotics}
\label{app:meanfield-strong}

We finally derive the strong-coupling behavior quoted in
Sec.~\ref{subsec:general-p}.  Choose coordinates in a neighborhood of the
selected director $n$ such that
\begin{equation}
    u
    =
    \left(
        \sqrt{1-|y|^2},
        y
    \right),
    \qquad
    y\in\mathbb R^{p-1},
    \qquad
    |y|\ll1.
    \label{eq:app-local-y}
\end{equation}
Then
\begin{equation}
    (n\cdot u)^2
    =
    1-|y|^2,
\end{equation}
and therefore
\begin{equation}
\begin{split}
    P_p(n\cdot u)
    &=
    \frac{
        p(1-|y|^2)-1
    }{
        p-1
    }
\\
    &=
    1
    -
    \frac{p}{p-1}
    |y|^2.
\end{split}
    \label{eq:app-P-local}
\end{equation}
Using
\begin{equation}
    h
    =
    \kappa\frac{p-1}{p}S,
\end{equation}
the Bingham weight becomes, to leading order,
\begin{equation}
    e^{hP_p}
    =
    e^h
    \exp\!\left[
        -\kappa S |y|^2
    \right]
    \left[
        1+O(|y|^4)
    \right].
    \label{eq:app-Gaussian-weight}
\end{equation}
Corrections from the spherical measure begin at higher order and do not
affect the leading result.

Thus the transverse fluctuations are asymptotically Gaussian in
$p-1$ dimensions:
\begin{equation}
    \langle |y|^2\rangle
    =
    \frac{
        p-1
    }{
        2\kappa S
    }
    +
    O(\kappa^{-2}).
    \label{eq:app-y2}
\end{equation}
Substituting into Eq.~\eqref{eq:app-P-local},
\begin{equation}
\begin{split}
    S
    &=
    \langle P_p\rangle
\\
    &=
    1
    -
    \frac{p}{p-1}
    \langle |y|^2\rangle
    +
    O(\kappa^{-2})
\\
    &=
    1
    -
    \frac{p}{2\kappa S}
    +
    O(\kappa^{-2}).
\end{split}
    \label{eq:app-S-implicit-strong}
\end{equation}
Solving iteratively gives
\begin{equation}
    \boxed{
    S
    =
    1
    -
    \frac{p}{2\kappa}
    +
    O(\kappa^{-2}).
    }
    \label{eq:app-S-strong}
\end{equation}

Using
\begin{equation}
    S^2
    =
    1-\frac{p}{\kappa}
    +
    O(\kappa^{-2}),
\end{equation}
the denominator of Eq.~\eqref{eq:app-reff-S} becomes
\begin{equation}
\begin{split}
    1+(p-1)S^2
    &=
    p
    -
    \frac{
        p(p-1)
    }{\kappa}
    +
    O(\kappa^{-2})
\\
    &=
    p
    \left[
        1
        -
        \frac{p-1}{\kappa}
        +
        O(\kappa^{-2})
    \right].
\end{split}
\end{equation}
Therefore
\begin{equation}
    \boxed{
    r_{\rm eff}
    =
    1
    +
    \frac{
        p-1
    }{\kappa}
    +
    O(\kappa^{-2}),
    }
    \label{eq:app-reff-strong}
\end{equation}
or equivalently
\begin{equation}
    \boxed{
    r_{\rm eff}-1
    \sim
    \frac{p-1}{\kappa}.
    }
    \label{eq:app-reff-strong-leading}
\end{equation}

The approach to effective rank one is thus controlled by ordinary
transverse equipartition about the selected projective direction.  Exact
rank one is reached only in the infinite-concentration limit of this
classical mean-field model.

\subsection{Finite-\texorpdfstring{$N$}{N} symmetry and the order of limits}
\label{app:meanfield-finiteN}

For completeness, we make explicit the finite-$N$ qualification discussed
in Sec.~\ref{subsec:effective-rank}.  At zero external field the Gibbs
measure is exactly $O(p)$ invariant.  Consequently,
\begin{equation}
    \langle T_N\rangle
    =
    R
    \langle T_N\rangle
    R^{\mathsf T}
\end{equation}
for every $R\in O(p)$, and Schur invariance together with
$\Tr T_N=1$ implies
\begin{equation}
    \boxed{
    \langle T_N\rangle
    =
    \frac1p\mathbbm{1}.
    }
    \label{eq:app-finiteN-T}
\end{equation}
This statement does not imply that typical ordered configurations have
isotropic spectra.

One may characterize the ordered phase using the empirical eigenvalues of
$T_N$, a symmetry-breaking boundary condition, or a quasi-average.  If an
infinitesimal field selecting a director $n$ is denoted by $h_{\rm sb}$,
the symmetry-broken order parameter corresponds schematically to the order
of limits
\begin{equation}
    \lim_{h_{\rm sb}\to0^+}
    \lim_{N\to\infty}
    \langle T_N\rangle_{h_{\rm sb}},
    \label{eq:app-quasiaverage}
\end{equation}
rather than
\begin{equation}
    \lim_{N\to\infty}
    \langle T_N\rangle_{h_{\rm sb}=0}.
\end{equation}
All mean-field ordered tensors in the main text should be understood in
this standard thermodynamic sense.

\section{Geometry of the polarized statistical family}
\label{app:geometry}

This appendix collects the geometric calculations used in
Sec.~\ref{sec:geometry}.  We first derive the Fisher metric of the
$p=2$ polarized family and the geodesic curvature of its orientation
orbits.  We then determine the strong-polarization asymptotics and compare
the result with two other natural metrics on the same family.  Finally, we
show explicitly how Fisher extensivity produces an apparent suppression of
extrinsic curvature that is unrelated to dynamical autonomy.

\subsection{Fisher metric of the von Mises family}
\label{app:fisher-metric}

For $p=2$, the projective angle may be represented by the doubled coordinate
$x=2\theta$.  The ordered one-site family is
\begin{equation}
    \rho(x|r,\psi)
    =
    \frac{
        e^{r\cos(x-\psi)}
    }{
        2\pi I_0(r)
    },
    \label{eq:app-fisher-density}
\end{equation}
where $r\geq0$ is the concentration and $\psi$ is the common orientation.
Define
\begin{equation}
    A(r)
    =
    \frac{I_1(r)}{I_0(r)}.
    \label{eq:app-A-def}
\end{equation}
Then
\begin{equation}
    \langle\cos(x-\psi)\rangle=A(r).
\end{equation}

The log likelihood is
\begin{equation}
    \ell(x;r,\psi)
    =
    r\cos(x-\psi)
    -
    \log\!\left[
        2\pi I_0(r)
    \right].
    \label{eq:app-loglikelihood}
\end{equation}
The two score functions are
\begin{align}
    \partial_r\ell
    &=
    \cos(x-\psi)-A(r),
    \label{eq:app-score-r}
    \\
    \partial_\psi\ell
    &=
    r\sin(x-\psi).
    \label{eq:app-score-psi}
\end{align}

The Fisher metric is
\begin{equation}
    g_{ab}
    =
    \left\langle
        \partial_a\ell\,
        \partial_b\ell
    \right\rangle.
    \label{eq:app-Fisher-def}
\end{equation}
Reflection symmetry about the director immediately gives
\begin{equation}
    g_{r\psi}=0.
\end{equation}

For the radial component,
\begin{equation}
\begin{split}
    g_{rr}
    &=
    \left\langle
        \left(
            \cos(x-\psi)-A
        \right)^2
    \right\rangle
\\
    &=
    \frac{dA}{dr}.
\end{split}
\end{equation}
The standard Bessel identity gives
\begin{equation}
    \boxed{
    A'(r)
    =
    1-\frac{A(r)}{r}-A(r)^2.
    }
    \label{eq:app-Aprime}
\end{equation}

For the angular component,
\begin{equation}
    g_{\psi\psi}
    =
    r^2
    \left\langle
        \sin^2(x-\psi)
    \right\rangle.
\end{equation}
Using
\begin{equation}
    \left\langle
        \cos^2(x-\psi)
    \right\rangle
    =
    A'(r)+A(r)^2
\end{equation}
and Eq.~\eqref{eq:app-Aprime},
\begin{equation}
    \left\langle
        \sin^2(x-\psi)
    \right\rangle
    =
    \frac{A(r)}{r}.
\end{equation}
Hence
\begin{equation}
    g_{\psi\psi}
    =
    rA(r).
\end{equation}

The Fisher metric is therefore
\begin{equation}
    \boxed{
    ds_F^2
    =
    A'(r)\,dr^2
    +
    rA(r)\,d\psi^2.
    }
    \label{eq:app-Fisher-metric}
\end{equation}

\subsection{Geodesic curvature of an orientation orbit}
\label{app:orbit-curvature}

Consider more generally a two-dimensional metric
\begin{equation}
    ds^2
    =
    E(r)\,dr^2
    +
    G(r)\,d\psi^2.
    \label{eq:app-diagonal-metric}
\end{equation}
The fixed-$r$ orientation orbit has unit tangent
\begin{equation}
    t
    =
    \frac{1}{\sqrt{G}}
    \partial_\psi.
    \label{eq:app-unit-tangent}
\end{equation}
The only Christoffel symbol required for its normal acceleration is
\begin{equation}
    \Gamma^r_{\psi\psi}
    =
    -\frac{G'}{2E}.
    \label{eq:app-Christoffel-rpsipsi}
\end{equation}
Thus
\begin{equation}
\begin{split}
    \nabla_t t
    &=
    \frac{1}{G}
    \Gamma^r_{\psi\psi}
    \partial_r
\\
    &=
    -
    \frac{G'}{2EG}
    \partial_r.
\end{split}
\end{equation}
Taking its norm gives
\begin{equation}
    \boxed{
    k_g
    =
    \|\nabla_t t\|
    =
    \frac{
        |G'(r)|
    }{
        2G(r)\sqrt{E(r)}
    }.
    }
    \label{eq:app-general-kg}
\end{equation}

For the Fisher metric,
\begin{equation}
    E(r)=A'(r),
    \qquad
    G(r)=rA(r),
\end{equation}
and therefore
\begin{equation}
    \boxed{
    k_F(r)
    =
    \frac{
        A(r)+rA'(r)
    }{
        2rA(r)\sqrt{A'(r)}
    }.
    }
    \label{eq:app-Fisher-kg}
\end{equation}
For $r>0$, the numerator is positive, so the absolute value has been
suppressed.

Because the orientation orbit is one dimensional, the norm of its second
fundamental form in the two-dimensional ambient statistical manifold is
identical to $k_F$.

\subsection{Gaussian curvature}
\label{app:Fisher-Gaussian}

For the diagonal metric
Eq.~\eqref{eq:app-diagonal-metric}, introduce proper radial distance
\begin{equation}
    dR
    =
    \sqrt{E(r)}\,dr
\end{equation}
and write
\begin{equation}
    f(R)
    =
    \sqrt{G(r(R))}.
\end{equation}
The metric then takes the surface-of-revolution form
\begin{equation}
    ds^2
    =
    dR^2
    +
    f(R)^2d\psi^2.
\end{equation}
Its Gaussian curvature is
\begin{equation}
    K
    =
    -\frac{1}{f}
    \frac{d^2f}{dR^2}.
    \label{eq:app-K-proper}
\end{equation}
Returning to the coordinate $r$ gives
\begin{equation}
    \boxed{
    K
    =
    -
    \frac{1}{
        2\sqrt{EG}
    }
    \frac{d}{dr}
    \left[
        \frac{
            G'
        }{
            \sqrt{EG}
        }
    \right].
    }
    \label{eq:app-general-Gaussian}
\end{equation}

For the Fisher family,
\begin{equation}
    \boxed{
    K_F(r)
    =
    -
    \frac{1}{
        2\sqrt{
            rA(r)A'(r)
        }
    }
    \frac{d}{dr}
    \left[
        \frac{
            A(r)+rA'(r)
        }{
            \sqrt{
                rA(r)A'(r)
            }
        }
    \right].
    }
    \label{eq:app-Fisher-Gaussian}
\end{equation}
Equations~\eqref{eq:app-Fisher-kg} and
\eqref{eq:app-Fisher-Gaussian} determine the extrinsic and intrinsic
curvature of the orientation foliation exactly.

\subsection{Strong-concentration asymptotics}
\label{app:Fisher-asymptotics}

At large $r$, the modified Bessel functions give
\begin{equation}
    A(r)
    =
    1
    -
    \frac{1}{2r}
    -
    \frac{1}{8r^2}
    -
    \frac{1}{8r^3}
    +
    O(r^{-4}),
    \label{eq:app-A-large}
\end{equation}
and hence
\begin{equation}
    A'(r)
    =
    \frac{1}{2r^2}
    +
    \frac{1}{4r^3}
    +
    \frac{3}{8r^4}
    +
    O(r^{-5}).
    \label{eq:app-Aprime-large}
\end{equation}

Substituting into Eq.~\eqref{eq:app-Fisher-kg} gives
\begin{equation}
    k_F(r)
    =
    \frac{1}{\sqrt2}
    \left[
        1+\frac{1}{4r}
        +O(r^{-2})
    \right],
    \label{eq:app-kF-large}
\end{equation}
and therefore
\begin{equation}
    \boxed{
    \lim_{r\to\infty}
    k_F(r)
    =
    \frac1{\sqrt2}.
    }
    \label{eq:app-kF-limit}
\end{equation}

Similarly,
Eq.~\eqref{eq:app-Fisher-Gaussian} gives
\begin{equation}
    \boxed{
    K_F(r)
    =
    -\frac12
    +
    O(r^{-2}),
    }
    \label{eq:app-KF-large}
\end{equation}
so that
\begin{equation}
    \boxed{
    \frac{
        k_F
    }{
        \sqrt{-K_F}
    }
    \longrightarrow1.
    }
    \label{eq:app-horocycle-ratio}
\end{equation}

The limiting relation
\begin{equation}
    k_g=\sqrt{-K}
\end{equation}
is the characteristic curvature relation for a horocycle in a hyperbolic
plane of constant negative curvature.  Thus the strongly concentrated end of
the Fisher manifold becomes asymptotically hyperbolic, while the
orientation orbits approach horocycles rather than geodesics.

The same asymptotics make the radial-distance behavior transparent:
\begin{equation}
\begin{split}
    \sqrt{A'(r)}
    &=
    \frac{1}{\sqrt2\,r}
    \left[
        1+\frac{1}{4r}
        +O(r^{-2})
    \right],
\end{split}
\end{equation}
and hence
\begin{equation}
\begin{split}
    L_F(r)
    &=
    \int^r
    dt\,
    \sqrt{A'(t)}
\\
    &=
    \frac{1}{\sqrt2}
    \log r
    +
    c_F
    +
    O(r^{-1}),
\end{split}
    \label{eq:app-Fisher-distance}
\end{equation}
where $c_F$ depends on the lower endpoint used to define the radial
coordinate.  The rank-one boundary therefore lies at infinite Fisher
distance.

\subsection{Hilbert--Schmidt geometry}
\label{app:HS-geometry}

The same $p=2$ family may instead be regarded as a submanifold of the
Euclidean vector space of real symmetric matrices.  Write
\begin{equation}
    T(m,\psi)
    =
    \frac12
    \begin{pmatrix}
        1+m\cos\psi
        &
        m\sin\psi
        \\
        m\sin\psi
        &
        1-m\cos\psi
    \end{pmatrix}.
    \label{eq:app-T-m-psi}
\end{equation}
Its differential is
\begin{equation}
\begin{split}
    dT
    =
    \frac12
    \begin{pmatrix}
        \cos\psi\,dm
        -
        m\sin\psi\,d\psi
        &
        \sin\psi\,dm
        +
        m\cos\psi\,d\psi
        \\
        \sin\psi\,dm
        +
        m\cos\psi\,d\psi
        &
        -\cos\psi\,dm
        +
        m\sin\psi\,d\psi
    \end{pmatrix}.
\end{split}
\end{equation}
A direct calculation gives
\begin{equation}
    \Tr(dT^2)
    =
    \frac12
    \left(
        dm^2+m^2d\psi^2
    \right).
\end{equation}
Thus
\begin{equation}
    \boxed{
    ds_{\rm HS}^2
    =
    \frac12
    \left(
        dm^2+m^2d\psi^2
    \right).
    }
    \label{eq:app-HS-metric}
\end{equation}

Applying Eq.~\eqref{eq:app-general-kg} with
\begin{equation}
    E=\frac12,
    \qquad
    G=\frac{m^2}{2},
\end{equation}
gives
\begin{equation}
    \boxed{
    k_{\rm HS}
    =
    \frac{\sqrt2}{m}.
    }
    \label{eq:app-HS-kg}
\end{equation}
Hence
\begin{equation}
    \boxed{
    k_{\rm HS}
    \longrightarrow
    \sqrt2
    \qquad
    (m\to1).
    }
    \label{eq:app-HS-limit}
\end{equation}
The Hilbert--Schmidt geometry therefore does not make the orientation orbit
asymptotically geodesic.

\subsection{Bures geometry}
\label{app:Bures-geometry}

The positive trace-one tensor
Eq.~\eqref{eq:app-T-m-psi} can also be treated as a real two-level density
matrix.  In Bloch coordinates its radius is $m$, and on the real
two-dimensional disk the Bures line element is
\begin{equation}
    \boxed{
    ds_B^2
    =
    \frac14
    \left[
        \frac{
            dm^2
        }{
            1-m^2
        }
        +
        m^2d\psi^2
    \right].
    }
    \label{eq:app-Bures-metric}
\end{equation}
Thus
\begin{equation}
    E(m)
    =
    \frac{
        1
    }{
        4(1-m^2)
    },
    \qquad
    G(m)
    =
    \frac{
        m^2
    }{4}.
\end{equation}
Equation~\eqref{eq:app-general-kg} gives
\begin{equation}
    \boxed{
    k_B
    =
    \frac{
        2\sqrt{1-m^2}
    }{
        m
    }.
    }
    \label{eq:app-Bures-kg}
\end{equation}
Therefore
\begin{equation}
    \boxed{
    k_B\longrightarrow0
    \qquad
    (m\to1).
    }
    \label{eq:app-Bures-limit}
\end{equation}

The effective rank satisfies
\begin{equation}
    r_{\rm eff}
    =
    \frac{
        2
    }{
        1+m^2
    },
\end{equation}
so
\begin{equation}
    r_{\rm eff}-1
    =
    \frac{
        1-m^2
    }{
        1+m^2
    }.
    \label{eq:app-reff-minus-one}
\end{equation}
It follows that
\begin{equation}
\begin{split}
    k_B^2
    &=
    \frac{
        4(1-m^2)
    }{
        m^2
    }
\\
    &=
    4
    \frac{
        1+m^2
    }{
        m^2
    }
    \left(
        r_{\rm eff}-1
    \right),
\end{split}
    \label{eq:app-Bures-reff}
\end{equation}
and hence
\begin{equation}
    \boxed{
    k_B^2
    \sim
    8
    \left(
        r_{\rm eff}-1
    \right)
    }
    \qquad
    (m\to1).
    \label{eq:app-Bures-reff-asymptotic}
\end{equation}
In the Bures geometry, vanishing extrinsic curvature is therefore directly
slaved to spectral polarization.

\subsection{Comparison of metric dependence}
\label{app:metric-comparison}

The three metrics considered above act on the same family of collective
states,
\begin{equation}
    T(m,\psi),
\end{equation}
and the same physical strong-polarization limit
\begin{equation}
    m\to1,
    \qquad
    r_{\rm eff}\to1.
\end{equation}
Nevertheless,
\begin{equation}
    \boxed{
    k_{\rm HS}
    \longrightarrow
    \sqrt2,
    \qquad
    k_F
    \longrightarrow
    \frac1{\sqrt2},
    \qquad
    k_B
    \longrightarrow
    0.
    }
    \label{eq:app-three-metrics}
\end{equation}

An overall constant rescaling
\begin{equation}
    g\longrightarrow c\,g
\end{equation}
rescales all inverse lengths according to
\begin{equation}
    k_g
    \longrightarrow
    c^{-1/2}k_g,
    \label{eq:app-kg-rescaling}
\end{equation}
so the numerical values of the two nonzero limits depend on convention.
The qualitative distinction
\begin{equation}
    k_g\to0
    \qquad\text{versus}\qquad
    k_g\not\to0
\end{equation}
does not.

Consequently, the vanishing of the second fundamental form cannot be
deduced from the polarized tensor alone.  It depends on the additional
choice of an ambient metric.

\subsection{Fisher extensivity}
\label{app:Fisher-extensivity}

The Fisher metric contains a second instructive ambiguity.  Consider $N$
statistically independent copies with product density
\begin{equation}
    \rho_N(x_1,\ldots,x_N|\vartheta)
    =
    \prod_{i=1}^{N}
    \rho(x_i|\vartheta),
    \label{eq:app-product-density}
\end{equation}
where $\vartheta^a$ denotes the parameters of the one-site family.  The
log likelihood is
\begin{equation}
    \ell_N
    =
    \sum_{i=1}^{N}
    \ell_i.
\end{equation}
Its score is therefore
\begin{equation}
    \partial_a\ell_N
    =
    \sum_{i=1}^{N}
    \partial_a\ell_i.
\end{equation}
Since each one-site score has zero expectation and distinct sites are
independent,
\begin{equation}
\begin{split}
    g_{ab}^{(N)}
    &=
    \left\langle
        \partial_a\ell_N
        \partial_b\ell_N
    \right\rangle
\\
    &=
    \sum_{i,j}
    \left\langle
        \partial_a\ell_i
        \partial_b\ell_j
    \right\rangle
\\
    &=
    N
    g_{ab}^{(1)}.
\end{split}
\end{equation}
Thus
\begin{equation}
    \boxed{
    g_F^{(N)}
    =
    N g_F^{(1)}.
    }
    \label{eq:app-Fisher-N}
\end{equation}

Because the rescaling factor is constant, the Levi--Civita connection in
the same coordinates is unchanged.  The unit tangent, however, scales as
\begin{equation}
    t^{(N)}
    =
    N^{-1/2}t^{(1)}.
\end{equation}
It follows that
\begin{equation}
    \boxed{
    k_F^{(N)}
    =
    \frac{
        1
    }{
        \sqrt N
    }
    k_F^{(1)}.
    }
    \label{eq:app-Fisher-kg-N}
\end{equation}
Likewise, sectional and Gaussian curvatures scale as
\begin{equation}
    \boxed{
    K_F^{(N)}
    =
    \frac1N
    K_F^{(1)}.
    }
    \label{eq:app-Fisher-K-N}
\end{equation}

The dimensionless ratio
\begin{equation}
    \boxed{
    \frac{
        k_F^{(N)}
    }{
        \sqrt{|K_F^{(N)}|}
    }
    =
    \frac{
        k_F^{(1)}
    }{
        \sqrt{|K_F^{(1)}|}
    }
    }
    \label{eq:app-Fisher-ratio-N}
\end{equation}
is therefore independent of $N$.

For the strongly polarized von Mises family,
\begin{equation}
    \frac{
        k_F^{(N)}
    }{
        \sqrt{-K_F^{(N)}}
    }
    \longrightarrow1
\end{equation}
for every $N$, even though
\begin{equation}
    k_F^{(N)}
    \longrightarrow0
\end{equation}
as $N\to\infty$ at fixed concentration.

Thus an absolute reduction of Fisher geodesic curvature with system size
does not signal increasing physical autonomy.  It can arise solely from the
extensive normalization of the statistical metric.

\subsection{Geometric versus dynamical normal acceleration}
\label{app:geometry-dynamics-distinction}

The calculations above also clarify when an extrinsic-curvature criterion
would acquire physical significance.

For a curve $X^A(s)$ evolving as a geodesic of a physical kinetic metric
$G_{AB}$,
\begin{equation}
    \ddot X^A
    +
    \Gamma^A_{BC}
    \dot X^B\dot X^C
    =
    0,
\end{equation}
the second fundamental form of an embedded submanifold controls the normal
acceleration required to remain on that submanifold.  In such a theory the
extrinsic geometry of the kinetic metric is physically meaningful.

The metric used in Sec.~\ref{sec:geometry}, however, was not derived from
the rotor dynamics.  Fisher, Hilbert--Schmidt, and Bures geometry are
different geometrically natural structures on the same family of states,
and they need not coincide with the kinetic geometry of the physical
collective theory.

For the adapted kinetic metric used later,
\begin{equation}
    G_{\rm kin}
    =
    \begin{pmatrix}
        {\cal I}(y) & 0\\
        0 & M_\perp
    \end{pmatrix},
\end{equation}
with coordinates $(\psi,y^\alpha)$, the relevant Christoffel symbol at the
reference orbit is
\begin{equation}
    \boxed{
    \Gamma^\alpha_{\psi\psi}
    =
    -\frac12
    (M_\perp^{-1})^{\alpha\beta}
    \partial_\beta{\cal I}.
    }
    \label{eq:app-kinetic-Christoffel}
\end{equation}
Thus the forcing term
$\tfrac12(\partial_\alpha{\cal I})\dot\psi^2$ in the normal equation is the
covariant normal acceleration associated with the orbit's extrinsic bending
in the physical kinetic metric.  The dynamical leakage criterion therefore
combines this physical geometry with the restoring operator $K_\perp$.

For a general dynamical system, the linearized transverse equation instead
takes the form
\begin{equation}
    M_{\alpha\beta}
    \ddot y^\beta
    +
    \Gamma_{\alpha\beta}
    \dot y^\beta
    +
    K_{\alpha\beta}
    y^\beta
    =
    F_\alpha,
    \label{eq:app-general-normal-dynamics}
\end{equation}
where $M$, $\Gamma$, and $K$ are determined by the actual microscopic or
effective dynamics.  In the adiabatic regime,
\begin{equation}
    y^\alpha
    \simeq
    (K^{-1})^{\alpha\beta}
    F_\beta.
    \label{eq:app-general-normal-response}
\end{equation}

The relevant criterion for low-energy autonomy is therefore suppression of
this physical normal response.  Sections~\ref{sec:twoharmonic}--
\ref{sec:dynamics} construct such a criterion explicitly.

\section{Two-harmonic rate function and fluctuation Hessians}
\label{app:hessians}

This appendix derives the fluctuation formulas used in
Sec.~\ref{sec:twoharmonic}.  The central point is that the first harmonic
$M_1$ determines the collective tensor $T$ in the $p=2$ model, whereas the
second harmonic $M_2$ contains additional shape information.  The physical
stiffnesses are obtained from the Hessian of the thermodynamic free energy,
not from an independently chosen state-space metric.

We first formulate the reflection-symmetric problem as a two-parameter
exponential family, then enlarge it to four real moment coordinates in
order to treat phase fluctuations.  Finally, we construct the
fixed-polarization families used in Sec.~\ref{subsec:fixed-polarization}.

\subsection{Two-harmonic exponential family}
\label{app:twoharmonic-family}

Define
\begin{equation}
    M_1
    =
    \left\langle
        e^{ix}
    \right\rangle,
    \qquad
    M_2
    =
    \left\langle
        e^{2ix}
    \right\rangle .
    \label{eq:app-M1-M2}
\end{equation}
On the aligned reflection-symmetric branch,
\begin{equation}
    M_1=m,
    \qquad
    M_2=q,
    \label{eq:app-aligned-mq}
\end{equation}
with $m$ and $q$ real.

The corresponding maximum-entropy distribution at fixed $(m,q)$ belongs
to the exponential family
\begin{equation}
    \rho_{a,b}(x)
    =
    \frac{
        \exp[
            a\cos x+b\cos2x
        ]
    }{
        Z(a,b)
    },
    \label{eq:app-twoharmonic-rho}
\end{equation}
where all integrals in this appendix use the normalized measure
$dx/(2\pi)$ and
\begin{equation}
    Z(a,b)
    =
    \int_0^{2\pi}
    \frac{dx}{2\pi}
    \exp[
        a\cos x+b\cos2x
    ].
    \label{eq:app-twoharmonic-Z}
\end{equation}
Define
\begin{equation}
    \Phi(a,b)
    =
    \log Z(a,b).
    \label{eq:app-twoharmonic-Phi}
\end{equation}
Differentiation gives
\begin{equation}
    \boxed{
    m
    =
    \partial_a\Phi,
    \qquad
    q
    =
    \partial_b\Phi.
    }
    \label{eq:app-mq-gradient}
\end{equation}

The Hessian of $\Phi$ is the covariance matrix of the sufficient
statistics $\cos x$ and $\cos2x$:
\begin{equation}
    \boxed{
    C_e
    =
    \nabla^2\Phi
    =
    \begin{pmatrix}
        \operatorname{Var}(\cos x)
        &
        \operatorname{Cov}(\cos x,\cos2x)
        \\[1mm]
        \operatorname{Cov}(\cos x,\cos2x)
        &
        \operatorname{Var}(\cos2x)
    \end{pmatrix}.
    }
    \label{eq:app-Ce}
\end{equation}
For a nondegenerate distribution in the interior of the moment domain,
$C_e$ is positive definite.

It is sometimes convenient to express these entries in terms of the
Fourier moments
\begin{equation}
    c_n
    =
    \langle\cos nx\rangle.
    \label{eq:app-cn-def}
\end{equation}
Since
\begin{equation}
    c_1=m,
    \qquad
    c_2=q,
\end{equation}
elementary trigonometric identities give
\begin{equation}
    \boxed{
    C_e
    =
    \begin{pmatrix}
        \dfrac{1+q}{2}-m^2
        &
        \dfrac{m+c_3}{2}-mq
        \\[3mm]
        \dfrac{m+c_3}{2}-mq
        &
        \dfrac{1+c_4}{2}-q^2
    \end{pmatrix}.
    }
    \label{eq:app-Ce-Fourier}
\end{equation}

\subsection{Legendre duality and the even Hessian}
\label{app:Legendre-even}

The entropy cost of imposing moments $(m,q)$ is the Legendre transform
\begin{equation}
    \boxed{
    I(m,q)
    =
    \sup_{a,b}
    \left[
        am+bq-\Phi(a,b)
    \right].
    }
    \label{eq:app-I-mq}
\end{equation}
Inside the moment domain the supremum is attained at the unique
$(a,b)$ satisfying Eq.~\eqref{eq:app-mq-gradient}.  Consequently,
\begin{equation}
    \nabla I
    =
    \begin{pmatrix}
        a\\
        b
    \end{pmatrix}.
    \label{eq:app-I-gradient}
\end{equation}

Differentiating once more,
\begin{equation}
    \frac{\partial(m,q)}
         {\partial(a,b)}
    =
    C_e,
\end{equation}
so by inversion,
\begin{equation}
    \boxed{
    \nabla^2 I
    =
    C_e^{-1}.
    }
    \label{eq:app-I-Hessian}
\end{equation}

The dimensionless two-harmonic free energy at fixed moments is
\begin{equation}
    f(m,q)
    =
    I(m,q)
    -
    \frac{\kappa_1}{2}m^2
    -
    \frac{\kappa_2}{2}q^2.
    \label{eq:app-twoharmonic-f}
\end{equation}
Its first derivatives are
\begin{equation}
    \frac{\partial f}{\partial m}
    =
    a-\kappa_1m,
    \qquad
    \frac{\partial f}{\partial q}
    =
    b-\kappa_2q.
\end{equation}
Stationarity therefore gives
\begin{equation}
    \boxed{
    a=\kappa_1m,
    \qquad
    b=\kappa_2q.
    }
    \label{eq:app-twoharmonic-saddle}
\end{equation}

The Hessian in the reflection-even sector is
\begin{equation}
    \boxed{
    K_e
    =
    C_e^{-1}
    -
    \begin{pmatrix}
        \kappa_1&0\\
        0&\kappa_2
    \end{pmatrix}.
    }
    \label{eq:app-Ke}
\end{equation}
Thus for an even fluctuation
\begin{equation}
    \delta y_e
    =
    \begin{pmatrix}
        \delta m\\
        \delta q
    \end{pmatrix},
\end{equation}
the quadratic free-energy variation is
\begin{equation}
    \delta^2f_e
    =
    \frac12
    \delta y_e^{\mathsf T}
    K_e
    \delta y_e.
    \label{eq:app-even-quadratic}
\end{equation}

Local stability against even fluctuations requires
\begin{equation}
    K_e>0.
    \label{eq:app-even-positive}
\end{equation}
The two eigenvectors of $K_e$ are in general mixtures of changes in the
first-harmonic polarization $m$ and changes in the higher-harmonic shape
$q$.  They need not coincide with either coordinate axis.

\subsection{Full moment space and the odd block}
\label{app:full-moment-space}

To treat phase fluctuations, introduce the four real moments
\begin{equation}
\begin{aligned}
    c_1 &= \langle\cos x\rangle,
    &\qquad c_2 &= \langle\cos 2x\rangle,\\
    s_1 &= \langle\sin x\rangle,
    &\qquad s_2 &= \langle\sin 2x\rangle.
\end{aligned}
\label{eq:app-four-moments}
\end{equation}
Then
\begin{equation}
    M_1=c_1+is_1,
    \qquad
    M_2=c_2+is_2.
    \label{eq:app-complex-moments}
\end{equation}

Consider the four-parameter exponential family
\begin{equation}
    \rho(x)
    =
    \frac{1}{Z_4}
    \exp\!
    \left[
        a_1\cos x
        +
        a_2\cos2x
        +
        b_1\sin x
        +
        b_2\sin2x
    \right].
    \label{eq:app-four-parameter-rho}
\end{equation}
Let
\begin{equation}
    \Phi_4
    =
    \log Z_4.
\end{equation}
Writing
\begin{equation}
    \theta
    =
    (a_1,a_2,b_1,b_2)^{\mathsf T},
    \qquad
    z
    =
    (c_1,c_2,s_1,s_2)^{\mathsf T},
\end{equation}
we have
\begin{equation}
    z=\nabla_\theta\Phi_4.
\end{equation}
The corresponding four-dimensional rate function is
\begin{equation}
    I_4(z)
    =
    \sup_\theta
    \left[
        \theta^{\mathsf T}z
        -
        \Phi_4(\theta)
    \right].
    \label{eq:app-I4}
\end{equation}
Therefore
\begin{equation}
    \boxed{
    \nabla_z^2 I_4
    =
    C^{-1},
    }
    \label{eq:app-I4-Hessian}
\end{equation}
where $C$ is the covariance matrix of the four sufficient statistics.

At an aligned state,
\begin{equation}
    (a_1,a_2,b_1,b_2)
    =
    (a,b,0,0),
\end{equation}
the distribution is invariant under
\begin{equation}
    x\longrightarrow-x.
\end{equation}
The cosine observables are even and the sine observables odd, so every
even--odd covariance vanishes.  Hence
\begin{equation}
    \boxed{
    C
    =
    \begin{pmatrix}
        C_e&0\\
        0&C_o
    \end{pmatrix},
    }
    \label{eq:app-C-blocks}
\end{equation}
where
\begin{equation}
    \boxed{
    C_o
    =
    \begin{pmatrix}
        \langle\sin^2x\rangle
        &
        \langle\sin x\,\sin2x\rangle
        \\[1mm]
        \langle\sin x\,\sin2x\rangle
        &
        \langle\sin^22x\rangle
    \end{pmatrix}.
    }
    \label{eq:app-Co}
\end{equation}

In terms of the cosine Fourier moments,
\begin{equation}
    \boxed{
    C_o
    =
    \begin{pmatrix}
        \dfrac{1-q}{2}
        &
        \dfrac{m-c_3}{2}
        \\[3mm]
        \dfrac{m-c_3}{2}
        &
        \dfrac{1-c_4}{2}
    \end{pmatrix}.
    }
    \label{eq:app-Co-Fourier}
\end{equation}

The full interaction contribution is
\begin{equation}
    -\frac{\kappa_1}{2}
    \left(
        c_1^2+s_1^2
    \right)
    -
    \frac{\kappa_2}{2}
    \left(
        c_2^2+s_2^2
    \right).
    \label{eq:app-full-interaction}
\end{equation}
Thus the aligned free-energy Hessian is block diagonal, with
\begin{equation}
    \boxed{
    K_e
    =
    C_e^{-1}
    -
    D,
    \qquad
    K_o
    =
    C_o^{-1}
    -
    D,
    }
    \label{eq:app-Ke-Ko}
\end{equation}
where
\begin{equation}
    D
    =
    \begin{pmatrix}
        \kappa_1&0\\
        0&\kappa_2
    \end{pmatrix}.
    \label{eq:app-D-def}
\end{equation}

The odd block therefore governs fluctuations of the imaginary parts of
$M_1$ and $M_2$, equivalently their phases.

\subsection{Rotational Ward identity}
\label{app:Ward-identity}

The global $O(2)$ rotation
\begin{equation}
    x\longrightarrow x+\alpha
    \label{eq:app-global-rotation}
\end{equation}
acts on the complex moments as
\begin{equation}
    M_1
    \longrightarrow
    e^{i\alpha}M_1,
    \qquad
    M_2
    \longrightarrow
    e^{2i\alpha}M_2.
    \label{eq:app-moment-rotation}
\end{equation}
At the aligned state,
\begin{equation}
    M_1=m,
    \qquad
    M_2=q,
\end{equation}
an infinitesimal rotation gives
\begin{equation}
    \delta s_1=m\,\delta\alpha,
    \qquad
    \delta s_2=2q\,\delta\alpha.
\end{equation}
The tangent vector to the rotational orbit in the odd moment sector is
therefore
\begin{equation}
    \boxed{
    v_G
    =
    \begin{pmatrix}
        m\\
        2q
    \end{pmatrix}.
    }
    \label{eq:app-vG}
\end{equation}

Since the free energy is exactly invariant under
Eq.~\eqref{eq:app-global-rotation}, its Hessian must annihilate this tangent
direction at a stationary point:
\begin{equation}
    \boxed{
    K_o v_G=0.
    }
    \label{eq:app-Ward}
\end{equation}
This is the Goldstone Ward identity quoted in
Sec.~\ref{subsec:physical-hessians}.

The identity can also be verified directly from linear response.  A small
rotation of the aligned density generates the odd natural-parameter vector
\begin{equation}
    \delta h_o
    =
    \delta\alpha
    \begin{pmatrix}
        a\\
        2b
    \end{pmatrix}.
\end{equation}
The corresponding odd moment variation is
\begin{equation}
    \delta s
    =
    C_o\,\delta h_o.
\end{equation}
But the exact rotational transformation requires
\begin{equation}
    \delta s
    =
    \delta\alpha
    \begin{pmatrix}
        m\\
        2q
    \end{pmatrix}.
\end{equation}
Hence
\begin{equation}
    C_o
    \begin{pmatrix}
        a\\
        2b
    \end{pmatrix}
    =
    \begin{pmatrix}
        m\\
        2q
    \end{pmatrix}.
    \label{eq:app-Co-identity}
\end{equation}
Using the saddle conditions
\begin{equation}
    a=\kappa_1m,
    \qquad
    b=\kappa_2q,
\end{equation}
this becomes
\begin{equation}
    C_oD
    \begin{pmatrix}
        m\\
        2q
    \end{pmatrix}
    =
    \begin{pmatrix}
        m\\
        2q
    \end{pmatrix}.
\end{equation}
Multiplication by $C_o^{-1}$ immediately yields
Eq.~\eqref{eq:app-Ward}.

This provides a useful exact check on any numerical evaluation of $C_o$ and
$K_o$.

\subsection{Relative phase as the non-Goldstone odd mode}
\label{app:relative-phase}

Write the moments more generally as
\begin{equation}
    M_1
    =
    m e^{i\psi_1},
    \qquad
    M_2
    =
    q e^{i\psi_2}.
\end{equation}
The invariant relative phase is
\begin{equation}
    \chi
    =
    \psi_2-2\psi_1.
    \label{eq:app-chi}
\end{equation}
For infinitesimal phase fluctuations,
\begin{equation}
    \delta s_1
    =
    m\,\delta\psi_1,
    \qquad
    \delta s_2
    =
    q\,\delta\psi_2.
    \label{eq:app-phase-odd}
\end{equation}

Define
\begin{equation}
    \eta
    =
    \frac{2q}{m},
    \qquad
    w
    =
    \begin{pmatrix}
        -\eta\\
        1
    \end{pmatrix}.
    \label{eq:app-w-def}
\end{equation}
Then
\begin{equation}
    w^{\mathsf T}v_G=0,
\end{equation}
and
\begin{equation}
\begin{split}
    w^{\mathsf T}
    \begin{pmatrix}
        \delta s_1\\
        \delta s_2
    \end{pmatrix}
    &=
    -\frac{2q}{m}
    \left(
        m\,\delta\psi_1
    \right)
    +
    q\,\delta\psi_2
\\
    &=
    q
    \left(
        \delta\psi_2
        -
        2\delta\psi_1
    \right)
\\
    &=
    q\,\delta\chi.
\end{split}
    \label{eq:app-w-relative-phase}
\end{equation}
Thus $w$ isolates the relative-phase deformation while annihilating the
Goldstone rotation.

Because $K_o$ is a symmetric $2\times2$ matrix with the exact null vector
$v_G$, it has rank at most one.  In a locally stable aligned phase with one
nonzero odd restoring force,
\begin{equation}
    \boxed{
    K_o
    =
    \gamma\,ww^{\mathsf T}
    }
    \label{eq:app-Ko-gamma}
\end{equation}
for some
\begin{equation}
    \gamma>0.
\end{equation}
With the choice that the second component of $w$ equals unity,
\begin{equation}
    \gamma=(K_o)_{22}.
    \label{eq:app-gamma-K22}
\end{equation}
The unique nonzero eigenvalue is
\begin{equation}
    \boxed{
    \lambda_o^+
    =
    \gamma
    \left(
        1+\eta^2
    \right).
    }
    \label{eq:app-odd-eigenvalue}
\end{equation}

Equivalently, the odd quadratic free energy can be written directly in
terms of the relative phase:
\begin{equation}
\begin{split}
    \delta^2f_o
    &=
    \frac12
    \delta s^{\mathsf T}
    K_o
    \delta s
\\
    &=
    \frac{\gamma}{2}
    \left(
        w^{\mathsf T}\delta s
    \right)^2
\\
    &=
    \boxed{
    \frac{\gamma q^2}{2}
    (\delta\chi)^2
    }.
\end{split}
    \label{eq:app-relative-phase-energy}
\end{equation}
The common orientation remains exactly flat, while $\chi$ measures the
single physical odd deformation.

\subsection{Why the collective tensor does not determine the Hessian}
\label{app:T-vs-Hessian}

For $p=2$, the tensor depends only on the first complex moment:
\begin{equation}
    T
    =
    \frac12
    \begin{pmatrix}
        1+\Re M_1
        &
        \Im M_1
        \\
        \Im M_1
        &
        1-\Re M_1
    \end{pmatrix}.
    \label{eq:app-T-M1}
\end{equation}
Its eigenvalues are
\begin{equation}
    \lambda_\pm
    =
    \frac{
        1\pm|M_1|
    }{2},
\end{equation}
and hence
\begin{equation}
    \boxed{
    r_{\rm eff}
    =
    \frac{
        2
    }{
        1+|M_1|^2
    }.
    }
    \label{eq:app-reff-M1}
\end{equation}
No component of $M_2$ appears.

By contrast, both covariance matrices
\begin{equation}
    C_e,
    \qquad
    C_o
\end{equation}
depend on the complete equilibrium distribution and therefore on
higher Fourier moments such as $q=c_2$, $c_3$, and $c_4$.  Through
Eq.~\eqref{eq:app-Ke-Ko}, the physical fluctuation Hessian inherits this
additional information.

Consequently,
\begin{equation}
    \boxed{
    T
    \ \text{does not determine}\
    K_e\ \text{or}\ K_o.
    }
    \label{eq:app-T-not-Hessian}
\end{equation}

The fixed-polarization construction below makes this independence explicit.

\subsection{Fixed-polarization branches}
\label{app:fixed-m-branches}

Fix
\begin{equation}
    m=m_\star,
    \qquad
    0<m_\star<1.
    \label{eq:app-fixed-m}
\end{equation}
For each finite $b$, determine $a=a(b)$ from
\begin{equation}
    \boxed{
    \partial_a\Phi(a(b),b)
    =
    m_\star.
    }
    \label{eq:app-fixed-m-equation}
\end{equation}
Since
\begin{equation}
    \partial_a^2\Phi
    =
    \operatorname{Var}(\cos x)
    =
    (C_e)_{11}>0
    \label{eq:app-m-monotone-a}
\end{equation}
for every nondegenerate distribution, $m$ is strictly increasing as a
function of $a$ at fixed $b$.  The solution $a(b)$ is therefore unique.

The associated second harmonic is
\begin{equation}
    q(b)
    =
    \partial_b\Phi(a(b),b).
    \label{eq:app-q-of-b}
\end{equation}
Differentiating the fixed-$m$ condition gives
\begin{equation}
    0
    =
    (C_e)_{11}
    \frac{da}{db}
    +
    (C_e)_{12},
\end{equation}
and hence
\begin{equation}
    \boxed{
    \frac{da}{db}
    =
    -
    \frac{
        (C_e)_{12}
    }{
        (C_e)_{11}
    }.
    }
    \label{eq:app-da-db}
\end{equation}
Differentiating $q(b)$ then gives
\begin{align}
    \frac{dq}{db}
    &=
    (C_e)_{12}
    \frac{da}{db}
    +
    (C_e)_{22}
    \nonumber\\
    &=
    (C_e)_{22}
    -
    \frac{
        (C_e)_{12}^2
    }{
        (C_e)_{11}
    }.
\end{align}
Therefore
\begin{equation}
    \boxed{
    \frac{dq}{db}
    =
    \frac{
        \det C_e
    }{
        (C_e)_{11}
    }
    >0.
    }
    \label{eq:app-dq-db}
\end{equation}
The hidden shape variable $q$ therefore varies monotonically along the
fixed-polarization branch.

Each point on this branch can be realized as a saddle of a physical
two-harmonic mean-field model by defining
\begin{equation}
    \boxed{
    \kappa_1(b)
    =
    \frac{
        a(b)
    }{
        m_\star
    },
    \qquad
    \kappa_2(b)
    =
    \frac{
        b
    }{
        q(b)
    }.
    }
    \label{eq:app-fixed-m-kappas}
\end{equation}
These definitions make
\begin{equation}
    a=\kappa_1m_\star,
    \qquad
    b=\kappa_2q
\end{equation}
identically true.

Meanwhile,
\begin{equation}
    \boxed{
    r_{\rm eff}
    =
    \frac{
        2
    }{
        1+m_\star^2
    }
    }
    \label{eq:app-fixed-m-reff}
\end{equation}
is exactly constant.

Thus the family varies
\begin{equation}
    q,\quad
    C_e,\quad
    C_o,\quad
    K_e,\quad
    K_o
\end{equation}
while leaving the collective tensor spectrum unchanged.

\subsection{Numerical evaluation of the fixed-polarization family}
\label{app:fixed-m-numerics}

The numerical examples in Sec.~\ref{subsec:fixed-polarization} are obtained
directly from one-dimensional quadrature.

For a specified pair $(a,b)$, compute
\begin{equation}
    c_n(a,b)
    =
    \frac{
        \displaystyle
        \int_0^{2\pi}
        \frac{dx}{2\pi}
        \cos(nx)
        e^{a\cos x+b\cos2x}
    }{
        \displaystyle
        \int_0^{2\pi}
        \frac{dx}{2\pi}
        e^{a\cos x+b\cos2x}
    }.
    \label{eq:app-cn-integral}
\end{equation}
The required quantities are
\begin{equation}
    m=c_1,
    \qquad
    q=c_2,
    \qquad
    c_3,
    \qquad
    c_4.
\end{equation}
Equations~\eqref{eq:app-Ce-Fourier} and
\eqref{eq:app-Co-Fourier} then give the covariance matrices without numerical
differentiation.

For fixed $m_\star$, one solves
\begin{equation}
    c_1(a,b)=m_\star
\end{equation}
for $a$ at each chosen $b$, constructs
\begin{equation}
    \kappa_1=\frac{a}{m_\star},
    \qquad
    \kappa_2=\frac{b}{q},
\end{equation}
and evaluates
\begin{equation}
    K_e=C_e^{-1}-D,
    \qquad
    K_o=C_o^{-1}-D.
\end{equation}

For the benchmark
\begin{equation}
    m_\star=0.9,
\end{equation}
three stable examples used in
Table~\ref{tab:fixed-m-spectrum} are
\begin{align}
    b&=1.0:
    &
    a&\simeq2.99755356,
    &
    q&\simeq0.69253495,
    \label{eq:app-example-b1}
    \\
    b&=2.7:
    &
    a&\simeq1.88164353,
    &
    q&\simeq0.82463346,
    \label{eq:app-example-b27}
    \\
    b&=6.0:
    &
    a&\simeq1.62073537,
    &
    q&\simeq0.91807981.
    \label{eq:app-example-b6}
\end{align}
The corresponding physical couplings are
\begin{align}
    (\kappa_1,\kappa_2)
    &\simeq
    (3.33061507,1.44397045),
    \\
    &\simeq
    (2.09071503,3.27418195),
    \\
    &\simeq
    (1.80081708,6.53537954).
\end{align}
All three have
\begin{equation}
    \boxed{
    r_{\rm eff}
    =
    \frac{2}{1+0.9^2}
    \simeq
    1.10497238.
    }
    \label{eq:app-m09-reff}
\end{equation}

The even Hessian eigenvalues are respectively
\begin{align}
    \operatorname{spec}K_e
    &\simeq
    (3.85554,70.91490),
    \\
    &\simeq
    (6.64555,15.97607),
    \\
    &\simeq
    (4.89777,68.29301),
    \label{eq:app-Ke-numerical}
\end{align}
while the unique positive odd eigenvalues are
\begin{equation}
    \lambda_o^+
    \simeq
    62.82718,
    \qquad
    79.05558,
    \qquad
    166.24290.
    \label{eq:app-Ko-numerical}
\end{equation}
The second odd eigenvalue vanishes to numerical precision in every case, as
required by Eq.~\eqref{eq:app-Ward}.

These examples establish the finite-coupling independence used in the main
text:
\begin{equation}
    \boxed{
    r_{\rm eff}
    =
    \mathrm{constant}
    \quad\not\Longrightarrow\quad
    \operatorname{spec}K_\perp
    =
    \mathrm{constant}.
    }
    \label{eq:app-fixed-rank-not-spectrum}
\end{equation}

They do not by themselves establish a parametrically infinite separation of
scales.  The locking construction of Appendix~\ref{app:stability} and
Sec.~\ref{sec:locking} supplies that stronger result.

\subsection{Numerical global-minimum check}
\label{app:global-minimum-check}

The Hessian analysis above establishes local stability of the three
fixed-polarization states used in Table~\ref{tab:fixed-m-spectrum}.
For completeness, we also check numerically that these states minimize the
full two-harmonic mean-field free energy rather than merely representing
locally stable saddles.

For positive $\kappa_1$ and $\kappa_2$, the mean-field variational problem
may be represented by the auxiliary free energy
\begin{equation}
    f_{\rm aux}(m,q,\chi)
    =
    \frac{\kappa_1}{2}m^2
    +
    \frac{\kappa_2}{2}q^2
    -
    \log Z_{\rm aux}(m,q,\chi),
    \label{eq:app-global-faux}
\end{equation}
where rotational invariance has been used to fix the phase of $M_1$, while
\begin{equation}
    Z_{\rm aux}(m,q,\chi)
    =
    \int_0^{2\pi}
    \frac{dx}{2\pi}
    \exp\!
    \left[
        \kappa_1 m\cos x
        +
        \kappa_2 q\cos(2x-\chi)
    \right].
    \label{eq:app-global-Zaux}
\end{equation}
Here
\begin{equation}
    0\leq m\leq1,
    \qquad
    0\leq q\leq1,
    \qquad
    0\leq\chi<2\pi.
    \label{eq:app-global-domain}
\end{equation}
The relative phase $\chi$ is therefore allowed to vary freely; the
minimization is not restricted to the aligned reflection-symmetric branch.

We evaluated Eq.~\eqref{eq:app-global-Zaux} using
$4096$-point Gauss--Legendre quadrature and minimized
Eq.~\eqref{eq:app-global-faux} over the compact domain
Eq.~\eqref{eq:app-global-domain} using differential evolution.
For each coupling pair, five independent optimization seeds were used,
followed by local polishing.  We also minimized the four boundary faces
$m=0$, $m=1$, $q=0$, and $q=1$ separately.

The results are summarized in Table~\ref{tab:global-minimum-check}:
\begin{table}[htbp]
\centering
\begin{tabular}{c c c c c}
\hline
$b$
&
$(\kappa_1,\kappa_2)$
&
$(m_{\rm min},q_{\rm min},\chi_{\rm min})$
&
$f_{\rm min}$
&
$\min_{\partial\mathcal D}f-f_{\rm min}$
\\
\hline
$1.0$
&
$(3.33061507,1.44397045)$
&
$(0.900000,\ 0.692535,\ 0)$
&
$-0.4781332724$
&
$0.0145580$
\\
$2.7$
&
$(2.09071503,3.27418195)$
&
$(0.900000,\ 0.824633,\ 0)$
&
$-0.5037746894$
&
$0.00842123$
\\
$6.0$
&
$(1.80081708,6.53537954)$
&
$(0.900000,\ 0.918080,\ 0)$
&
$-1.6580058516$
&
$0.00682629$
\\
\hline
\end{tabular}
\caption{
Numerical global-minimum check for the three fixed-polarization states
used in Table~\ref{tab:fixed-m-spectrum}.
The final column gives the difference between the lowest minimum found on
the boundary of
$\mathcal D=[0,1]\times[0,1]\times S^1$
and the interior aligned minimum.
Repeated global searches reproduce the quoted minima to better than
$10^{-10}$ in free energy.
}
\label{tab:global-minimum-check}
\end{table}

In all three cases the global search returns, to numerical precision, the
same aligned state used in Table~\ref{tab:fixed-m-spectrum}.
The spread among the best free energies obtained from the five independent
global searches is below $5\times10^{-12}$, and no boundary or misaligned
configuration with lower free energy was found.

For the two cases with $\kappa_2>2$, the $m=0$ boundary also contains a
nonzero second-harmonic, or tetratic, stationary state.  These occur at
\begin{equation}
    q_{\rm tet}
    \simeq
    0.765794,
    \qquad
    0.911707,
\end{equation}
for $b=2.7$ and $b=6.0$, respectively, with free energies
\begin{equation}
    f_{\rm tet}
    \simeq
    -0.236407,
    \qquad
    -1.454065.
\end{equation}
Both lie above the corresponding aligned minima.

Thus the three examples used in the main text are not merely locally
stable within the fluctuation analysis: within the full two-harmonic
mean-field variational problem they are identified as global minima in our numerical searches.

\section{Locking interactions and stability}
\label{app:stability}

This appendix gives the detailed stability analysis underlying
Sec.~\ref{sec:locking}.  Starting from a stable aligned saddle of the
two-harmonic model, we introduce rotationally invariant locking interactions
whose value and first variation vanish on the complete orientation orbit of
the prescribed state.  Their second variation is nonzero in the transverse
directions.  We derive the resulting even and odd Hessians, determine the
complete strict quadratic stability domain, prove preservation of a global
minimum for nonnegative locking couplings, and establish the strong-locking
asymptotics used in the main text.

\subsection{Target orbit and locking potential}
\label{app:locking-target}

Let
\begin{equation}
    M_{1\star}
    =
    m_\star e^{i\psi},
    \qquad
    M_{2\star}
    =
    q_\star e^{2i\psi}
    \label{eq:app-locking-target}
\end{equation}
be an aligned ordered saddle of the unmodified two-harmonic free energy
$f_0$, with
\begin{equation}
    m_\star>0.
\end{equation}
Define
\begin{equation}
    \boxed{
    c_\star
    =
    \frac{q_\star}{m_\star^2}.
    }
    \label{eq:app-cstar}
\end{equation}
Every point on the orientation orbit then satisfies
\begin{equation}
    |M_1|^2=m_\star^2,
    \qquad
    M_2=c_\star M_1^2.
    \label{eq:app-target-constraints}
\end{equation}

We add the locking potential
\begin{equation}
    \boxed{
    f_{\rm lock}
    =
    \frac{\lambda_1}{8}
    \left(
        |M_1|^2-m_\star^2
    \right)^2
    +
    \frac{\lambda_2}{2}
    \left|
        M_2-c_\star M_1^2
    \right|^2.
    }
    \label{eq:app-locking-potential}
\end{equation}
The complete free energy is
\begin{equation}
    f
    =
    f_0+f_{\rm lock}.
    \label{eq:app-full-locking-f}
\end{equation}

The two residuals
\begin{equation}
    R_1
    =
    |M_1|^2-m_\star^2,
    \qquad
    R_2
    =
    M_2-c_\star M_1^2
    \label{eq:app-locking-residuals}
\end{equation}
both vanish identically on the target orbit.  Since
\begin{equation}
    f_{\rm lock}
    =
    \frac{\lambda_1}{8}R_1^2
    +
    \frac{\lambda_2}{2}|R_2|^2,
\end{equation}
its first variation is proportional to the residuals themselves:
\begin{equation}
\begin{split}
    \delta f_{\rm lock}
    &=
    \frac{\lambda_1}{4}
    R_1\,\delta R_1
\\
    &\quad
    +
    \lambda_2
    \Re\!\left(
        R_2^\ast\,\delta R_2
    \right).
\end{split}
    \label{eq:app-locking-first-variation}
\end{equation}
Therefore
\begin{equation}
    \boxed{
    f_{\rm lock}|_\star=0,
    \qquad
    \delta f_{\rm lock}|_\star=0.
    }
    \label{eq:app-locking-vanishing}
\end{equation}

The prescribed state is consequently a stationary point of the locked
theory for every $\lambda_1$ and $\lambda_2$.  In particular,
\begin{equation}
    \boxed{
    m_\star,\qquad
    q_\star,\qquad
    T_\star,\qquad
    r_{\rm eff}(T_\star)
    }
    \label{eq:app-locking-fixed}
\end{equation}
are independent of the locking couplings.

The potential also preserves the global rotational symmetry.  Under
\begin{equation}
    M_1\longrightarrow e^{i\alpha}M_1,
    \qquad
    M_2\longrightarrow e^{2i\alpha}M_2,
\end{equation}
one has
\begin{equation}
    R_1\longrightarrow R_1,
    \qquad
    R_2\longrightarrow e^{2i\alpha}R_2,
\end{equation}
so $f_{\rm lock}$ is invariant.  The Goldstone orientation mode therefore
cannot acquire a restoring force from the locking interaction.

\subsection{Quadratic expansion}
\label{app:locking-quadratic}

Choose the representative of the target orbit with
\begin{equation}
    \psi=0,
\end{equation}
and parameterize small moment fluctuations as
\begin{align}
    M_1
    &=
    m_\star+\delta m+i\xi_1,
    \label{eq:app-M1-fluct}
    \\
    M_2
    &=
    q_\star+\delta q+i\xi_2.
    \label{eq:app-M2-fluct}
\end{align}
The even fluctuations are
\begin{equation}
    \delta y_e
    =
    \begin{pmatrix}
        \delta m\\
        \delta q
    \end{pmatrix},
\end{equation}
while the odd fluctuations are
\begin{equation}
    \delta y_o
    =
    \begin{pmatrix}
        \xi_1\\
        \xi_2
    \end{pmatrix}.
\end{equation}

For the first residual,
\begin{equation}
\begin{split}
    R_1
    &=
    |M_1|^2-m_\star^2
\\
    &=
    2m_\star\delta m
    +
    \delta m^2
    +
    \xi_1^2.
\end{split}
    \label{eq:app-R1-expansion}
\end{equation}
Only the term linear in the fluctuations contributes when $R_1^2$ is
expanded to quadratic order.  Thus
\begin{equation}
    \frac{\lambda_1}{8}R_1^2
    =
    \frac{\lambda_1m_\star^2}{2}
    (\delta m)^2
    +
    O(\delta^3).
    \label{eq:app-first-lock-quadratic}
\end{equation}

For the second residual,
\begin{equation}
\begin{split}
    R_2
    &=
    M_2-c_\star M_1^2
\\
    &=
    \delta q
    -
    2c_\star m_\star\,\delta m
\\
    &\quad
    +
    i
    \left(
        \xi_2
        -
        2c_\star m_\star\,\xi_1
    \right)
    +
    O(\delta^2).
\end{split}
    \label{eq:app-R2-expansion}
\end{equation}
Define
\begin{equation}
    \boxed{
    \eta
    =
    2c_\star m_\star
    =
    \frac{2q_\star}{m_\star},
    }
    \label{eq:app-eta}
\end{equation}
and
\begin{equation}
    \boxed{
    w
    =
    \begin{pmatrix}
        -\eta\\
        1
    \end{pmatrix}.
    }
    \label{eq:app-locking-w}
\end{equation}
Then
\begin{equation}
    R_2
    =
    w^{\mathsf T}\delta y_e
    +
    i\,w^{\mathsf T}\delta y_o
    +
    O(\delta^2),
    \label{eq:app-R2-w}
\end{equation}
so
\begin{equation}
\begin{split}
    \frac{\lambda_2}{2}|R_2|^2
    &=
    \frac{\lambda_2}{2}
    \left(
        w^{\mathsf T}\delta y_e
    \right)^2
\\
    &\quad
    +
    \frac{\lambda_2}{2}
    \left(
        w^{\mathsf T}\delta y_o
    \right)^2
    +
    O(\delta^3).
\end{split}
    \label{eq:app-second-lock-quadratic}
\end{equation}

Thus the second locking interaction adds the same rank-one quadratic form
to the even and odd moment sectors.

\subsection{Locked even Hessian}
\label{app:locked-even}

Let
\begin{equation}
    A
    =
    K_e^{(0)}
    =
    \begin{pmatrix}
        A_{11}&A_{12}\\
        A_{12}&A_{22}
    \end{pmatrix}
    \label{eq:app-locking-A-def}
\end{equation}
be the even Hessian of the unmodified model, as derived in
Appendix~\ref{app:hessians}.  Equations
\eqref{eq:app-first-lock-quadratic} and
\eqref{eq:app-second-lock-quadratic} give
\begin{equation}
    \boxed{
    K_e
    =
    A
    +
    \lambda_1m_\star^2
        e_1e_1^{\mathsf T}
    +
    \lambda_2
        ww^{\mathsf T},
    }
    \label{eq:app-locked-Ke}
\end{equation}
where
\begin{equation}
    e_1
    =
    \begin{pmatrix}
        1\\0
    \end{pmatrix}.
\end{equation}
Explicitly,
\begin{equation}
    \boxed{
    K_e
    =
    \begin{pmatrix}
        A_{11}
        +
        \lambda_1m_\star^2
        +
        \eta^2\lambda_2
        &
        A_{12}-\eta\lambda_2
        \\[2mm]
        A_{12}-\eta\lambda_2
        &
        A_{22}+\lambda_2
    \end{pmatrix}.
    }
    \label{eq:app-locked-Ke-explicit}
\end{equation}

For a symmetric $2\times2$ matrix, strict positive definiteness is
equivalent to positivity of one diagonal pivot and of its Schur complement.
Using the lower-right entry gives
\begin{equation}
    A_{22}+\lambda_2>0
    \label{eq:app-even-condition-one}
\end{equation}
together with
\begin{equation}
\begin{split}
    0<
    &
    A_{11}
    +
    \lambda_1m_\star^2
    +
    \eta^2\lambda_2
\\
    &-
    \frac{
        (A_{12}-\eta\lambda_2)^2
    }{
        A_{22}+\lambda_2
    }.
\end{split}
    \label{eq:app-even-Schur}
\end{equation}
Solving for $\lambda_1$ gives
\begin{equation}
    \boxed{
    \lambda_1
    >
    \lambda_{1,\mathrm{crit}}(\lambda_2),
    }
    \label{eq:app-lambda1-condition}
\end{equation}
where
\begin{equation}
    \boxed{
    \lambda_{1,\mathrm{crit}}(\lambda_2)
    =
    \frac{1}{m_\star^2}
    \left[
        \frac{
            (A_{12}-\eta\lambda_2)^2
        }{
            A_{22}+\lambda_2
        }
        -
        A_{11}
        -
        \eta^2\lambda_2
    \right].
    }
    \label{eq:app-lambda1crit}
\end{equation}

Equations~\eqref{eq:app-even-condition-one} and
\eqref{eq:app-lambda1-condition} are necessary and sufficient for strict
stability of the even amplitude--shape sector.

At the boundary
\begin{equation}
    \lambda_1
    =
    \lambda_{1,\mathrm{crit}}(\lambda_2),
\end{equation}
one eigenvalue of $K_e$ vanishes.  Crossing this curve changes the sign of
the corresponding amplitude--shape restoring force.

\subsection{Locked odd Hessian and Goldstone preservation}
\label{app:locked-odd}

Let
\begin{equation}
    O
    =
    K_o^{(0)}
    \label{eq:app-O-def}
\end{equation}
be the odd Hessian of the unmodified model.  From
Appendix~\ref{app:hessians},
\begin{equation}
    O v_G=0,
    \qquad
    v_G
    =
    \begin{pmatrix}
        m_\star\\
        2q_\star
    \end{pmatrix}.
    \label{eq:app-O-Goldstone}
\end{equation}
The locking vector obeys
\begin{equation}
\begin{split}
    w^{\mathsf T}v_G
    &=
    -\frac{2q_\star}{m_\star}
    m_\star
    +
    2q_\star
\\
    &=0.
\end{split}
    \label{eq:app-w-vG-zero}
\end{equation}
Thus the new odd quadratic form also annihilates the Goldstone direction.

Since $O$ is symmetric and its support in the two-dimensional odd sector is
orthogonal to $v_G$, it may be written
\begin{equation}
    O
    =
    \gamma
    ww^{\mathsf T},
    \label{eq:app-O-gamma}
\end{equation}
where, with the chosen parametrization
$w=(-\eta,1)^{\mathsf T}$,
\begin{equation}
    \gamma=O_{22}.
    \label{eq:app-gamma-def}
\end{equation}
The locked odd Hessian is therefore
\begin{equation}
    \boxed{
    K_o
    =
    O+\lambda_2ww^{\mathsf T}
    =
    (\gamma+\lambda_2)
    ww^{\mathsf T}.
    }
    \label{eq:app-locked-Ko}
\end{equation}

The Goldstone eigenvalue remains exactly zero:
\begin{equation}
    K_o v_G=0.
\end{equation}
The single non-Goldstone odd eigenvalue is
\begin{equation}
    \boxed{
    k_o^+
    =
    (\gamma+\lambda_2)
    \left(
        1+\eta^2
    \right).
    }
    \label{eq:app-locked-odd-eigenvalue}
\end{equation}
Hence strict stability of the relative-phase direction requires
\begin{equation}
    \boxed{
    \lambda_2>-\gamma.
    }
    \label{eq:app-odd-stability}
\end{equation}

At
\begin{equation}
    \lambda_2=-\gamma,
\end{equation}
the relative-phase restoring force vanishes while the Goldstone mode remains
present by symmetry.

\subsection{Complete strict quadratic stability domain}
\label{app:complete-local-domain}

Combining the even and odd conditions gives the complete strict local
stability region around the prescribed ordered orbit.

The even lower-right pivot requires
\begin{equation}
    \lambda_2>-A_{22},
\end{equation}
while the odd mode requires
\begin{equation}
    \lambda_2>-\gamma.
\end{equation}
Therefore
\begin{equation}
    \boxed{
    \lambda_2
    >
    \max\{-A_{22},-\gamma\}.
    }
    \label{eq:app-lambda2-combined}
\end{equation}
For every such $\lambda_2$, the remaining even condition is
\begin{equation}
    \boxed{
    \lambda_1
    >
    \lambda_{1,\mathrm{crit}}(\lambda_2).
    }
    \label{eq:app-full-local-domain-lambda1}
\end{equation}

Together,
\begin{equation}
    \boxed{
    \lambda_2
    >
    \max\{-A_{22},-\gamma\},
    \qquad
    \lambda_1
    >
    \lambda_{1,\mathrm{crit}}(\lambda_2)
    }
    \label{eq:app-complete-local-domain}
\end{equation}
are necessary and sufficient for strict positivity of every
non-Goldstone quadratic mode of the locked moment theory.

The adjective ``quadratic'' is important.  For arbitrary negative locking
couplings, Eq.~\eqref{eq:app-complete-local-domain} establishes local
stability of the chosen saddle but does not imply that this saddle is the
global minimum of the full free energy.  The globally controlled region
needed for the strong-locking construction is described next.

\subsection{Preservation of a global minimum}
\label{app:locking-global}

Suppose that $\rho_\star$ is a global minimum of the unmodified free-energy
functional:
\begin{equation}
    f_0[\rho]
    \geq
    f_0[\rho_\star]
    \qquad
    \text{for all normalized }\rho.
    \label{eq:app-base-global}
\end{equation}
For
\begin{equation}
    \lambda_1\geq0,
    \qquad
    \lambda_2\geq0,
    \label{eq:app-nonnegative-locking}
\end{equation}
each term in Eq.~\eqref{eq:app-locking-potential} is nonnegative:
\begin{equation}
    f_{\rm lock}[\rho]\geq0.
\end{equation}
At the target,
\begin{equation}
    f_{\rm lock}[\rho_\star]=0.
\end{equation}
Therefore
\begin{align}
    f[\rho]
    &=
    f_0[\rho]+f_{\rm lock}[\rho]
    \nonumber\\
    &\geq
    f_0[\rho]
    \nonumber\\
    &\geq
    f_0[\rho_\star]
    \nonumber\\
    &=
    f[\rho_\star].
\end{align}
Hence
\begin{equation}
    \boxed{
    \rho_\star
    \text{ global minimum of }f_0
    \quad\Longrightarrow\quad
    \rho_\star
    \text{ global minimum of }f
    }
    \label{eq:app-global-theorem}
\end{equation}
for every
\begin{equation}
    \lambda_1,\lambda_2\geq0.
\end{equation}

The argument does not require uniqueness of the minimum.  Additional
configurations may also minimize the locked theory if they simultaneously
saturate the appropriate constraints and have the same base free energy.
What is guaranteed is that increasing nonnegative locking couplings cannot
raise the prescribed ordered state above some competing configuration.

If the base state is also a strict local minimum,
\begin{equation}
    A>0,
    \qquad
    \gamma>0,
    \label{eq:app-base-strict}
\end{equation}
then the local statement follows independently from the Hessian:
\begin{equation}
    \lambda_1m_\star^2
    e_1e_1^{\mathsf T}
    \succeq0,
    \qquad
    \lambda_2ww^{\mathsf T}
    \succeq0.
\end{equation}
Thus
\begin{equation}
    \boxed{
    \lambda_1,\lambda_2\geq0
    }
\end{equation}
is a guaranteed stable quadrant, both locally and, when
Eq.~\eqref{eq:app-base-global} holds, globally.

\subsection{Strong-locking trajectory}
\label{app:strong-locking}

We now consider the one-parameter family
\begin{equation}
    \boxed{
    \lambda_1=\alpha_1\lambda,
    \qquad
    \lambda_2=\alpha_2\lambda,
    \qquad
    \alpha_1>0,
    \quad
    \alpha_2>0,
    }
    \label{eq:app-locking-ray}
\end{equation}
with
\begin{equation}
    \lambda\longrightarrow\infty.
\end{equation}
This trajectory lies entirely inside the guaranteed nonnegative quadrant.

By Eq.~\eqref{eq:app-locking-fixed}, the target state remains exactly fixed
for every $\lambda$:
\begin{equation}
    \boxed{
    m=m_\star,
    \qquad
    q=q_\star,
    \qquad
    T=T_\star,
    \qquad
    r_{\rm eff}=r_{\rm eff}(T_\star).
    }
    \label{eq:app-fixed-along-locking-ray}
\end{equation}

The even Hessian has the asymptotic form
\begin{equation}
    K_e
    =
    \lambda L_e+A,
    \label{eq:app-Ke-large-lambda}
\end{equation}
where
\begin{equation}
    \boxed{
    L_e
    =
    \alpha_1 m_\star^2
    e_1e_1^{\mathsf T}
    +
    \alpha_2
    ww^{\mathsf T}.
    }
    \label{eq:app-Le}
\end{equation}
Explicitly,
\begin{equation}
    L_e
    =
    \begin{pmatrix}
        \alpha_1 m_\star^2+\alpha_2\eta^2
        &
        -\alpha_2\eta
        \\[1mm]
        -\alpha_2\eta
        &
        \alpha_2
    \end{pmatrix}.
    \label{eq:app-Le-explicit}
\end{equation}

Its determinant is
\begin{equation}
\begin{split}
    \det L_e
    &=
    \alpha_2
    \left(
        \alpha_1 m_\star^2+\alpha_2\eta^2
    \right)
    -
    \alpha_2^2\eta^2
\\
    &=
    \boxed{
    \alpha_1\alpha_2 m_\star^2
    }>0,
\end{split}
    \label{eq:app-Le-det}
\end{equation}
while
\begin{equation}
    \Tr L_e
    =
    \alpha_1 m_\star^2
    +
    \alpha_2
    \left(
        1+\eta^2
    \right)
    >0.
    \label{eq:app-Le-trace}
\end{equation}
Thus
\begin{equation}
    \boxed{
    L_e>0.
    }
    \label{eq:app-Le-positive}
\end{equation}

Let
\begin{equation}
    \Sigma_L
    =
    \alpha_1 m_\star^2
    +
    \alpha_2
    \left(
        1+\eta^2
    \right).
    \label{eq:app-mathcalT}
\end{equation}
The two eigenvalues of $L_e$ are
\begin{equation}
    \boxed{
    \ell_\pm
    =
    \frac12
    \left[
        \Sigma_L
        \pm
        \sqrt{
            \Sigma_L^2
            -
            4\alpha_1\alpha_2 m_\star^2
        }
    \right].
    }
    \label{eq:app-Le-eigenvalues}
\end{equation}
Since
\begin{equation}
    \alpha_1\alpha_2 m_\star^2>0,
\end{equation}
both satisfy
\begin{equation}
    0<\ell_-\leq\ell_+.
\end{equation}

Standard eigenvalue perturbation then gives
\begin{equation}
    \boxed{
    k_{e,\pm}
    =
    \lambda\ell_\pm
    +
    O(1).
    }
    \label{eq:app-even-large-lambda}
\end{equation}
In particular,
\begin{equation}
    k_{e,\min}
    =
    \lambda\ell_-+O(1)
    \longrightarrow\infty.
    \label{eq:app-even-min-diverges}
\end{equation}

The odd sector follows directly from
Eq.~\eqref{eq:app-locked-Ko}:
\begin{equation}
    K_o
    =
    \left(
        \gamma+\alpha_2\lambda
    \right)
    ww^{\mathsf T}.
\end{equation}
Its Goldstone eigenvalue remains zero, while the physical relative-phase
stiffness is
\begin{equation}
\begin{split}
    k_o^+
    &=
    \left(
        \gamma+\alpha_2\lambda
    \right)
    \|w\|^2
\\
    &=
    \left(
        \gamma+\alpha_2\lambda
    \right)
    \left(
        1+\eta^2
    \right).
\end{split}
\end{equation}
Hence
\begin{equation}
    \boxed{
    k_o^+
    =
    \alpha_2
    \left(
        1+\eta^2
    \right)
    \lambda
    +
    O(1).
    }
    \label{eq:app-odd-large-lambda}
\end{equation}

All three non-Goldstone restoring coefficients therefore scale linearly
with the locking parameter:
\begin{equation}
    \boxed{
    k_{\perp,\alpha}
    =
    \Theta(\lambda),
    \qquad
    \lambda\to\infty.
    }
    \label{eq:app-all-stiffness-large}
\end{equation}

Together with
Eq.~\eqref{eq:app-fixed-along-locking-ray}, this proves the static
parametric-separation result:
\begin{equation}
    \boxed{
    \begin{gathered}
        T_\star=\mathrm{constant},
        \qquad
        r_{\rm eff}(T_\star)=\mathrm{constant},
        \\[1mm]
        k_{\perp,\alpha}
        \longrightarrow\infty
        \qquad
        \text{for every non-Goldstone mode}.
    \end{gathered}
    }
    \label{eq:app-static-parametric-theorem}
\end{equation}

The result is stronger than finite independence of the Hessian from
$r_{\rm eff}$.  The same equilibrium tensor can belong to a one-parameter
family whose complete normal restoring spectrum becomes arbitrarily stiff.

\subsection{Strongly polarized benchmark}
\label{app:locking-benchmark}

For the numerical benchmark in
Sec.~\ref{subsec:locking-benchmark}, choose
\begin{equation}
    m_\star=0.99
\end{equation}
on the one-harmonic base branch
\begin{equation}
    \kappa_2=0.
\end{equation}
The saddle equation
\begin{equation}
    m_\star
    =
    \frac{
        I_1(a_\star)
    }{
        I_0(a_\star)
    }
\end{equation}
gives
\begin{equation}
    a_\star
    \simeq
    50.2538474.
    \label{eq:app-benchmark-a}
\end{equation}
Therefore
\begin{equation}
    \kappa_1
    =
    \frac{
        a_\star
    }{
        m_\star
    }
    \simeq
    50.7614620.
    \label{eq:app-benchmark-kappa}
\end{equation}

The second harmonic of the same equilibrium distribution is
\begin{equation}
    q_\star
    =
    \frac{
        I_2(a_\star)
    }{
        I_0(a_\star)
    }
    \simeq
    0.960600032,
    \label{eq:app-benchmark-q}
\end{equation}
so
\begin{equation}
    c_\star
    =
    \frac{
        q_\star
    }{
        m_\star^2
    }
    \simeq
    0.980104103
    \label{eq:app-benchmark-cstar}
\end{equation}
and
\begin{equation}
    \eta
    =
    \frac{
        2q_\star
    }{
        m_\star
    }
    \simeq
    1.94060612.
    \label{eq:app-benchmark-eta}
\end{equation}

The effective rank is
\begin{equation}
    \boxed{
    r_{\rm eff}
    =
    \frac{
        2
    }{
        1+0.99^2
    }
    \simeq
    1.01005000.
    }
    \label{eq:app-benchmark-reff}
\end{equation}

For the symmetric locking ray
\begin{equation}
    \alpha_1=\alpha_2=1,
    \qquad
    \lambda_1=\lambda_2=\lambda,
\end{equation}
the leading even matrix is
\begin{equation}
    L_e
    =
    \begin{pmatrix}
        m_\star^2+\eta^2
        &
        -\eta
        \\
        -\eta
        &
        1
    \end{pmatrix}.
\end{equation}
Its eigenvalues are
\begin{equation}
    \boxed{
    \ell_-
    \simeq
    0.175958,
    \qquad
    \ell_+
    \simeq
    5.570095.
    }
    \label{eq:app-benchmark-even-eigenvalues}
\end{equation}
The odd coefficient is
\begin{equation}
    \boxed{
    1+\eta^2
    \simeq
    4.765952.
    }
    \label{eq:app-benchmark-odd-coefficient}
\end{equation}
Thus
\begin{align}
    k_{e,-}
    &=
    0.175958\,\lambda
    +
    O(1),
    \\
    k_{e,+}
    &=
    5.570095\,\lambda
    +
    O(1),
    \\
    k_o^+
    &=
    4.765952\,\lambda
    +
    O(1).
    \label{eq:app-benchmark-stiffnesses}
\end{align}

In particular,
\begin{equation}
    \boxed{
    k_{\perp,\min}
    =
    0.175958\,\lambda
    +
    O(1),
    }
    \label{eq:app-benchmark-min}
\end{equation}
while Eq.~\eqref{eq:app-benchmark-reff} remains unchanged for every
$\lambda$.

This benchmark illustrates the central point in an already strongly
polarized regime: increasing the locking scale does not further reduce the
effective rank.  It changes an independent property of the ordered state,
namely the restoring cost of leaving its low-dimensional sector.

\section{Dynamical response and infrared suppression}
\label{app:dynamics}

This appendix supplies the dynamical derivations used in
Sec.~\ref{sec:dynamics}.  Starting from a simple collective kinetic
realization, we identify the physical transverse normal frequencies, derive
the normal forcing induced by motion along the ordered orbit, and establish
the strong-locking scaling of the corresponding response.  We then prove a
uniform bound on the normal susceptibility over an infrared frequency
window and derive the associated suppression of equilibrium fluctuations.

Throughout this appendix, $s$ denotes the microscopic evolution parameter
of the underlying dynamical theory,
\begin{equation}
    \dot X
    \equiv
    \frac{dX}{ds}.
\end{equation}
No identification of $s$ with an emergent clock variable is assumed.

\subsection{Collective kinetic structure}
\label{app:collective-kinetics}

A simple collective kinetic realization for the first two complex moments is
\begin{equation}
    \boxed{
    L_{\rm kin}
    =
    \frac{N}{2}
    \left[
        \mu_1|\dot M_1|^2
        +
        \mu_2|\dot M_2|^2
    \right],
    }
    \label{eq:app-collective-kinetic}
\end{equation}
with
\begin{equation}
    \mu_1>0,
    \qquad
    \mu_2>0.
\end{equation}
Writing
\begin{equation}
    M_1
    =
    m e^{i\psi},
    \qquad
    M_2
    =
    q e^{i(2\psi+\chi)},
    \label{eq:app-dynamical-polar}
\end{equation}
gives
\begin{equation}
\begin{split}
    L_{\rm kin}
    =
    \frac{N}{2}
    \big[
        &\mu_1\dot m^2
        +
        \mu_2\dot q^2
\\
        &+
        \mu_1m^2\dot\psi^2
        +
        \mu_2q^2
        (2\dot\psi+\dot\chi)^2
    \big].
\end{split}
    \label{eq:app-kinetic-expanded}
\end{equation}

On the aligned slice $\chi=0$, the coefficient of the tangent velocity
$\dot\psi^2$ is
\begin{equation}
    \boxed{
    {\cal I}(m,q)
    =
    \mu_1m^2
    +
    4\mu_2q^2.
    }
    \label{eq:app-orientation-inertia}
\end{equation}
We refer to ${\cal I}$ as the orientation inertia.

The phase-sector kinetic matrix in the coordinates $(\psi,\chi)$ is
\begin{equation}
    G_{\rm phase}
    =
    \begin{pmatrix}
        \mu_1m^2+4\mu_2q^2
        &
        2\mu_2q^2
        \\[1mm]
        2\mu_2q^2
        &
        \mu_2q^2
    \end{pmatrix}.
    \label{eq:app-phase-metric}
\end{equation}
Its determinant is
\begin{equation}
    \boxed{
    \det G_{\rm phase}
    =
    \mu_1\mu_2m^2q^2>0
    }
    \label{eq:app-phase-det}
\end{equation}
for $m,q\neq0$.  Thus the Goldstone and relative-phase directions both have
finite kinetic norm.

At a fixed ordered state the tangent and normal phase coordinates may be
orthogonalized with respect to
Eq.~\eqref{eq:app-phase-metric}.  For example, defining locally
\begin{equation}
    \widetilde\psi
    =
    \psi
    +
    \frac{
        2\mu_2q_\star^2
    }{
        {\cal I}_\star
    }
    \chi
    \label{eq:app-psi-orthogonal}
\end{equation}
eliminates the quadratic
$\dot{\widetilde\psi}\dot\chi$ term at the reference state.  The resulting
relative-phase inertia is
\begin{equation}
    \boxed{
    \mu_\chi
    =
    \frac{
        \mu_1\mu_2m_\star^2q_\star^2
    }{
        {\cal I}_\star
    }
    >0.
    }
    \label{eq:app-relative-phase-inertia}
\end{equation}

This explicit construction illustrates the more general local form used
below.  From this point onward we use the kinetically orthogonal tangent
coordinate $\widetilde\psi$ of Eq.~\eqref{eq:app-psi-orthogonal} and,
to avoid clutter, relabel it again as $\psi$.  This diagonalizes the kinetic
quadratic form at the reference state; it does not remove normal components
of an external force when that force is specified in the original phase
coordinates.  Those components are included in $F_\perp^{\rm ext}$ below.
After choosing coordinates adapted to the ordered orbit, let
\begin{equation}
    X^A
    =
    (\psi,y^\alpha),
\end{equation}
where $y^\alpha$ label the three non-Goldstone collective directions.  To
quadratic order in normal velocities,
\begin{equation}
    \boxed{
    L_{\rm coll}
    =
    \frac{N}{2}
    \left[
        {\cal I}(y)\dot\psi^2
        +
        \dot y^{\mathsf T}
        M_\perp
        \dot y
    \right]
    -
    N f(y)
    +
    \cdots ,
    }
    \label{eq:app-general-Lcoll}
\end{equation}
where
\begin{equation}
    M_\perp>0.
\end{equation}
The omitted terms contain higher powers of the normal coordinates or
velocities.  Dissipation may be included through the Rayleigh function
\begin{equation}
    {\cal R}
    =
    \frac{N}{2}
    \dot y^{\mathsf T}
    \Gamma_\perp
    \dot y,
    \qquad
    \Gamma_\perp\succeq0.
    \label{eq:app-Rayleigh}
\end{equation}

Unlike the statistical metrics studied in Appendix~\ref{app:geometry},
$M_\perp$ and ${\cal I}$ belong to the specified physical kinetic theory.

\subsection{Normal frequencies}
\label{app:normal-frequencies}

Expand the free energy about the target orbit:
\begin{equation}
    f(y)
    =
    f_\star
    +
    \frac12
    y^{\mathsf T}K_\perp y
    +
    O(y^3).
    \label{eq:app-normal-f}
\end{equation}
The Goldstone coordinate $\psi$ has already been separated, so
$K_\perp$ acts only on the three non-Goldstone directions.

In the absence of damping and external forcing, the linearized normal
equation is
\begin{equation}
    M_\perp\ddot y
    +
    K_\perp y
    =
    0.
    \label{eq:app-free-normal-eom}
\end{equation}
Seeking
\begin{equation}
    y(s)
    =
    e_\alpha e^{i\omega s}
\end{equation}
gives the generalized eigenvalue problem
\begin{equation}
    \boxed{
    K_\perp e_\alpha
    =
    \omega_{\perp,\alpha}^2
    M_\perp e_\alpha.
    }
    \label{eq:app-generalized-normal-eigen}
\end{equation}

Equivalently, define the mass-normalized stiffness
\begin{equation}
    \widehat K_\perp
    =
    M_\perp^{-1/2}
    K_\perp
    M_\perp^{-1/2}.
    \label{eq:app-Khat}
\end{equation}
Its ordinary eigenvalues are
\begin{equation}
    \omega_{\perp,\alpha}^2.
\end{equation}
We define the transverse gap by
\begin{equation}
    \boxed{
    \Delta_\perp
    =
    \min_\alpha
    \omega_{\perp,\alpha}.
    }
    \label{eq:app-transverse-gap}
\end{equation}

Along the strong-locking trajectory of
Appendix~\ref{app:stability},
\begin{equation}
    K_\perp(\lambda)
    =
    \lambda L_\perp
    +
    O(1),
    \qquad
    L_\perp>0.
    \label{eq:app-Kperp-large}
\end{equation}
Assume that
\begin{equation}
    M_\perp(\lambda)
    \longrightarrow
    M_\perp^{(\infty)}>0.
    \label{eq:app-Mperp-limit}
\end{equation}
Then
\begin{equation}
    \widehat K_\perp
    =
    \lambda\widehat L_\perp
    +
    O(1),
    \label{eq:app-Khat-large}
\end{equation}
where
\begin{equation}
    \widehat L_\perp
    =
    \left(
        M_\perp^{(\infty)}
    \right)^{-1/2}
    L_\perp
    \left(
        M_\perp^{(\infty)}
    \right)^{-1/2}
    >0.
    \label{eq:app-Lhat}
\end{equation}

Let the eigenvalues of $\widehat L_\perp$ be
\begin{equation}
    \nu_\alpha>0.
\end{equation}
Then
\begin{equation}
    \boxed{
    \omega_{\perp,\alpha}^2
    =
    \lambda\nu_\alpha
    +
    O(1),
    }
    \label{eq:app-normal-frequency-scaling}
\end{equation}
and therefore
\begin{equation}
    \boxed{
    \Delta_\perp
    =
    \Theta(\lambda^{1/2}).
    }
    \label{eq:app-gap-scaling}
\end{equation}

Together with the fixed-polarization theorem of
Appendix~\ref{app:stability},
\begin{equation}
    \boxed{
    r_{\rm eff}
    =
    \mathrm{constant},
    \qquad
    \Delta_\perp
    \longrightarrow\infty.
    }
    \label{eq:app-rank-gap-independence}
\end{equation}

\subsection{Tangential motion as a normal force}
\label{app:tangent-force}

The kinetic coefficient of the Goldstone direction generally depends on the
normal coordinates.  Expand
\begin{equation}
    {\cal I}(y)
    =
    {\cal I}_\star
    +
    g^{\mathsf T}y
    +
    O(y^2),
    \label{eq:app-I-linear}
\end{equation}
where
\begin{equation}
    \boxed{
    g_\alpha
    =
    \left.
    \frac{\partial{\cal I}}
         {\partial y^\alpha}
    \right|_\star.
    }
    \label{eq:app-g-def}
\end{equation}

The Euler--Lagrange equation following from
Eq.~\eqref{eq:app-general-Lcoll}, including Rayleigh dissipation and an
external normal force $F_\perp^{\rm ext}$, is
\begin{equation}
    \boxed{
    M_\perp\ddot y
    +
    \Gamma_\perp\dot y
    +
    K_\perp y
    =
    \frac12
    g\,\dot\psi^2
    +
    F_\perp^{\rm ext}
    +
    \cdots .
    }
    \label{eq:app-normal-eom}
\end{equation}

The term
\begin{equation}
    \frac12g\,\dot\psi^2
\end{equation}
is the physical normal force induced by tangent motion.  It may be viewed as
a collective centrifugal force associated with variation of the orientation
inertia across the normal directions.

For the explicit kinetic ansatz
Eq.~\eqref{eq:app-collective-kinetic},
\begin{equation}
    {\cal I}(m,q)
    =
    \mu_1m^2+4\mu_2q^2,
\end{equation}
so in the even $(m,q)$ sector
\begin{equation}
    \boxed{
    g_e
    =
    \begin{pmatrix}
        2\mu_1m_\star\\
        8\mu_2q_\star
    \end{pmatrix}.
    }
    \label{eq:app-explicit-g}
\end{equation}
The centrifugal term has no linear relative-phase component in these adapted
coordinates.  A generic external phase drive need not share this property.
With
\begin{equation}
    c_{\rm kin}
    =
    \frac{2\mu_2q_\star^2}{{\cal I}_\star},
    \label{eq:app-ckin}
\end{equation}
the local transformation
$\widetilde\psi=\psi_1+c_{\rm kin}\chi$ implies
\begin{equation}
    \delta W
    =
    \tau\,\delta\widetilde\psi
    -
    c_{\rm kin}\tau\,\delta\chi,
\end{equation}
for a generalized torque $\tau$ coupled to the original $M_1$ phase.
Therefore
\begin{equation}
    \boxed{
    F_\chi^{\rm ext}
    =
    -c_{\rm kin}\tau .
    }
    \label{eq:app-phase-drive-force}
\end{equation}
Only a force matched to rigid motion along the kinetic Goldstone direction
has no such normal projection.

Equation~\eqref{eq:app-normal-eom} is the dynamical replacement for a
metric-dependent extrinsic-curvature criterion.  Both its forcing and its
restoring operator are properties of the physical collective theory.

\subsection{Steady adiabatic leakage}
\label{app:steady-leakage}

Consider uniform motion along the orientation orbit,
\begin{equation}
    \dot\psi=\Omega,
    \qquad
    \ddot\psi=0,
    \label{eq:app-uniform-orientation}
\end{equation}
maintained, if necessary, by an external tangent torque.

At leading order in the normal displacement,
Eq.~\eqref{eq:app-normal-eom} has the stationary solution
\begin{equation}
    \boxed{
    y_{\rm leak}
    =
    \frac{\Omega^2}{2}
    K_\perp^{-1}g.
    }
    \label{eq:app-static-leakage}
\end{equation}
In a low-frequency expansion this is the leading adiabatic response.

From
\begin{equation}
    K_\perp
    =
    \lambda L_\perp+O(1)
\end{equation}
with $L_\perp>0$,
\begin{equation}
    K_\perp^{-1}
    =
    \frac1\lambda
    L_\perp^{-1}
    +
    O(\lambda^{-2}).
    \label{eq:app-Kinverse}
\end{equation}
Therefore, for fixed $\Omega$,
\begin{equation}
    \boxed{
    y_{\rm leak}
    =
    O\!\left(
        \frac{\Omega^2}{\lambda}
    \right).
    }
    \label{eq:app-leak-lambda}
\end{equation}
Since
\begin{equation}
    \Delta_\perp^2
    =
    \Theta(\lambda),
\end{equation}
this may equivalently be written
\begin{equation}
    \boxed{
    y_{\rm leak}
    =
    O\!\left(
        \frac{\Omega^2}
             {\Delta_\perp^2}
    \right).
    }
    \label{eq:app-leak-gap}
\end{equation}

Thus the same strong-locking limit that raises the normal gap suppresses the
actual displacement caused by motion along the ordered direction.

\subsection{Adiabatic elimination and the tangent effective theory}
\label{app:adiabatic-elimination}

The same result may be expressed directly at the level of the infrared
effective action.

To leading nontrivial order, the part of
Eq.~\eqref{eq:app-general-Lcoll} involving the normal coordinates is
\begin{equation}
    L_y
    =
    N
    \left[
        \frac12
        g^{\mathsf T}y\,\dot\psi^2
        -
        \frac12
        y^{\mathsf T}K_\perp y
    \right]
    +
    \cdots ,
    \label{eq:app-Ly-adiabatic}
\end{equation}
where normal derivative terms have been omitted in the adiabatic limit.

The algebraic equation of motion is
\begin{equation}
    y
    =
    \frac12
    K_\perp^{-1}g\,\dot\psi^2.
    \label{eq:app-y-integrated}
\end{equation}
Substituting back into
Eq.~\eqref{eq:app-Ly-adiabatic} gives
\begin{equation}
\begin{split}
    L_{\rm eff}
    =
    &
    \frac{N}{2}
    {\cal I}_\star
    \dot\psi^2
\\
    &+
    \frac{N}{8}
    \left(
        g^{\mathsf T}
        K_\perp^{-1}
        g
    \right)
    \dot\psi^4
    +
    \cdots .
\end{split}
    \label{eq:app-effective-tangent-L}
\end{equation}
Hence
\begin{equation}
    \boxed{
    g^{\mathsf T}
    K_\perp^{-1}
    g
    =
    O(\lambda^{-1}).
    }
    \label{eq:app-effective-correction-scaling}
\end{equation}

The leading correction generated by virtual displacement of the normal
sector therefore vanishes as the locking scale increases:
\begin{equation}
    \boxed{
    \Delta L_{\rm eff}
    =
    O(\lambda^{-1})
    \dot\psi^4.
    }
    \label{eq:app-EFT-correction}
\end{equation}

This provides an effective-theory formulation of dynamical autonomy.
Integrating out the heavy transverse collective modes leaves a tangent
theory whose induced nonlinear corrections are parametrically suppressed.

\subsection{Frequency-dependent normal response}
\label{app:frequency-response}

For a general harmonic normal force
\begin{equation}
    F_\perp(s)
    =
    \Re\!\left[
        F_\nu e^{i\nu s}
    \right],
\end{equation}
seek a response
\begin{equation}
    y(s)
    =
    \Re\!\left[
        y_\nu e^{i\nu s}
    \right].
\end{equation}
Equation~\eqref{eq:app-normal-eom} gives
\begin{equation}
    \left[
        K_\perp
        -
        \nu^2M_\perp
        +
        i\nu\Gamma_\perp
    \right]
    y_\nu
    =
    F_\nu.
\end{equation}
Define the normal susceptibility
\begin{equation}
    \boxed{
    {\cal G}_\perp(\nu)
    =
    \left[
        K_\perp
        -
        \nu^2M_\perp
        +
        i\nu\Gamma_\perp
    \right]^{-1}.
    }
    \label{eq:app-normal-susceptibility}
\end{equation}
Then
\begin{equation}
    y_\nu
    =
    {\cal G}_\perp(\nu)F_\nu.
    \label{eq:app-normal-response}
\end{equation}

For a periodic rigid tangent drive with
$F_\perp^{\rm ext}=0$ in the adapted coordinates,
\begin{equation}
    \psi(s)
    =
    \epsilon\cos\omega s,
    \label{eq:app-periodic-drive}
\end{equation}
one has
\begin{equation}
    \dot\psi^2
    =
    \frac{
        \epsilon^2\omega^2
    }{2}
    \left[
        1-\cos(2\omega s)
    \right].
    \label{eq:app-drive-square}
\end{equation}
The induced normal force therefore contains a static component and a
component at frequency $2\omega$.  The latter is
\begin{equation}
    F_{2\omega}
    =
    -
    \frac{
        \epsilon^2\omega^2
    }{4}
    g,
\end{equation}
and hence
\begin{equation}
    \boxed{
    y_{2\omega}
    =
    -
    \frac{
        \epsilon^2\omega^2
    }{4}
    \left[
        K_\perp
        -
        4\omega^2M_\perp
        +
        2i\omega\Gamma_\perp
    \right]^{-1}
    g.
    }
    \label{eq:app-periodic-response}
\end{equation}

For a generic externally imposed phase drive, there can additionally be a
normal force at the fundamental frequency.  In particular,
Eq.~\eqref{eq:app-phase-drive-force} gives
\begin{equation}
    \chi_\omega
    =
    {\cal G}_\chi(\omega)
    \left(
        -c_{\rm kin}\tau_\omega
    \right).
    \label{eq:app-linear-phase-response}
\end{equation}
If the torque required to impose an inertial phase modulation satisfies
$\tau_\omega=O(\epsilon\omega^2)$, then throughout a fixed infrared window
\begin{equation}
    \chi_\omega
    =
    O\!\left(
        \frac{\epsilon\omega^2}{\lambda}
    \right).
    \label{eq:app-linear-phase-scaling}
\end{equation}
This term can dominate the $O(\epsilon^2)$ centrifugal response at small
amplitude, but it obeys the same inverse-locking suppression.

In the weakly damped regime, enhanced response occurs when
\begin{equation}
    2\omega
    \simeq
    \omega_{\perp,\alpha}.
    \label{eq:app-twoomega-resonance}
\end{equation}
The tangent sector is therefore not exactly closed at arbitrary frequency.
Its autonomy is an infrared property controlled by the transverse gap.

\subsection{Uniform infrared susceptibility bound}
\label{app:uniform-IR-bound}

We now prove the uniform low-frequency estimate used in
Sec.~\ref{subsec:operational-autonomy}.

Introduce mass-normalized variables
\begin{equation}
    \widehat y
    =
    M_\perp^{1/2}y,
    \qquad
    \widehat F
    =
    M_\perp^{-1/2}F,
    \label{eq:app-mass-normalized-vars}
\end{equation}
and define
\begin{align}
    \widehat K_\perp
    &=
    M_\perp^{-1/2}
    K_\perp
    M_\perp^{-1/2},
    \\
    \widehat\Gamma_\perp
    &=
    M_\perp^{-1/2}
    \Gamma_\perp
    M_\perp^{-1/2}.
\end{align}
The mass-normalized susceptibility is
\begin{equation}
    \boxed{
    \widehat{\cal G}_\perp(\nu)
    =
    \left[
        \widehat K_\perp
        -
        \nu^2\mathbbm{1}
        +
        i\nu\widehat\Gamma_\perp
    \right]^{-1}.
    }
    \label{eq:app-Ghat}
\end{equation}

From Eq.~\eqref{eq:app-Khat-large}, let
\begin{equation}
    c_\star
    =
    \lambda_{\min}
    \left(
        \widehat L_\perp
    \right)
    >0.
    \label{eq:app-cstar-dynamical}
\end{equation}
For sufficiently large $\lambda$,
\begin{equation}
    \widehat K_\perp
    \succeq
    \frac{
        c_\star\lambda
    }{2}
    \mathbbm{1}.
    \label{eq:app-Khat-lower-bound}
\end{equation}

Consider first a fixed infrared window
\begin{equation}
    |\nu|
    \leq
    \Omega_{\rm IR},
    \label{eq:app-fixed-window}
\end{equation}
where $\Omega_{\rm IR}$ does not scale with $\lambda$.  For sufficiently
large $\lambda$,
\begin{equation}
    \widehat K_\perp
    -
    \nu^2\mathbbm{1}
    \succeq
    \frac{
        c_\star\lambda
    }{4}
    \mathbbm{1}
    \label{eq:app-real-part-bound}
\end{equation}
uniformly throughout the window.

Let
\begin{equation}
    {\cal A}_\nu
    =
    \widehat K_\perp
    -
    \nu^2\mathbbm{1}
    +
    i\nu\widehat\Gamma_\perp.
\end{equation}
For any complex vector $z$,
\begin{equation}
\begin{split}
    \|{\cal A}_\nu z\|\,\|z\|
    &\geq
    \left|
        z^\dagger{\cal A}_\nu z
    \right|
\\
    &\geq
    \Re
    \left(
        z^\dagger{\cal A}_\nu z
    \right)
\\
    &=
    z^\dagger
    \left(
        \widehat K_\perp
        -
        \nu^2\mathbbm{1}
    \right)
    z
\\
    &\geq
    \frac{
        c_\star\lambda
    }{4}
    \|z\|^2.
\end{split}
    \label{eq:app-resolvent-proof}
\end{equation}
Thus the smallest singular value of ${\cal A}_\nu$ obeys
\begin{equation}
    \sigma_{\min}({\cal A}_\nu)
    \geq
    \frac{
        c_\star\lambda
    }{4},
\end{equation}
and therefore
\begin{equation}
    \boxed{
    \sup_{
        |\nu|\leq\Omega_{\rm IR}
    }
    \left\|
        \widehat{\cal G}_\perp(\nu)
    \right\|
    \leq
    \frac{
        4
    }{
        c_\star\lambda
    }
    =
    O(\lambda^{-1}).
    }
    \label{eq:app-uniform-fixed-window}
\end{equation}

No small-damping approximation is required for this bound; positive
semidefinite damping cannot reduce the positive real part supplied by the
restoring operator.

More generally, the same conclusion holds for a frequency window
$\Omega_{\rm IR}(\lambda)$ that grows with $\lambda$ provided
\begin{equation}
    \frac{
        \Omega_{\rm IR}(\lambda)
    }{
        \Delta_\perp(\lambda)
    }
    \longrightarrow0.
    \label{eq:app-growing-window-condition}
\end{equation}
Equivalently,
\begin{equation}
    \Omega_{\rm IR}^2=o(\lambda).
\end{equation}
Then
\begin{equation}
    \boxed{
    \sup_{
        |\nu|\leq\Omega_{\rm IR}(\lambda)
    }
    \left\|
        \widehat{\cal G}_\perp(\nu)
    \right\|
    =
    O(\lambda^{-1}).
    }
    \label{eq:app-uniform-growing-window}
\end{equation}

This is the strongest low-energy autonomy statement obtained in the present
model: the entire transverse response operator vanishes uniformly over any
frequency window that remains parametrically below the normal gap.

\subsection{Application to tangent-induced leakage}
\label{app:uniform-tangent-leakage}

For the periodic tangent drive of
Eq.~\eqref{eq:app-periodic-drive}, the mass-normalized amplitude of the
oscillatory normal response satisfies
\begin{equation}
\begin{split}
    \|\widehat y_{2\omega}\|
    &\leq
    \frac{
        \epsilon^2\omega^2
    }{4}
    \left\|
        \widehat{\cal G}_\perp(2\omega)
    \right\|
    \|\widehat g\|,
\end{split}
    \label{eq:app-periodic-bound}
\end{equation}
where
\begin{equation}
    \widehat g
    =
    M_\perp^{-1/2}g.
\end{equation}
For any fixed low-frequency window,
Eq.~\eqref{eq:app-uniform-fixed-window} therefore gives
\begin{equation}
    \boxed{
    \|\widehat y_{2\omega}\|
    =
    O\!\left(
        \frac{
            \epsilon^2\omega^2
        }{
            \lambda
        }
    \right).
    }
    \label{eq:app-periodic-leak-scaling}
\end{equation}

The suppression is thus not confined to exactly steady motion.  It holds
uniformly for finite-frequency tangent driving as long as the drive remains
well below the moving transverse resonances.

\subsection{Equilibrium normal fluctuations}
\label{app:normal-fluctuations}

The strong-locking limit also suppresses equilibrium fluctuations away from
the ordered orbit.

Because the free energy per rotor is dimensionless, the large-$N$
equilibrium probability near the minimum is
\begin{equation}
    P(y)
    \propto
    \exp\!
    \left[
        -\frac{N}{2}
        y^{\mathsf T}
        K_\perp
        y
    \right].
    \label{eq:app-Gaussian-P}
\end{equation}
The Gaussian covariance is therefore
\begin{equation}
    \boxed{
    \left\langle
        yy^{\mathsf T}
    \right\rangle
    =
    \frac1N
    K_\perp^{-1}.
    }
    \label{eq:app-normal-covariance}
\end{equation}
Using Eq.~\eqref{eq:app-Kinverse},
\begin{equation}
    \boxed{
    \left\langle
        yy^{\mathsf T}
    \right\rangle
    =
    O\!\left(
        \frac{
            1
        }{
            N\lambda
        }
    \right).
    }
    \label{eq:app-covariance-scaling}
\end{equation}
Consequently, a typical root-mean-square normal displacement scales as
\begin{equation}
    \boxed{
    y_{\rm rms}
    =
    O\!\left(
        \frac{
            1
        }{
            \sqrt{N\lambda}
        }
    \right).
    }
    \label{eq:app-rms-scaling}
\end{equation}

The $N$-dependence here has a different status from the Fisher-metric
extensivity discussed in Appendix~\ref{app:geometry}.  In the present case
the suppression arises from the physical Gibbs weight of normal
fluctuations, while the additional factor $\lambda^{-1/2}$ is generated by
the physical locking stiffness.

\subsection{Summary of the dynamical strong-locking limit}
\label{app:dynamical-summary}

Combining Appendices~\ref{app:stability} and
\ref{app:dynamics}, the strong-locking ray satisfies
\begin{equation}
    \boxed{
    \begin{gathered}
        T_\star
        =
        \mathrm{constant},
        \qquad
        r_{\rm eff}(T_\star)
        =
        \mathrm{constant},
        \\[1mm]
        K_\perp
        =
        \lambda L_\perp+O(1),
        \qquad L_\perp>0,
        \\[1mm]
        \Delta_\perp
        =
        \Theta(\lambda^{1/2}),
        \\[1mm]
        K_\perp^{-1}
        =
        O(\lambda^{-1}),
        \\[1mm]
        \sup_{
            |\nu|\leq\Omega_{\rm IR}
        }
        \left\|
            \widehat{\cal G}_\perp(\nu)
        \right\|
        =
        O(\lambda^{-1}),
        \\[1mm]
        y_{\rm leak}
        =
        O(\lambda^{-1})
        \quad
        \text{for fixed low-frequency driving}.
    \end{gathered}
    }
    \label{eq:app-dynamical-summary}
\end{equation}

The hierarchy of implications is therefore
\begin{equation}
    \boxed{
    \text{positive locking}
    \ \Longrightarrow\
    K_\perp\uparrow
    \ \Longrightarrow\
    \Delta_\perp\uparrow
    \ \Longrightarrow\
    {\cal G}_\perp\downarrow .
    }
    \label{eq:app-lock-gap-response}
\end{equation}

The first arrow is the equilibrium theorem of
Appendix~\ref{app:stability}; the remaining arrows require the physical
kinetic structure analyzed here.

The transverse modes have not disappeared from the microscopic theory.
Rather, their characteristic frequencies move to parametrically high scales
and their response to infrared forcing vanishes.  This is the sense in which
the selected collective direction becomes dynamically autonomous at low
energy.






\bibliographystyle{JHEP}
	\cleardoublepage

\renewcommand*{\bibname}{References}

\bibliography{references}
\end{document}